%% file: ecal_ep.tex
\documentclass[12pt]{article}  
\usepackage{epsfig,subfigure}

\usepackage{pstricks}
\usepackage{epsfig}
\usepackage{pifont}
\usepackage{amssymb}
\usepackage{amsmath}
\usepackage{mathrsfs}
\usepackage{rotating}
\usepackage{longtable}
\usepackage{cite}
\usepackage{varioref}
\usepackage{times}
\usepackage{calrsfs}
\usepackage{a4}
\input{./defs.tex}
\newcommand{\bit}{\begin{itemize}}
\newcommand{\eit}{\end{itemize}}
\newcommand{\beq}{\begin{equation}}
\newcommand{\eeq}{\end{equation}}

\begin{document}

\begin{titlepage}
{\center\large
EUROPEAN LABORATORY FOR PARTICLE PHYSICS \\
}
\vspace*{2mm}
\begin{flushright}
CERN-PH-EP-2004-032 \\
CERN-AB-2004-030 OP \\
27 July 2004 \\

\end{flushright}

\begin{center}
{\Large  \bf 
Calibration of centre-of-mass energies at LEP 2 \\ 
\vspace{0.2cm} for a precise measurement of the W boson mass}

\vspace*{0.4cm}

The LEP Energy Working Group\\

\vspace*{0.8cm}
R.~Assmann$^{ 1}$,
E.~Barbero Soto$^{ 1}$,
D.~Cornuet$^{1}$,
B.~Dehning$^{  1}$, 
M.~Hildreth$^{ 1a}$,
J.~Matheson$^{ 1b}$, 
G.~Mugnai$^{  1}$, 
A.~M\"uller$^{   1c}$, 
E.~Peschardt$^{  1}$, 
M.~Placidi$^{  1}$,
J.~Prochnow$^{ 1}$,
F.~Roncarolo$^{1,2}$, 
P.~Renton$^{  3}$, 
E.~Torrence$^{ 1,4d}$, 
P.~S.~Wells$^{  1}$,
J.~Wenninger$^{  1}$, 
G.~Wilkinson$^{  3}$

\end{center}

\begin{flushleft}
\footnotesize
$^{  1}$CERN, European Organisation for Particle Physics,
CH-1211 Geneva 23, Switzerland \\
$^{  2}$University of Lausanne, CH-1015 Lausanne, Switzerland \\
$^{  3}$Department of Physics,  University of Oxford,  Keble Road,  Oxford
OX1 3RH,  UK\\
$^{  4}$Enrico Fermi Institute and Department of Physics, 
University of Chicago, Chicago IL 60637, USA \\
$^{  a}$Now at: University of Notre Dame, Notre Dame, Indiana 47405, USA \\
$^{  b}$Now at: CCLRC Rutherford Appleton Laboratory, Chilton, Didcot,
Oxfordshire, OX11 OQX, UK  \\
$^{  c}$Now at: ISS, Forschungszentrum Karlsruhe, Karlsruhe, Germany \\
$^{  d}$Now at: University of Oregon, Department of Physics, 
Eugene OR 97403, USA \\
\end{flushleft}

\vspace*{0.8cm}

\begin{abstract}

The determination of the centre-of-mass energies for all LEP~2 running
is presented.   Accurate knowledge of these energies
is of primary importance to set the absolute energy scale
for the measurement of the W boson mass.
The beam energy between 80 and 104~GeV is derived from continuous
measurements of the magnetic bending field by
16 NMR probes situated in a number of the LEP dipoles. 
The relationship between the fields measured
by the probes and the beam energy is defined in
the NMR model, which is calibrated against 
precise measurements of the average 
beam energy between 41 and 61~GeV made using
the resonant depolarisation technique.
The validity of the NMR model is verified by three independent methods:
the flux-loop, which is sensitive to the bending field of all the dipoles
of LEP;  the spectrometer,  which determines the energy
through measurements of the deflection of the beam in a magnet
of known integrated field; and an analysis of the variation of the
synchrotron tune with the total RF voltage.  
To obtain the centre-of-mass energies, corrections are then
applied to account for sources of bending field external to the
dipoles,  and variations in the local beam energy at each interaction point.
The relative error on the centre-of-mass energy determination for the 
majority of LEP~2 running is $1.2 \times 10^{-4}$,  
which is sufficiently precise so as not 
to introduce a dominant uncertainty on the W mass measurement.

\end{abstract}

\vspace{5mm}     

\begin{center}
\large
To be submitted to Eur.\ Phys.\ J.\ C.
\end{center}
\end{titlepage}

\input{introduction}

\vspace*{0.2cm}
\input{lep2_prog}

\input{mag_extrap}

\input{eb_model}

\input{ecm_ip}

\input{fl_ana}

\input{spec_setup}

\newpage
\input{spec_ana}

\input{qs_ana}

\input{eb_comb}

\input{syst_sum}

\input{ecm_spread}

\input{conclusions}

\end{document}

%% file: defs.tex
\def\ifmath#1{\relax\ifmmode #1\else $#1$\fi}%

\renewcommand {\rm} {\mathrm}

\newcommand  {\Zz}   {\ifmath{{\mathrm{Z}}}}

\newcommand {\Mw}   {\ifmath{M_{\mathrm{W}}}}

\newcommand {\ee}    {\mathrm{e}^+\mathrm{e}^-}

\newcommand{\xarc}{\ifmath{x_{\mathrm{arc}}}}
\newcommand  {\Qs}   {\ifmath{Q_{\mathrm{s}}}}

\newcommand{\Eb}{\ifmath{E_{\mathrm{b}}}}
\newcommand{\EbO}{\ifmath{E_{\mathrm{b}}^\rm{d}}}
\newcommand{\EbOm}{\ifmath{E_{\mathrm{b}}^\rm{d\,M}}}

\newcommand{\Ecm}{\ifmath{E_{\mathrm{CM}}}}
\newcommand{\Ecmn}{\ifmath{E_{\mathrm{CM}}^{\mathrm{nom}}}}
\newcommand{\Ecmnmr}{\ifmath{E_{\mathrm{CM}}^{\mathrm{MOD}}}}
\newcommand{\Epol}{\ifmath{E_\mathrm{b}^{\mathrm{RDP}}}}
\newcommand{\Efl}{\ifmath{E_\mathrm{b}^{\mathrm{FL}}}}
\newcommand{\Espect}{\ifmath{E_\mathrm{b}^{\mathrm{\, SPEC}}}}
\newcommand{\Etrue}{\ifmath{E_\mathrm{b}^{\mathrm{\, TRUE}}}}
\newcommand{\Enmr}{\ifmath{E_\mathrm{b}^{\mathrm{MOD}}}}

\newcommand{\Enmri}{\ifmath{E_\mathrm{b}^{\mathrm{MOD}\, i}}}
\newcommand{\EOnmr}{\ifmath{E_\mathrm{b}^{\mathrm{NMR}}}}
\newcommand{\EOnmri}{\ifmath{E_\mathrm{b}^{\mathrm{NMR} \, i}}}
\newcommand{\Eqs}{\ifmath{E_\mathrm{b}^{Q_s}}}
\newcommand{\Ebr}{\ifmath{E_{\mathrm{b}}^{\mathrm{ref}}}}

\newcommand{\Bnmri}{\ifmath{B^{\mathrm{NMR} \, i}}}
\newcommand{\Bfl}{\ifmath{B^{\mathrm{FL}}}}
\newcommand{\BL}{\ifmath{\int B \,\rm{dl}}}
\newcommand{\BLr}{\ifmath{\int B \,\rm{dl^{\, ref}}}}
\newcommand{\Bflp}{\ifmath{B^\mathrm{NMR}_\mathrm{FL}}}
\newcommand{\Bflpi}{\ifmath{B^{\mathrm{NMR}\,i}_\mathrm{FL}}}

\catcode`\"=\active                  
\newcommand"[1]{{\accent127#1}}             

\newcommand{\nin}{\noindent}

\newcommand{\bi}{\begin{itemize}}
\newcommand{\ei}{\end{itemize}}

\newcommand\barr[0]{\begin{array}}
\newcommand\earr[0]{\end{array}}

\newcommand{\microns}{\ifmath{\mathrm{\mu m}}}

\newcommand{\elc}{\ifmath{\mathrm{e}^-}}
\newcommand{\pos}{\ifmath{\mathrm{e}^+}}
\newcommand{\mTRS}{\ifmath{\mathrm{\langle TRS \rangle}}}
\newcommand{\fRFc}{\ifmath{\mathrm{f_c^{RF}}}}
\newcommand{\fRF}{\ifmath{\mathrm{f^{RF}}}}

\newcommand {\um}   {\microns}

\newcommand {\ys} {$\bullet$}

\catcode`@=11 
\def \gsim{\mathrel{\mathpalette\@versim>}}
\def \lsim{\mathrel{\mathpalette\@versim<}}
\def \@versim#1#2{\lower0.4ex\vbox{\baselineskip\z@skip\lineskip\z@skip
     \lineskiplimit\z@\ialign{$\m@th#1\hfil##\hfil$%
     \crcr#2\crcr\sim\crcr}}}
\catcode`@=12 

%% file: introduction.tex
\section{Introduction}

The operation of the large electron-positron (LEP) collider 
in the years 1996 to 2000 (LEP~2) saw the delivery
of almost 700~${\rm{pb^{-1}}}$ of integrated luminosity to each experiment
at $\ee$ collision energies above the W-pair production threshold.
A primary physics
motivation for the LEP~2 programme was the precision 
measurement of the W boson mass,  $\Mw \, \approx \, 80.4~\rm{GeV}/c^{\rm{2}}$.
The centre-of-mass energy, \Ecm, establishes the absolute energy scale
for this measurement, and any uncertainty in this
quantity leads to an uncertainty of 
$\Delta \Mw / \Mw \approx \Delta \Ecm / \Ecm$.   The statistical
precision on the full LEP~2 data set is around 30~MeV~\cite{EWWGREP}.
To avoid a significant contribution to the total error, this sets
a target of $\Delta \Ecm / \Ecm \, = \, 1-2 \times 10^{-4}$.   
This paper reports on the determination of the centre-of-mass
energies for all LEP~2 operation.  The results supersede those in
an earlier publication concerning the 1996 and 1997 LEP runs~\cite{LEP2ECAL}.

In the following section the main concepts which will be used in
the subsequent analysis are introduced,  together with a brief year-by-year
description of LEP~2 operation.  The method of the energy determination
is then presented.

The starting point of the energy determination is a set of 
precise calibrations of the mean beam energy around the ring, \Eb,  
performed with the {\it resonant depolarisation} (RDP) technique at energies
of $41 < \Eb < 61~\rm{GeV}$.   The {\it NMR magnetic model} relates
these calibrations to field measurements
made by NMR probes in selected dipoles.
The model is then used to set the absolute energy scale for
physics running in the regime $81 < \Eb < 104~\rm{GeV}$.   
RDP and the calibration of the 
magnetic model are explained in section~\ref{sec:rdpmag}.
Corrections
are applied to this energy estimate to account for variations with time
in the dipole strength during data-taking, and additional sources of bending
field, such as those arising from non-central
orbits in the quadrupoles.  These corrections are described in 
section~\ref{sec:eb_model}.
The NMR estimate together with these corrections
forms the full {\it \Eb\ model}.

In calculating the centre-of-mass energy at each experimental 
interaction point 
it is necessary to know the local beam energy, which differs significantly
from \Eb\ around the ring due to losses from synchrotron 
radiation and the boosts provided by the RF system.   Other potential
corrections to \Ecm\ come from the correlated effects of dispersion and
collision offsets, and any difference in energy between the electron
and positron beams.  These issues are discussed in section~\ref{sec:ecm_ip}. 

The most important uncertainty in the energy determination is that
associated with the NMR magnetic model.   This error is assigned from
the results of three complementary approaches, which in different
manners attempt to quantify the agreement between the model and the
true energy in the physics regime.

\begin{enumerate}
\item{
The {\it flux-loop}
was a sequence of copper loops which were embedded in the dipole cores and 
connected in series and which sensed the change of flux as the
magnets were ramped.   The number of NMR-equipped dipoles 
used in the magnetic model was limited, 
but comparison with the flux-loop data allows the
representability of this sampling to be assessed.  Flux-loop data
were accumulated in dedicated measurements throughout LEP~2 operation
which can be used to constrain the model, as is explained in 
section~\ref{sec:fltest}.
}
\item{
The {\it spectrometer} was a device installed and commissioned in 1999
and used throughout the 2000 run.  It consisted of a steel dipole with
precisely known integrated field, and triplets of {\it beam-position monitors}
(BPMs) on either side which enabled the beam deflection to be measured,
and thus the energy to be determined.  The spectrometer apparatus
and calibration is outlined in section~\ref{sec:specsetup}, and
the data analysis is presented in section~\ref{sec:specana}.
}
\item{
In a machine such as LEP
the {\it synchrotron tune}, $Q_s$, depends on the beam energy,  the energy loss
per turn, and the total RF voltage, $V_\rm{RF}$.  Since the energy loss
itself depends on the beam energy,  
an analysis of  the variation of $Q_s$  with
$V_\rm{RF}$ can be used to infer \Eb.  Experiments were conducted
in 1998, 1999 and 2000 to exploit this method.  A full description
is given in section~\ref{sec:qs}.
}
\end{enumerate}

The results of the three approaches can be assessed for compatibility.
If consistent,  they may be combined to set both a correction and an
associated uncertainty for the magnetic model. 
Such an analysis is
presented in section~\ref{sec:ebcom}. The resulting uncertainty, together
with the uncertainties from other sources,  is used in section~\ref{sec:systsum}
to assign the total error  on the collision energies.

The spread in the collision energies is relevant in the analysis of
the W boson width.   The understanding of the energy spread is
described in section~\ref{sec:ecmspread}.
The conclusions of the energy analysis can be found 
in section~\ref{sec:conclude}.

%% file: lep2_prog.tex
\section{The LEP Machine and the LEP~2 Programme}

\subsection{LEP Beam Energy and Synchroton Energy Loss}
\label{sec:basiceqn}

The energy, \Eb, of a beam of ultra-relativistic electrons
or positrons in a closed orbit is directly proportional to
the bending field, $B$, integrated around the beam
trajectory, $s$:

\begin{equation}
\Eb\ = \frac{ec}{2\pi} \oint  B \, \mathrm{ds}.
\label{eq:ebdef}
\end{equation}

\noindent For LEP 98\% of the nominal bending field was provided by 
3280 concrete-reinforced dipole magnets, of approximate length 5.8~m
and field of 1070~G at $\Eb = 100~\rm{GeV}$.   The remaining 2\% 
was dominated by steel-cored dipoles in the injection region,
with a small contribution coming from the special weak dipoles
designed to match the arcs to the straight sections.
There were other possible sources of effective dipole field,
such as the quadrupole magnets on the occasions when the mean 
beam trajectory was not centred.
Expression~\ref{eq:ebdef} is assumed in constructing the  NMR magnetic
model and is fundamental to the LEP~2 energy calibration.

As the beams circulate they lose energy through synchrotron
radiation.  The energy loss per turn, $U_0$, is given by:

\begin{equation}
U_0 = \frac{ C_\gamma \, (ec)^2 } {2 \pi} \oint \Eb^2 B^2 \mathrm{ds},
\label{eq:u0def}
\end{equation}

\noindent where the constant
$C_\gamma \equiv e^2/3\epsilon_0(m_e c^2)^3 = 
8.86 \times 10^{-5} \, \rm{(GeV)^{-3}}$.
This relation, together with expression~\ref{eq:ebdef} gives:

\begin{equation}
U_0 = C_\gamma \frac{\Eb^4}{\rho}.
\label{eq:u0def2}
\end{equation}

\noindent Here $\rho$ is the {\it effective bending radius},  
which in the case of 
LEP was approximately 3026~m.  Expression~\ref{eq:u0def2} 
gives an energy loss per turn of
2.9~GeV at beam energies of 100~GeV.

The synchrotron energy loss is replenished by the RF system.   In the LEP~2 era
this consisted of stations of super-conducting cavities situated on
either side of the four experimental interaction points.
The installation of new cavities,  and increases to the field
gradient of the existing klystrons, enabled the voltage of the
RF system to be augmented each year of LEP~2 operation.
Understanding the variation in beam energy around the ring from 
synchrotron losses and RF boosts is an important 
ingredient in the energy model.  Furthermore, the measurement of 
quantities sensitive to the energy loss,  such as the synchrotron tune,
can be used to determine the beam energy itself.

\subsection{LEP~2 Datasets and Operation}
\label{sec:lep2}

The LEP~2 programme began in 1996 when the collision energy of the beams
was first ramped to the $W^+W^-$ production threshold of 161~GeV,
and approximately $10~\rm{pb^{-1}}$ of integrated luminosity
was collected by each experiment.
Later in that year LEP was run at 172~GeV, and a dataset
of similar size was accumulated.
In each of the four subsequent years of operation the 
collision energy was raised to successively higher values,
such that almost half the integrated luminosity was 
delivered at nominal collision energies of 200~GeV and above.
The motivation for this policy was to improve the sensitivity
in the search for the Higgs boson and other new particles.
The step-by-step nature of the energy increase was 
dictated by the evolving capabilities of the
RF system.
The nominal energy points of operation, \Ecmn, are
listed in table~\ref{tab:lep2sum},  together with the
approximate integrated luminosities delivered to
each experiment.

\begin{table}
\begin{center}
\begin{tabular}{|l|c|c|c|c|c|c|c|c|c|c|} \hline
Year                  &
\multicolumn{2}{|c|}{1996} & 
\multicolumn{1}{|c|}{1997} & 
\multicolumn{1}{|c|}{1998} & 
\multicolumn{4}{|c|}{1999} & 
\multicolumn{2}{|c|}{2000} \\ \hline
\Ecmn\  [GeV] &
\multicolumn{1}{|c|}{161} & 
\multicolumn{1}{|c|}{172} & 
\multicolumn{1}{|c|}{183} & 
\multicolumn{1}{|c|}{189} & 
\multicolumn{1}{|c|}{192} & 
\multicolumn{1}{|c|}{196} & 
\multicolumn{1}{|c|}{200} & 
\multicolumn{1}{|c|}{202} & 
\multicolumn{1}{|c|}{205} & 
\multicolumn{1}{|c|}{207} \\  \hline
$\int {\cal{L}}\, dt$  [$\rm{pb^{-1}}$] 
                      &   10 &   10 &   54 &  158 &   26 &   76 &   83 &   41 &   83 &  140 \\ \hline
Physics optics        &
\multicolumn{2}{|c|}{90/60} &
\multicolumn{1}{|c|}{90/60} & 
\multicolumn{1}{|c|}{102/90} & 
\multicolumn{4}{|c|}{102/90} & 
\multicolumn{2}{|c|}{102/90} \\
&
\multicolumn{2}{|c|}{(108/90)} &
\multicolumn{1}{|c|}{(102/90)} &
\multicolumn{1}{|c|}{      } & 
\multicolumn{4}{|c|}{      } & 
\multicolumn{2}{|c|}{      } \\ \hline
Polarisation optics   &
\multicolumn{2}{|c|}{90/60} &
\multicolumn{1}{|c|}{60/60} & 
\multicolumn{1}{|c|}{60/60}  & 
\multicolumn{4}{|c|}{60/60} & 
\multicolumn{2}{|c|}{101/45} \\ 
&
\multicolumn{2}{|c|}{      } &
\multicolumn{1}{|c|}{      } &
\multicolumn{1}{|c|}{      } & 
\multicolumn{4}{|c|}{(101/45)} & 
\multicolumn{2}{|c|}{      } \\ \hline
\end{tabular}
\end{center}
\caption[]{Summary of the LEP~2 running parameters and performance.
Shown for each year are the nominal collision energies;  
the integrated luminosities
collected by a typical experiment;  the choice of optics for
the majority of the 
physics running (`physics optics') and
the preferred optics used for RDP calibration (`polarisation optics').
(Alternative choices of optics used during the run are given
in parentheses.)  The values given for the optics signify the betatron
phase advance in degrees between the focusing quadrupoles 
in the horizontal/vertical planes.}
\label{tab:lep2sum}
\end{table}

During normal operation the machine would be filled
with four electron and four positron bunches
at $\Eb \approx 22~\rm{GeV}$, and the beams would then be
ramped to physics energy, at which point
they would be steered into collision and experimental
data-taking began.   The {\it fill} would last 
until the beam currents fell below a useful level,
or an RF cavity trip precipitated the loss of the beam.
The mean fill lengths ranged from 5 hours in 1996 
to 2 hours in 1999.   After de-gaussing the magnets
the cycle would be repeated.  Following the
experience gained at LEP~1~\cite{LEP1PAPER}, 
{\it bending modulations} were performed in the 
1997--1999 runs prior to
colliding the beams, in which the dipole current
was modulated with a sequence of very small square pulses.
The purpose of this exercise was to condition
the magnets and suppress the effects of parasitic
currents.

\begin{figure}
\begin{center}
\epsfig{file=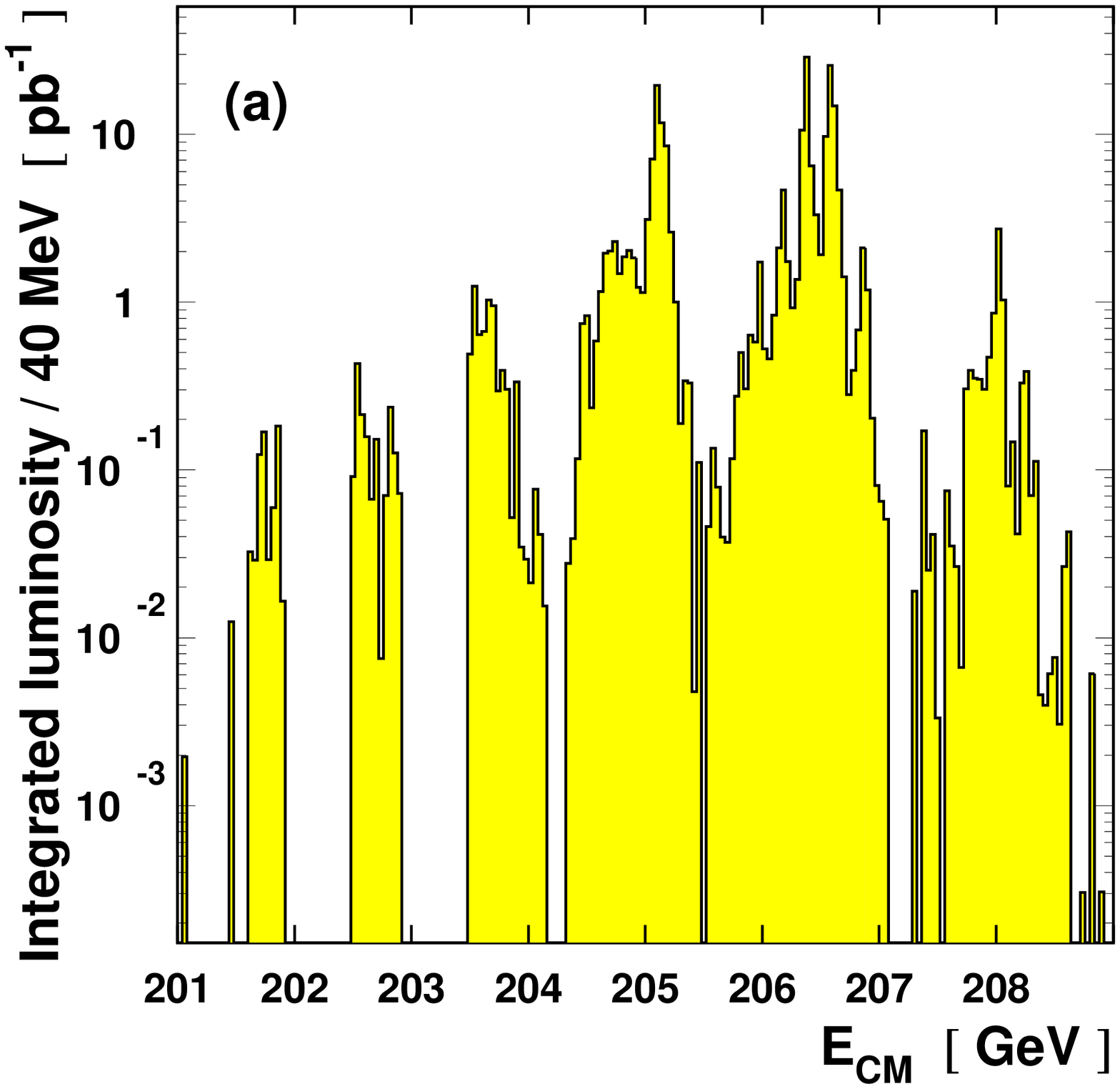,width=0.48\textwidth}
\epsfig{file=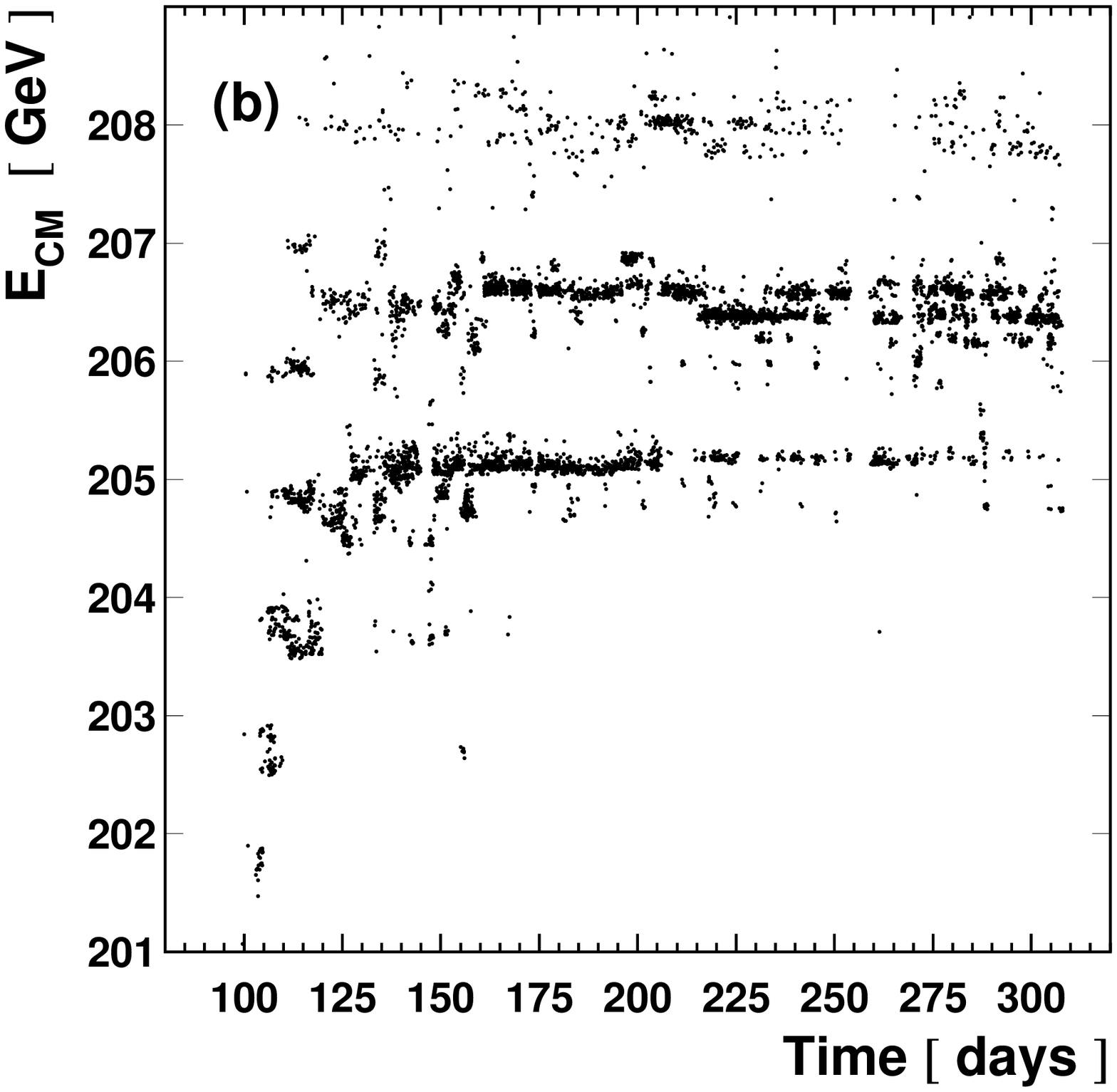,width=0.48\textwidth}
\caption{Distributions of collected luminosity for a single LEP experiment
in 2000.
(a) shows the integrated luminosity in bins of \Ecm.  
(b) shows the variation of \Ecm\ against day of year;
each entry corresponds to the mean energy for a data 
file of maximum length $\sim 30$ minutes.  The values of \Ecm\ have
been calculated using the full energy model.}
\label{fig:ecm_lumi}
\end{center}
\end{figure}

In 2000, the operation was modified in order 
to optimise still further the high-energy reach
of LEP~\cite{MINIRAMP}.
Fills were started at a beam energy safely
within the capabilities of the RF system. 
When the beam currents had decayed significantly,  typically after an hour,
the dipoles were ramped and luminosity was delivered at
a higher energy.  This procedure was repeated until the energy was at 
the limit of the RF, and data taken until the beam was lost through a 
klystron trip.  These {\it miniramps} lasted less than a minute,  and 
varied in step size with a mean value of
600~MeV.   Hardware signals
were used to flag the start and end of miniramps to the experiments,
which continued to take data throughout,
and this information was recorded with the logged triggers.
 The starting energy of fills, and the 
precise strategy of miniramping varied throughout the year,  
depending on the status of the RF system.
The luminosity in 2000 was therefore delivered through a near-continuum of
energies.  The {\it sub-fills} on either side of the miniramps 
can be seen in the `fine structure' of  figure~\ref{fig:ecm_lumi}~(a)
and \ref{fig:ecm_lumi}~(b),
which display the distribution of luminosity both with \Ecm\ and 
time for a single experiment.   The coarser bands in the 
plots arise through the choice of starting energy for the
fill, a decision dependent on the status of the RF system.
The two \Ecmn\ points listed in table~\ref{tab:lep2sum}
refer to the integrated totals delivered below and above an
arbitrary division value of 205.5~GeV.  The lower of these two bins
is dominated by data accumulated in the earlier part of the run.

Another aspect of operation which was unique to 2000, also deployed
to optimise the collision energy within the restrictions of the
available RF voltage,
was the coherent powering of corrector magnets to apply
a so-called {\it bending-field spreading} (BFS) boost.   The BFS is
discussed in section~\ref{sec:hcor}.

In addition to the high-energy running summarised 
in table~\ref{tab:lep2sum}, each year a number of fills
were performed at the \Zz\ resonance.
This was to provide calibration data for the
experiments.  During 1997, some data were also 
collected at nominal
centre-of-mass energies of 130-136~GeV,  to
investigate effects seen during operation at similar energies in 1995.
Finally, several fills were devoted to energy-calibration activities,
most notably RDP,
spectrometer and $Q_s$ measurements.
Most of these energy-calibration experiments were conducted with
single beams, and many of them spanned a variety of energy points.

Included in the information of table~\ref{tab:lep2sum} are the {\it machine
optics} which were used for physics operation (`physics optics') 
and for RDP measurements (`polarisation optics').
The values signify the betatron phase advance in degrees between the focusing 
quadrupoles of the LEP arcs in the horizontal/vertical planes.
The choice of optics evolved throughout the programme
in order to optimise the luminosity at each energy point.
Certain optics enhanced the build-up of polarisation, and thus
were favoured for RDP measurements.
As is explained in section~\ref{sec:eb_model}, the optics influences
\Eb\ in several ways, which must be accounted
for in the energy model.

%% file: mag_extrap.tex
\section{RDP and the NMR Magnetic Model}
\label{sec:rdpmag}

The LEP~2 energy scale is set by the {\it NMR magnetic 
model}.   Between beam energies of 41 and 61~GeV precise
measurements of \Eb\ are provided by {\it resonant depolarisation} 
(RDP).    Also
available are local measurements of the
bending field, made by NMR probes in selected dipoles.  Following 
expression~\ref{eq:ebdef}, and taking the probes to be
representative of the total bending field,
the magnetic model is calibrated through a linear fit 
between the RDP measurements and the NMR readings at
low energy.  Applying this calibrated model at
high-energy fixes \EOnmr, the dipole contribution to the beam energy in
physics operation.  Onto \EOnmr\ must be added corrections
coming from sources of bending field external to the dipoles.

Possible sources of error in the NMR model arise from
the limited sampling of the total bending field
provided by the probes,  and the consequences of 
any non-linearity in
the relationship between the field and \Eb, when
extrapolated up to high energy.

\subsection{RDP Measurements}

The best determination of the beam energy at a particular time is
by means of RDP.   The beam can build up a non-negligible 
transverse polarisation through the Sokolov-Ternov mechanism~\cite{SOKOLOV}.
The degree of polarisation can be measured by the angular distribution of 
Compton-scattered polarised laser light.   By exciting the beam with a 
transverse oscillating magnetic field, this polarisation can
be destroyed when the excitation frequency matches the spin 
precession frequency.  Determining the RDP frequency allows a precise
determination of \Eb\ through:
\begin{equation}
\Eb\ = \Epol\ \equiv \frac{ \nu_s \, . \, m_e \,c^2}{(g_e \, - \, 2)/ 2},
\label{eq:rdpeq}
\end{equation}

\noindent where $\nu_s$ is the `spin-tune', that is the number of electron-spin
precessions per turn,  $m_e$ is the electron mass and $(g_e - 2)/2$ is
the magnetic-moment anomaly of the electron.  
The beam energy measured by RDP is the average around the ring and 
over all particles.
The intrinsic precision
of RDP at the Z resonance is estimated to be 200~keV~\cite{RDPPAPER}.

At LEP~2 physics energies RDP cannot be performed.
The relative spread of the beam energy grows linearly with \Eb\ and 
consequently it becomes increasingly probable that the
beam will encounter machine imperfections which inhibit the
build-up of polarisation.
RDP measurements made at low energies are instead used to
calibrate the NMR model, which is then applied  in the
physics regime.   The systematic uncertainties in this
procedure can be minimised by making the span
of RDP measurements as wide as possible, in particular at
high energy.
Therefore during the LEP~2 programme techniques
were developed to reduce the machine imperfections and 
enhance the polarisation levels 
during RDP calibration.  These included the `k-modulation'
studies to measure the offsets between beam pick-ups
and quadrupole centres~\cite{KMOD}, the improved use
of magnet-position surveys, and the development of
dedicated polarisation optics~\cite{POLOPTICS}.
The maximum energy at which sufficient polarisation
was obtained for a reliable calibration measurement
was 61~GeV.    The time required for
a complete measurement at each energy point
was several hours.

\begin{table}
\begin{center}
\begin{tabular}{|c|r|c|c|c|c|c|r|}
\hline
 Fill  & \multicolumn{1}{|c|}{Date} & 
                      41 GeV & 45 GeV & 50 GeV & 55 GeV & 61 GeV & \multicolumn{1}{|c|}{Optics} \\
\hline
 3599  & 19 Aug '96 &        &        & \ys    &        &        &  90/60 \\
 3702  & 31 Oct '96 &        & \ys    &        &        &        &  90/60 \\
 3719  &  3 Nov '96 &        & \ys    & \ys    &        &        &  90/60 \\
\hline
 4000  & 17 Aug '97 &        & \ys    &        &        &        &  90/60\\
 4121  & 6 Sept '97 &        & \ys    & \ys    &        &        &  60/60\\
 4237  &30 Sept '97 &        & \ys    & \ys    &        &        &  60/60\\
 4242  &  2 Oct '97 & \ys    & \ys    & \ys    & \ys    &        &  60/60\\
 4274  & 10 Oct '97 &        & \ys    &        &        &        &  90/60\\
 4279  & 11 Oct '97 & \ys    & \ys    & \ys    & \ys    &        &  60/60\\
 4372  & 29 Oct '97 & \ys    & \ys    &        &        &        &  60/60 \\
\hline
 4666  &14 June '98 &        & \ys    & \ys    & \ys    &        &  60/60 \\
 4669  &18 June '98 &        & \ys    &        &        &        &  102/90 \\
 4843  &15 July '98 &        & \ys    &        & \ys    &        &  60/60 \\
 5137  & 6 Sept '98 &        & \ys    &        &        &        &  60/60 \\
 5141  & 7 Sept '98 & \ys    & \ys    & \ys    &        &        &  60/60 \\
 5214  &20 Sept '98 & \ys    & \ys    & \ys    & \ys    & \ys    &  60/60 \\
 5232  &29 Sept '98 &        & \ys    &        &        &        &  102/90 \\
 5337  & 18 Oct '98 & \ys    & \ys    & \ys    & \ys    & \ys    &  60/60 \\
\hline
 5670  & 7 June '99 & \ys    & \ys    &        &        &        &  60/60 \\
 5799  &25 June '99 & \ys    & \ys    & \ys    &        &        &  60/60 \\
 5969  &22 July '99 & \ys    &        &        &        &        &  60/60 \\
 5971  &22 July '99 &        & \ys    & \ys    &        &        &  60/60 \\
 6087  &  8 Aug '99 &        &        &        & \ys    &        &  60/60 \\ 
 6302  & 9 Sept '99 & \ys    & \ys    & \ys    & \ys    &        &  60/60 \\ 
 6371  &20 Sept '99 & \ys    & \ys    & \ys    & \ys    &        &  60/60 \\
 6397  &25 Sept '99 &        & \ys    & \ys    &        &        & 101/45 \\
 6404  &26 Sept '99 & \ys    & \ys    & \ys    & \ys    &        &  60/60 \\
 6432  &29 Sept '99 &        & \ys    & \ys    &        &        & 101/45 \\
 6509  &  9 Oct '99 & \ys    &        & \ys    &        & \ys    & 101/45 \\
 6627  & 27 Oct '99 &        & \ys    &        &        &        & 102/90 \\
\hline
 7129  & 11 May '00 &  \ys   & \ys    & \ys    &        &        & 101/45 \\
 7251  & 25 May '00 &  \ys   & \ys    & \ys    &        &        & 101/45 \\
 7519  &21 June '00 &  \ys   &        & \ys    &        &        & 101/45 \\
 7929  &26 July '00 &  \ys   &        &        &        &        & 101/45 \\
 8368  & 4 Sept '00 &  \ys   &        & \ys    &\ys     &        & 101/45 \\
 8446  &11 Sept '00 &  \ys   &        &        &        &        & 101/45 \\
 8556  &25 Sept '00 &        & \ys    & \ys    &\ys     &        & 101/45 \\
\hline
\end{tabular}
\end{center}
\caption{Successful RDP measurements at LEP~2.
Measured energy points are marked \ys.}
\label{tab:pollist}
\end{table}

The full list of LEP~2 RDP measurements is shown in
table~\ref{tab:pollist}, indicating the fill number,
date, nominal values of \Eb\ calibrated and optics used.
In total 86 energy points, distributed
through 37 fills, were calibrated.   The lowest energy measured
was 41~GeV,  a value dictated by the range of sensitivity
of the NMR probes.
Care was taken to 
perform a subset of the measurements with physics optics as well,
to allow for a cross-check against optics dependent
effects not foreseen in the the energy model.

\subsection{The NMR Probes and the NMR Magnetic Model}
\label{sec:nmrpol}

The NMR probes measured the local magnetic field with a relative
precision of $10^{-6}$.
Throughout LEP~2 operation a total of 16 probes were read
out during physics and RDP operation.
Time-integrated readings were logged every 5 minutes.
During the 2000 run additional records were logged in response
to rapid changes in field during   miniramps.
In the analysis the probes are designated by their octant 
location.
Each LEP octant had at least one probe, while octants 1 and 5
each had strings of five probes (1a--e; 5a--e).   Probes 1c and
1d were situated in the same dipole.
Other dipoles in the injection region, and the spectrometer,
were also instrumented with probes for limited periods of the 
programme, but these are not included in the NMR model.

The NMR probes were located underneath the vacuum chamber
above a steel field plate, installed to improve the uniformity
of the local field.  Radiation damage from synchrotron light 
led to a reduction in the probe locking efficiency, particularly
at low energy. In response to this problem the probes were
replaced, typically two to three times a year.  Precision mounts
first used in 1997 ensured that the replacement probes were 
installed to within 0.5~mm of their nominal positions.

In the NMR model the magnetic fields \Bnmri\ measured by 
each NMR $i=~\rm{1a},..,8$, after ramping to the excitation
current of interest,
are converted into an equivalent raw beam energy \EOnmri. The 
relation is assumed to be linear, of the 
form

\begin{equation}
\EOnmri =  a^i \,+\, b^i \Bnmri,
\label{eq:epol}
\end{equation}

\noindent with \EOnmr\ being used to signify the average over all \EOnmri.

Because 
the NMR probes are only sensitive to 
the dipole fields, it is necessary to account for 
the other sources of bending field in order to 
have the best possible model of the beam energy. 
Therefore \EOnmri\ is corrected to 

\begin{equation}
\Enmri = \EOnmri \,+\, \sum \Delta \Eb,
\label{eq:epolcor}
\end{equation}

\noindent where the sum runs over all the additional components
in the energy model detailed in section~\ref{sec:eb_model}.
These corrections are common to all NMRs and include energy
changes between the end of ramp and the time of interest.
In the calibration procedure the two parameters  
$a^i$ and $b^i$ for each probe are
determined by a fit to 
the energies measured by RDP.
Thereafter, all available values of \Enmri\ are averaged
together to give \Enmr,  which is taken as the
energy model's estimate of \Eb.   
At LEP~2 energies the 
error associated with \Enmr\ arising from the uncertainty in
the RDP measurements themselves is less than 0.5~MeV.

\subsection{NMR Residuals, High-Energy Scatter and Stability with Time}
\label{sec:banana}

The NMR model has been calibrated against the RDP data 
of each year separately, and the results of these
fits are used to define the energy model for that year.
As the calibration coefficients are observed to be
very stable for 1997 onwards, a calibration 
has also been made against the complete 1997--2000 dataset.
(This global fit can not be extended to 1996
because of differences in the exact probe locations
for this year.)    The mean (and RMS) coefficients averaged
over the 16 probes are found to be 
$\langle a \rangle = 91.17 (0.24)$~MeV/Gauss and $\langle b \rangle 
= 22 (61)$~MeV.
No significant difference is found between the fit
results for different optics.

\begin{figure}
\begin{center}
\epsfig{file=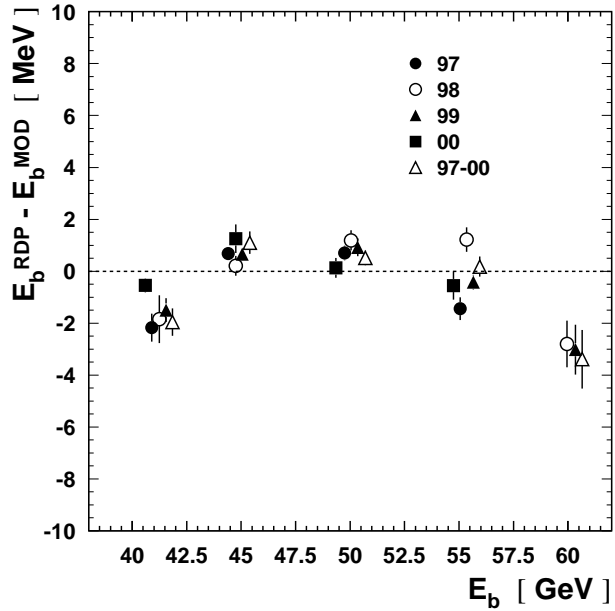,width=0.55\textwidth} 
\end{center}
\caption{Residuals of the fitted energy model to the RDP data in
1997-2000.  Each point is the mean over the available
measurements at that energy, with the error bar the statistical uncertainty
on this mean.  The horizontal positions of the points have been slightly 
adjusted to aid clarity.}
\label{fig:banana}
\end{figure}

Figure~\ref{fig:banana} shows the residuals of the separate 
fits to the RDP data, averaged over the probes, 
for the main datasets and those of
the global fit.   The error bars are the statistical
uncertainties on the mean of all the contributing 
measurements.  
There is a small, but characteristic,  non-linearity over the
sampled energy range.

\begin{figure}
\begin{center}
\epsfig{file=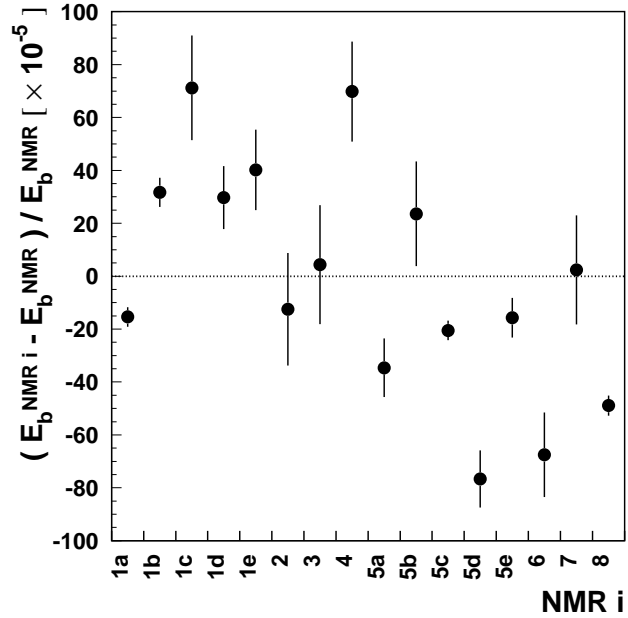,width=0.55\textwidth} 
\end{center}
\caption{Relative differences between the individual probe predictions
of the dipole energy and that of the average, evaluated at physics energy and
averaged over 1997--2000.  The error bars are half the difference between
the maximum and minimum values in these years.}
\label{fig:nmrscatt}
\end{figure}

The residuals of the individual NMRs entering
in figure~\ref{fig:banana}
agree to within a few MeV.  When the model is applied 
at high energy, however, the individual non-linearities
of each magnet, and the lever-arm over which
the calibration is extrapolated, lead to a significant
scatter in the prediction of \EOnmr.   Figure~\ref{fig:nmrscatt} shows
the relative differences between \EOnmri\ and \EOnmr\
evaluated during high-energy physics operation, averaged over
all 1997--2000 data.  The error bars are half the difference between the
maximum and minimum values in these years.   
There is no strong evidence of systematic structure in this distribution,
although the differences for those probes in octant~1 are
predominantly positive in sign, and those in octant~5 are
predominantly negative.
The RMS of the individual probe predictions is $43 \times 10^{-5}$.
If the measurements of the 16 probes are representative of the 3280 dipoles
in LEP, then the expected
precision of the dipole part of the model at $\Eb = 100\,\rm{GeV}$ is 11~MeV. 
The purpose of the flux-loop, spectrometer and $Q_s$ measurements 
presented in the following sections is to test this assumption
and to constrain further any offset between the model prediction
and the true energy.

Both figures~\ref{fig:banana} and~\ref{fig:nmrscatt} illustrate
the stability of the NMR calibration with time.   This can be
seen in a more quantitative fashion by using the calibration coefficients
of one year to evaluate \EOnmr\ during physics operation in another year.
The luminosity-weighted mean shifts in results are presented in 
table~\ref{tab:cal_stab}, and are always 4~MeV or less.
Larger shifts of around 30~MeV are seen when a later year's calibration
coefficients are applied to the 1996 data,  a difference attributable
to the change in probe locations after the 1996 run.

\begin{table}
\begin{center}
\begin{tabular}{|c|rrrr|} \hline
    & \multicolumn{4}{c|}{Dataset} \\ \hline
Fit      & '97 & '98 & '99 & '00 \\ \hline
'97      &    /&  4.1&  1.5&  0.9\\
'98      & -3.6&    /& -2.5& -0.3\\
'99      & -1.8&  1.9&    /&  2.1\\
'00      & -3.6& -0.1& -1.7&    /\\
'97-'00  & -2.0&  1.8& -0.6&  1.8\\ \hline
\end{tabular}
\end{center}
\caption{Shift in \EOnmr\ (MeV) observed in physics operation
when the data are reprocessed with NMR calibration coefficients
determined by a fit to the RDP data of another year.}
\label{tab:cal_stab}
\end{table}

%% file: eb_model.tex
\section{Other Components of the \bf{\Eb} Model}
\label{sec:eb_model}

The NMR fit gives the value of the energy from the dipole magnets
at start-of-fill.    The complete \Eb\ model adds to this
contributions coming from variations of the dipole magnet strength
during the fill,  as well as additional sources of bending field arising
from quadrupole effects,  horizontal correctors, and uncompensated 
currents flowing in the magnet power bars.   
These additional
model components,  represented by $\sum \Delta \Eb$ in 
expression~\ref{eq:epolcor},
are discussed in this section.
The relative importance of the model components during
physics running can be assessed from
table~\ref{tab:model_lumi}, which shows the 
luminosity-weighted contribution of each term to \Enmr\ at each 
high-energy point of the LEP~2 programme.

\begin{table}[htb]
\begin{center}
\begin{tabular}{|l|r|r|r|r|r|r|r|r|r|r|} \hline
Year          &\multicolumn{2}{|c|}{'96} & 
               \multicolumn{1}{|c|}{'97} &
               \multicolumn{1}{|c|}{'98} &
               \multicolumn{4}{|c|}{'99} & 
               \multicolumn{2}{|c|}{'00} \\ \hline
\Ecmn\      [GeV] &
\multicolumn{1}{|c|}{161} & 
\multicolumn{1}{|c|}{172} & 
\multicolumn{1}{|c|}{183} & 
\multicolumn{1}{|c|}{189} & 
\multicolumn{1}{|c|}{192} & 
\multicolumn{1}{|c|}{196} & 
\multicolumn{1}{|c|}{200} & 
\multicolumn{1}{|c|}{202} & 
\multicolumn{1}{|c|}{205} & 
\multicolumn{1}{|c|}{207} \\ \hline
\fRFc\                 &-13.8 &-14.2 &-20.2 &-27.5 &  1.2 &-27.8 &-40.0 &-24.4 &-32.3 &-40.2\\
\fRF\                  &  0.0 & -3.0 &-152.4&-187.0&-222.2&-229.7&-194.9&-129.8&-85.9 &-29.6\\
NMR rise               &  3.6 &  7.0 &  1.7 &  0.8 &  0.7 & -0.1 & -0.7 & -0.7 &  1.5 & 2.2 \\
Tides                  &  1.1 &  0.8 &  1.2 &  1.7 &  1.9 &  2.2 &  1.4 &  1.8 &  2.0 & 1.8 \\
Hcor / BFS             & -2.8 & -3.0 & -5.6 & -7.8 & -1.1 & -1.6 & -0.4 &  1.1 & 357.6&430.0\\
QFQD                   & -2.6 & -2.4 & -2.8 & -1.3 & -1.3 & -1.4 & -1.4 & -1.4 & -1.4 & -1.4\\
\hline
\end{tabular}
\end{center}
\caption{The luminosity-weighted corrections to \Enmr\ in MeV 
from each component
in the energy model at each nominal energy point.}
\label{tab:model_lumi}
\end{table}

\subsection{Quadrupole Effects}
\label{sec:xarc}

In a very high-energy synchrotron, such as LEP, the orbit length is
fixed by the {\it RF frequency}, \fRF.  The {\it central RF frequency}, \fRFc
corresponds to that orbit where the beam passes on average through the
centre of the quadrupoles. When
the RF frequency \fRF\ does not coincide with \fRFc, the beam
senses on average a dipole field in the quadrupoles, which
causes a change in the beam energy, $\Delta \Eb$, of:

\begin{equation}
\frac{\Delta \Eb}{\Eb} \: = \: - \frac{1}{\alpha_c} \, 
\frac{\fRF \, - \, \fRFc}{\fRF} ,
\label{eq:quadeffect}
\end{equation}

\noindent where $\alpha_c$ is the {\it momentum compaction factor}, 
the optics dependent values of which are listed in 
table~\ref{tab:alphac}, as calculated by the simulation program MAD~\cite{MAD}.
The nominal value of \fRFc\ is 352,254,170~Hz. 
The consequences of variations in both \fRF\ and \fRFc\ must be 
corrected for in the energy model.

\begin{table}
\begin{center}
\begin{tabular}{|l|c|c|} \hline
\multicolumn{2}{|c|}{Optics} &  $\alpha_c$ [ $\times 10^{-4}$ ] \\ \hline
             &90/60          & 1.86 \\
Physics      &108/90         & 1.43 \\ 
             &102/90         & 1.56 \\ \hline
             &60/60 (1997-98) & 3.87 \\
Polarisation &60/60 (1999)    & 3.77 \\
             &101/45         & 1.50 \\ \hline
\end{tabular}
\end{center}
\caption[]{Calculated values of the momentum compaction factor, $\alpha_c$,
for the physics and polarisation 
optics of the LEP~2 programme.  The
estimated relative uncertainties are 1\%.}
\label{tab:alphac}
\end{table}

\subsubsection{Central Frequency and Machine Circumference: $\Delta \Eb \, (\rm{f^{RF}_c})$}

\begin{figure}
\begin{center}
\epsfig{file=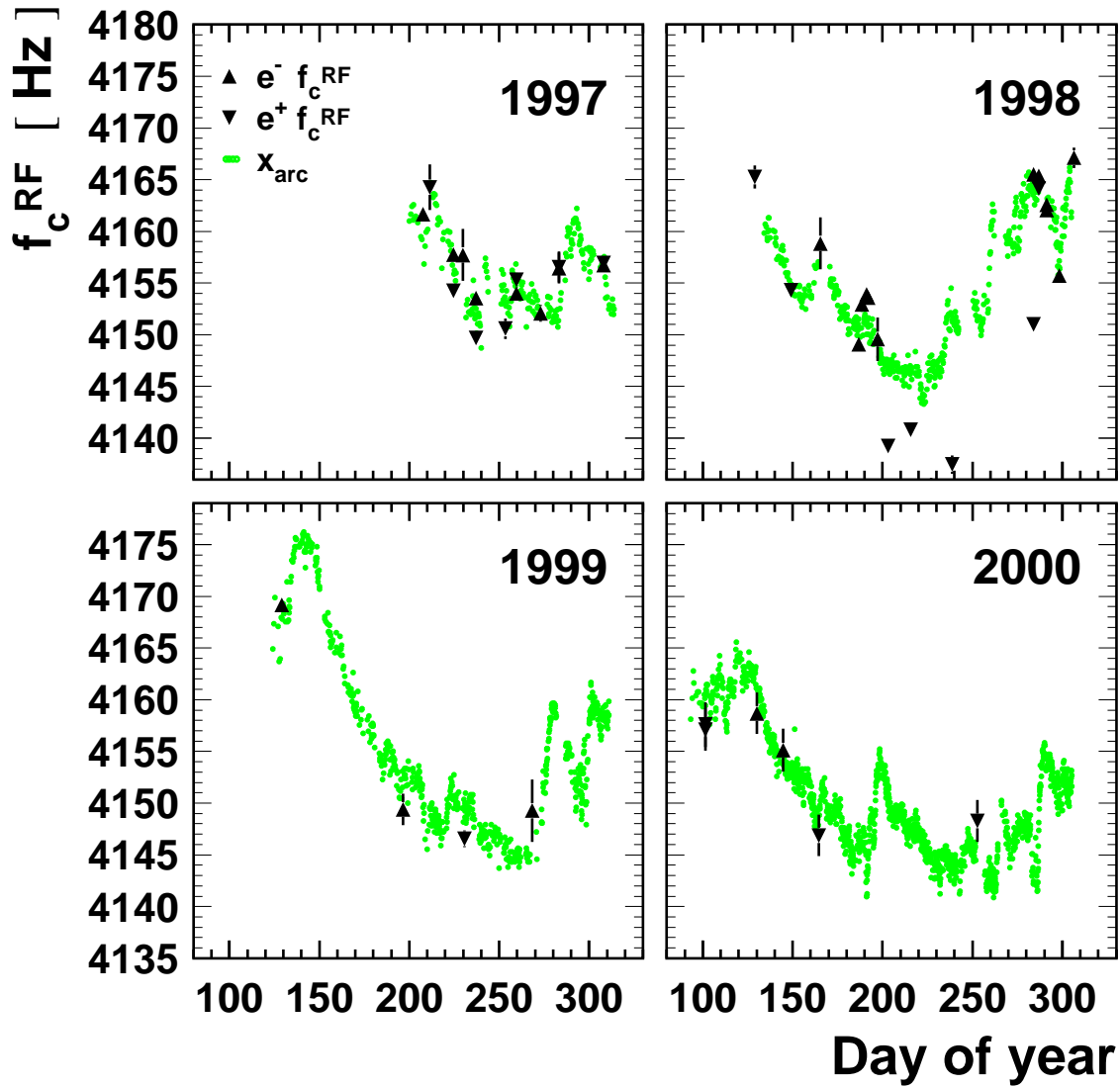,width=1.00\textwidth}
\caption[]{Evolution of the central frequency as a function of 
time for the four main datasets of the LEP~2 programme.
Shown are both the actual \fRFc measurements,  and the values
extracted from \xarc, after correction for tides.
 Note that the vertical
scale shows a variation in the last four digits of the LEP RF 
frequency, which is nominally $352 \, 254 \, 170$~Hz.}
\label{fig:xarc}
\end{center}
\end{figure}

The central frequency was measured only on a few occasions
during a year's running and required non-colliding beams~\cite{JORGFRFC}.
In between these measurements, \fRFc\ can be interpolated through
\xarc, the average horizontal beam position
in the LEP arcs as measured by the {beam-position monitors} (BPMs) 
at a defined RF frequency~\cite{JORGXARC}.
These measurements are shown in figure~\ref{fig:xarc}
for 1997-2000, where the \xarc\ data have been
normalised to the actual \fRFc\ measurements.   The central
frequency can be seen to change  by 20--30~Hz, and evolves
in a similar fashion over the course of each year.
The evolution indicates a change in the machine circumference,
one which is believed to be driven by a seasonal variation in the 
pressure of the water-table and the level of Lac L\'{e}man.

The \fRFc\ and \xarc\ measurements together 
allow the energy to be corrected
fill by fill.   The average values of this correction
are listed in table~\ref{tab:model_lumi} and are found
to be similar year to year.   The variation seen within
1999 comes about because the running at each energy
point was concentrated at different periods of the year,
rather than uniformly distributed.

The uncertainty on this correction is set by studying
the agreement between the direct \fRFc\ measurements
and \xarc.  In general these are consistent,  although
there are occasional discrepancies, such as for some of the \pos\ data
in 1998.   Globally the agreement is found to
be good to $\pm 2 \, \rm{Hz}$.
Any bias in \fRFc\ will apply to both the low-energy
calibration data and the high-energy running.
As the correction scales with energy, the effect of a bias
will be absorbed in the calibration coefficients and lead
to no net error at high energy.   This argument is
only valid, however,  when the optics,
and therefore $\alpha_c$, is the same for calibration
and physics operation.  This was the case in 1996,
and approximately so in 2000, but not in the other
years,  where the uncertainty in \fRFc\
induces  a residual $3 \times 10^{-5}$ error on \Enmr.

The LEP circumference was also distorted by the 
gravitational mechanism of earth tides, as discussed 
in section~\ref{sec:tide}.  These effects have been
subtracted in the \fRFc\ analysis. 
The \fRFc\ and \xarc\ measurements used in this analysis have
also been corrected for residual biases from horizontal corrector
effects~\cite{JORGFRFC}, discussed in section~\ref{sec:hcor}.

\subsubsection{RF Frequency Shifts: $\Delta \Eb \, (\fRF)$}

For 1997 and subsequent years the RF frequency was routinely increased 
by $\sim 100$~Hz from the nominal value in order to change the horizontal 
damping partition number.  This was done to squeeze the beam
more in the horizontal plane, which benefited both the
specific luminosity and the machine background at the experiments.
A side-effect of this strategy was that
the beam energy was reduced, following equation~\ref{eq:quadeffect}.
Since the 2000 run placed a premium on
reaching the highest possible energies, a smaller offset was chosen
in this year.

On the occasions when RF cavities tripped, the RF frequency was
temporarily decreased in order to keep the beam lifetime
high, and afterwards was raised to its previous value when full RF
voltage was restored.   This led to abrupt energy steps during a fill.
Therefore all \fRF\ manipulations were routinely logged, 
enabling the energy values at the experiments to be updated at each
change.

\vspace*{0.5cm}

Associated with the quadrupole-related energy corrections 
is an error arising from the 1\% uncertainty in
the momentum compaction factor,  which is conservatively
assumed to be in common between all optics.

\subsection{Continuous Energy Change During a Fill}

During the timescale of a fill the beam energy in general fluctuated by several
MeV, both because of variations in the dipole field and because of 
earth tides.  These effects are well understood
from LEP~1~\cite{LEP1PAPER}.

\subsubsection{Change in Dipole Field:  $\Delta \Eb \, (\rm{NMR \: rise})$}

The strength of the dipole magnets varied during the course of a fill,
both because of temperature effects and because of parasitic currents 
which flowed on the beampipe.    This evolution is 
included in the model by calculating
the field variation since start-of-fill averaged over
all available NMR probes, expressed as an energy change.
Measurements of the parasitic current show
different behaviour for octants 1,7 \& 8 compared
with octants 2 -- 6.   Therefore the average field
change is calculated with a weight for 
each NMR to reflect its octant location.

The size of the luminosity-weighted dipole change is less than 2~MeV for 
data-taking in 1997--1999.  
This is lower than in 1996 and 2000 because of the
routine use of bending modulations.
The difference in the size of the effect between 1996 and
2000 is directly attributable to the
short length of the sub-fills in the latter year.

\subsubsection{Earth Tides: $\Delta \Eb \, (\rm{Tides})$}
\label{sec:tide}

Tidal effects, due to the combined gravitational 
attraction of the Sun and Moon, can cause  
relative distortions of up to $10^{-8}$~\cite{MELCHIOR} in the
circumference of the LEP tunnel. 
During operation these distortions changed the positions 
of the quadrupoles with respect to the beam, and resulted 
in energy  variations through the same mechanism as is 
described in section~\ref{sec:xarc}.   
The amplitude of the ring distortions has been calibrated against
the LEP BPM system to a precision of 5\%~\cite{JORGTIDE}.

\vspace*{0.5cm}

\begin{figure}
\begin{center}
\epsfig{file=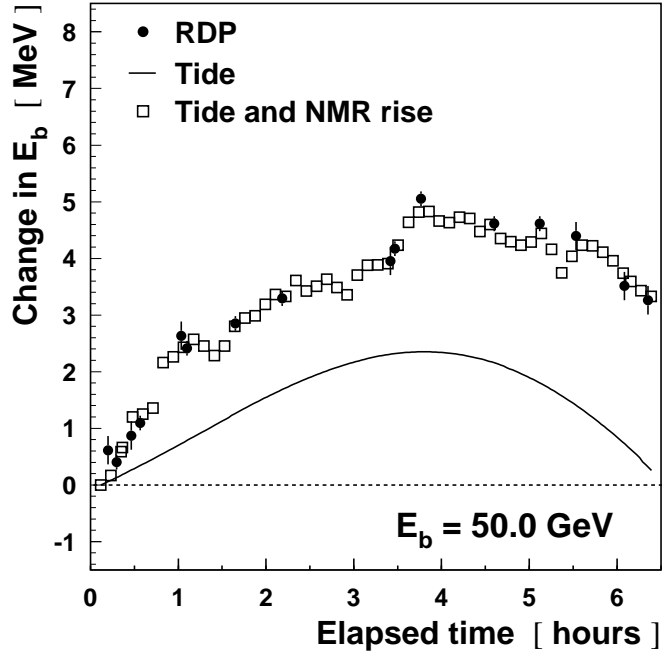,width=0.60\textwidth}
\caption[]{Comparison between the tide and NMR rise components of the 
energy model and RDP measurements
for fill 6432. The tide contribution is also
shown separately.}
\label{fig:tide_exp}
\end{center}
\end{figure}

Occasions when repeated RDP measurements were made over a period
of several hours can be used to test the modelling of the
energy change during a fill.    Figure~\ref{fig:tide_exp}
shows results from 50~GeV operation in fill 6432 during 1999.
Shown is the change in \Eb\ as measured by RDP and as
predicted by the model, plotted against elapsed time since
the start of the experiment.   The energy change of the
model receives contributions from the dipole change seen
in the NMRs, which rises by 4~MeV, 
and that from the earth tide, which first rises by 2~MeV and 
then falls to zero.   The model has been normalised to
the RDP over the first 30 minutes of the experiment;
throughout the following 6 hours excellent agreement is seen.

From such experiments the uncertainty on \Enmr\ 
from the combined modelling
of tide effects and dipole field change is
known to be very small.  A correlated
error of 0.5~MeV is assigned for all years,
independent of energy.

\subsection{Horizontal Corrector Effects}
\label{sec:hcor}

{\it Horizontal correctors} are small, independently-powered dipole 
magnets which were used to correct local deviations in the orbit.
The global effect of these corrections had 
the potential to influence \Eb\ and thus must be 
accounted for in the energy model.

In the last year of LEP operation the horizontal correctors were purposefully
powered in a coherent manner in order
to increase the fraction of bending field outside the 
main dipoles; this {\it bending-field spreading} (BFS)
significantly increased the beam energy attainable at
a given RF voltage and is described by an important model component
unique to the 2000 run.

\subsubsection{Horizontal Correctors Prior to 2000:  $\Delta \Eb \, (\rm{Hcor})$}

Each horizontal corrector, $i$,  provides an angular kick,
$\theta_x^i$, in a region where the local horizontal
dispersion is $D^i_x$.  Hence, summing over all 
magnets, there is a lengthening $\Delta L$ in the orbit
where
\begin{equation}
\Delta L \: = \: \sum_i \, {D^i_x \, \theta_x^i}.
\label{eq:deltal}
\end{equation}

\begin{figure}[htb]
\begin{center}
\epsfig{file=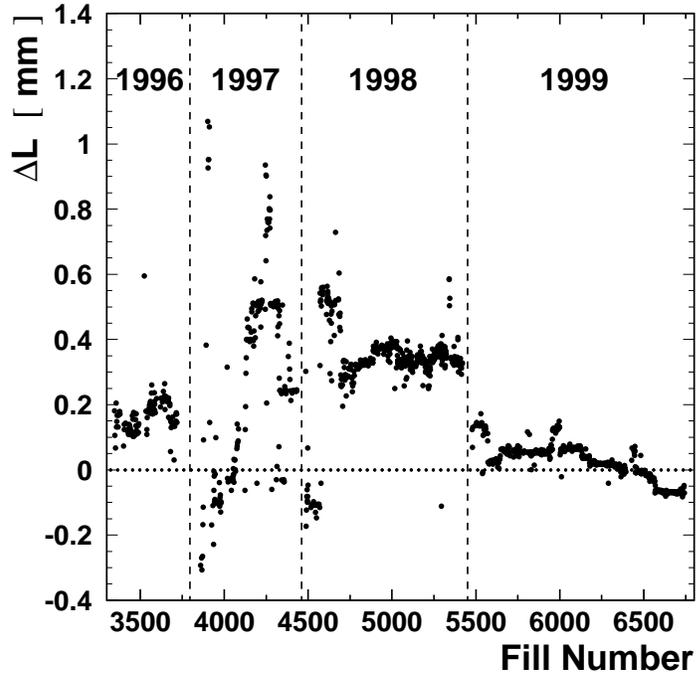,width=0.61\textwidth}
\caption[]{The orbit lengthening $\Delta L$ caused by the
horizontal correctors, plotted against fill number for 1996--1999.}
\label{fig:deltal}
\end{center}
\end{figure}

\noindent This orbit lengthening leads to an energy
change of

\begin{equation}
\frac{\Delta \Eb}{\Eb} \: = \: - \frac{\Delta L}{\alpha_c \, C},
\label{eq:hcoreffect}
\end{equation}

\noindent where $C$ is the LEP circumference.   The actual
value of $\Delta L$ is plotted against fill number in 
figure~\ref{fig:deltal} and can be seen to vary significantly
with time. 
Different corrector settings were required for each 
optics, as was day-by-day adjustment by the operators  in order 
to optimise the machine performance.   
A fill-by-fill mean value
of $\Delta L$ is used in calculating \Enmr\ during physics running.
The largest correction is -8~MeV for the 1998 run.
When analysing the RDP calibration data, individual corrector manipulations
within the fill are considered.

An alternative way to picture the effect of the correctors on \Eb\
is to assume that the fields responsible for the kicks 
sum to augment the total bending field of the ring.  
This model is naive, as some of the corrections
compensate the orbit distortions introduced by misaligned
quadrupoles.    
The results of dedicated measurements~\cite{LEP1PAPER,JORGHCOR}
favour the orbit lengthening model, but find both
descriptions to be compatible with the data.

The difference between the effects of the two models is 30~\%.
This value, applied to the high-energy correction, is taken as
an uncertainty.

\subsubsection{Synchrotron Energy Loss and Bending-Field Spreading: 
$\Delta \Eb \, (\rm{BFS})$}

Considering a machine with bending magnets and horizontal
orbit correctors only,  and neglecting for the moment 
the effects of orbit distortions,
equations~\ref{eq:ebdef} and~\ref{eq:u0def} lead to
the following expressions for the beam energy
and synchrotron energy loss per turn:

\begin{equation}
E_b \: = \: \frac{ec}{2 \pi} \, (B_d \, L_d \: + \: B_c \, L_c);
\end{equation}

\begin{equation}
U_0 \: = \: \frac{C_\gamma \, (ec)^2}{2 \pi} \, E_b^2 \, (B^2_d L_d \: + \: B_c^2 L_c).
\end{equation}

\noindent Here $B_d$ is the field, and $L_d$ the total (magnetic) length of 
the dipole magnets. $B_c$ and $L_c$ are the corresponding
quantities for the horizontal correctors.   Practically,
the maximum value of $U_0$ is dictated by the available
RF voltage.  Keeping this constant, and assuming 
$B_c L_c \, \ll \, B_d L_d$, allows the maximum attainable
energy, $E^\rm{M}_b$, to be written:

\begin{equation}
E^\rm{M}_b \: \approx \: \EbOm \, \left (1 \: + \: \frac{1}{2} 
\frac{B_c L_c}{B_d L_d} \left(1 \, - \, \frac{1}{2} \frac{B_c}{B_d} \right) \right),
\label{eq:bfs}
\end{equation}

\noindent where \EbOm\ is the maximum energy that can reached when
$B_c = 0$ and the dipoles alone are used to define the beam energy.
From expression~\ref{eq:bfs} it is clear that the beam energy may be
increased above \EbOm\ by using the correctors to {\it spread}
the bending over more magnets.   This method is 
referred to as bending-field spreading (BFS)~\cite{BFSREF}.

BFS was deployed in physics operation during the 2000 LEP run.  
In order to maximise its
effect $\sim 100$ additional corrector magnets which had 
previously never been cabled, or had been removed from the tunnel,
were connected or re-installed.   Including these, $B_c L_c \approx 6.5 \,\rm{Tm}$,
to be compared with $B_d L_d = 2092 \,\rm{Tm}$, at a nominal \Eb\ of 100~GeV.
Since $B_c/B_d \approx 1/2$,
the maximum additional energy  predicted
by expression~\ref{eq:bfs}  is 120~MeV.
(This calculation assumes that 20\% of the available bending field
of the correctors is reserved for orbit steering.)

A more complete analysis of BFS must account for
orbit distortions.  The kicks provided by the
horizontal correctors cause the beam to move
away from the central orbit in the defocusing 
quadrupoles, and this leads to an additional
source of bending field which approximately
doubles the energy boost.    The exact value
of boost is calculated from the simulation program, MAD~\cite{MAD}.
This has been done and then parameterised as a function
of corrector setting.  The luminosity-weighted
corrections to the energy model from BFS  
are included in table~\ref{tab:model_lumi}.  
(Note that these are the corrections to $E_b$ rather than 
\EbOm, and hence are larger than the 
values discussed above.)  The lower boost at the 205~GeV
energy point is because the BFS was not used at the 
start of the run, and then initially operated below its maximum
setting.

The LEP spectrometer was used in dedicated experiments
to measure the energy boost from the BFS.   This procedure 
is described in section~\ref{sec:bfsmeas}.
These measurements confirm the expected energy gain with a precision
of 3.5~\%, which is taken as the systematic uncertainty in
the model.

\subsection{Quadrupole Current Imbalance: $\Delta \Eb \, (\rm{QFQD})$}

Any different phase advance in the horizontal and vertical planes of the
LEP optics meant that in the quadrupole power bars running around the
LEP ring, at a distance of roughly 1~m from the vacuum chamber,
there was a current difference between the circuit feeding the
focusing (QF) and defocusing (QD) quadrupoles.   This imbalance 
resulted in an additional source of bending field seen by the beam,
which is accounted for by the {\it QFQD} component of 
the energy model.

The dependence of the QFQD energy correction on the quadrupole
current imbalance was calibrated at LEP~1 to a precision
of 25\%~\cite{LEP1PAPER}.

%% file: ecm_ip.tex
\section{Evaluation of \Ecm\ at the Interaction Points}
\label{sec:ecm_ip}

The estimate of the collision energy
at each experimental interaction point (IP)~\footnote{The four 
experimental interaction points were IP2 (L3), IP4 (ALEPH), IP6 (OPAL)
and IP8 (DELPHI).}, \Ecmnmr, is
given by

\begin{eqnarray}
\Ecmnmr & = & 2 \times \Enmr \, + \, \sum \Delta \Ecm,
\nonumber
\end{eqnarray}

\noindent where $\sum \Delta \Ecm$ represents the sum of several possible
corrections, which are in principle IP specific.  
The most important of these arises from 
the fact that the local beam energy at each IP differs from \Eb, the average
energy around the ring, because of the combined effects of synchrotron
radiation and the RF system.   
Dispersion effects and the possibility of an 
energy difference between the $e^-$ and $e^+$ 
beams must also be considered.   In practice no corrections are applied
for these latter terms, but the associated uncertainties 
are accounted for in the error
assignment.

\subsection{Corrections from the RF System}
\label{sec:rfcorr}

\begin{figure}
\begin{center}
\epsfig{file=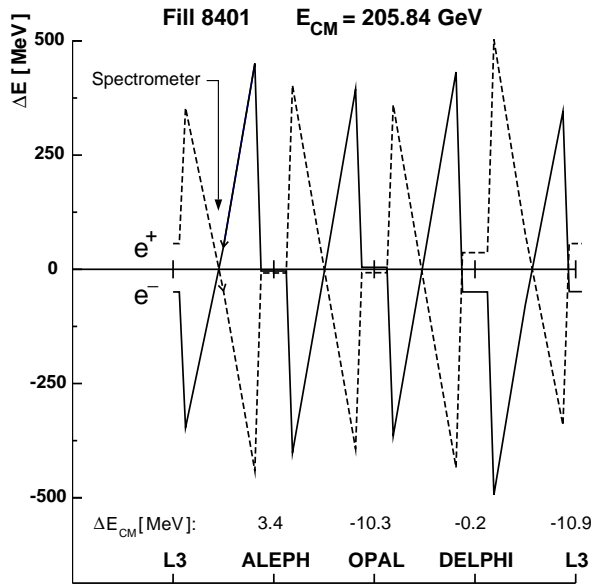,width=0.49\textwidth}
\end{center}
\caption{The RF sawtooth for a typical RF configuration in 2000.
The electron beam is represented by the left-going
solid line,  the positron beam by the right-going dotted line.  
The locations of the experiments and the LEP spectrometer
are indicated.
$\Delta E = 0$ corresponds to the average beam energy
of LEP.
$\Delta \Ecm$ is the correction to \Ecmnmr\ at each IP due to RF effects.}
\label{fig:sawtooth}
\end{figure}

As explained in section~\ref{sec:basiceqn},  the 
energy loss of the beams due to
synchrotron radiation was replenished by stations of super-conducting 
RF cavities
situated on either side of the experimental IPs~\footnote{Several
copper cavities, retained from LEP~1, also contributed $\sim 3\%$ 
to the overall voltage.}.  
It is necessary to model the variation
in energy around the ring in order
to calculate \Ecmnmr.   The calculated variation is shown in 
figure~\ref{fig:sawtooth}  for both  $e^-$ and $e^+$ for a typical 
fill in 2000.  The continuous loss from the synchrotron radiation
and the localised boosts from the RF stations lead to a characteristic
{\it sawtooth} distribution in both the energy loss and in the horizontal
displacement between the two beams.   The variation in horizontal
displacement is measured by
an array of 500 BPMs distributed throughout the ring.

The sawtooth variations are to first order 
anti-symmetric between the two beams, hence the correction to \Ecmnmr\
is in general small.   The calculation of the sawtooth is however
rendered challenging by the instability of the RF system, the configuration
of which varied from fill to fill as units broke and were repaired, 
and within fills, as units tripped.  Additional inputs to
the calculation come from knowledge of the absolute voltage calibration
scale,  and the alignment and phasing of the cavities.

During 2000 (and late in the 1999 run)  dedicated fills
were taken with single beams in order to perform \Eb\ measurements with
the energy spectrometer.  Knowledge of the sawtooth is required to 
relate  the local energy at the spectrometer, close to IP3,
with \Enmr.  The demands placed on the RF modelling are more
exacting for these single-beam fills,  as the result 
for an individual spectrometer measurement is
directly sensitive to the absolute knowledge of the sawtooth.
The spectrometer apparatus and analysis are discussed in sections~\ref{sec:specsetup} 
and~\ref{sec:specana}.


\subsubsection{Modelling the Sawtooth}

The modelling of the energy corrections from the RF system
is carried out by the iterative calculation of the stable RF phase angle $\psi_s$
which proceeds by setting the total energy gain,
$V_{\mathrm{RF}} \sin \psi_s$, of the 
beams as they travel around the machine equal to the sum of all known energy losses.
Here $V_{\mathrm{RF}}$ is the total RF accelerating voltage
which is calculated using detailed measurements of the RF cavities,
such as their voltage calibrations and their longitudinal misalignments.
When available, the measured value of the synchrotron tune, $Q_s$, 
and the difference in horizontal displacement  between the beams as they enter 
and leave the experimental IPs, are used to constrain energy
variations due to the overall RF voltage scale and RF phase errors.  
(A full discussion of the synchrotron tune and its 
relationship to energy loss is given
in section~\ref{sec:qs}.)

\begin{figure}
\begin{center}
\epsfig{file=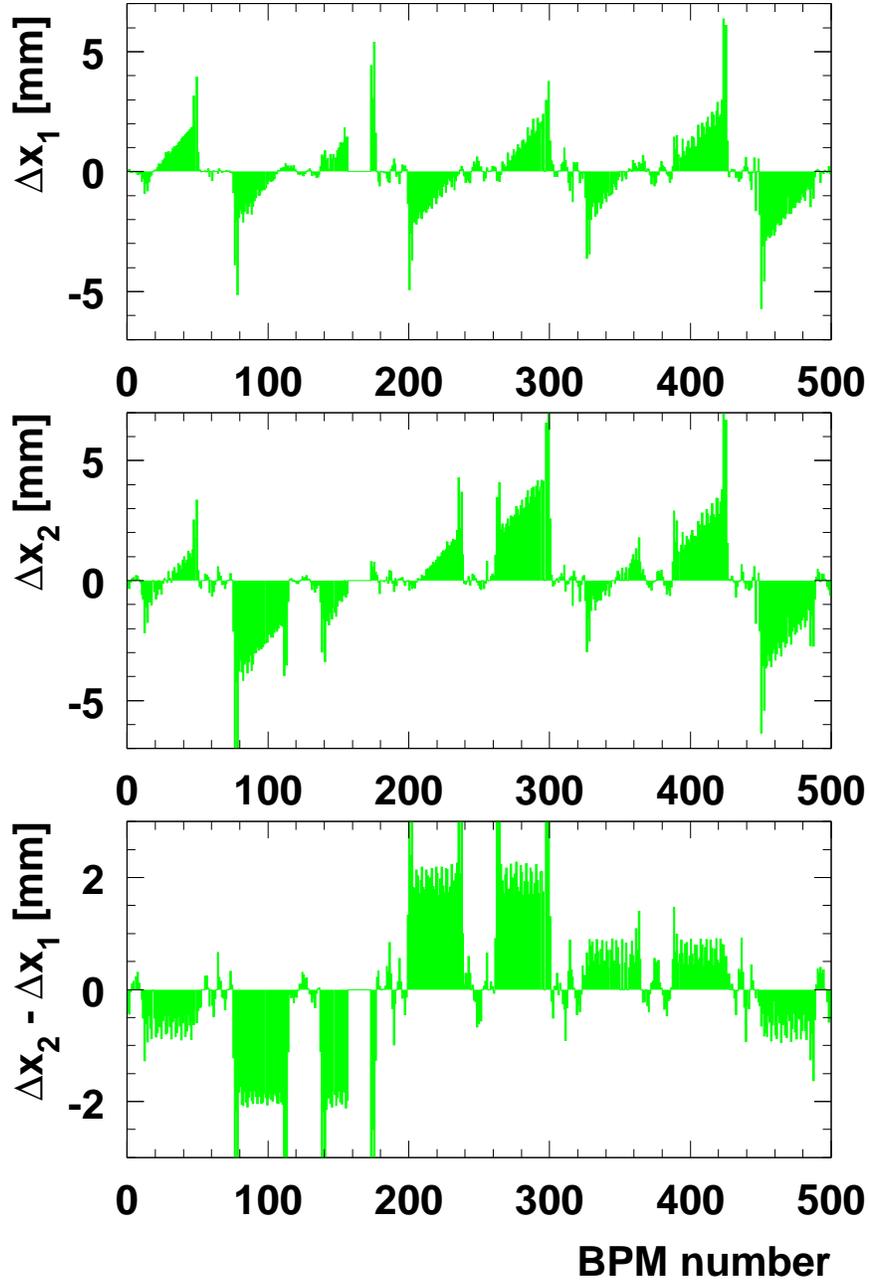,width=0.75\textwidth}
\end{center}
\caption{An example of the difference between electron and positron orbits
($\Delta x_i$ vs BPM number) 
for two different RF configurations (configuration 1 top; configuration 2 middle).
IPs 2,4,6 and 8 are situated close to BPMs 60, 190, 310 and 440
respectively.
The energy sawtooth
is clearly visible.  The difference of the two orbit differences is displayed on the bottom
plot. The step changes reflect the
different energy gains at the four IPs for the two configurations.
Note that the orbit information is missing between BPMs 160 and 176.}
\label{fig:JorgRFfig}
\end{figure}

The model of the RF system described above has been used to calculate the centre-of-mass
energy corrections due to the RF system parameters for the whole of LEP~2 running.  
Its calculation of the energy loss in the LEP arcs, however, treats each arc as a 
single entity, rather than considering each magnetic component individually.
For the spectrometer studies, a more detailed model has been developed
based on the MAD program~\cite{MAD}.  
This model incorporates the detailed measurements
of the RF cavities,
on top of the complete specification of the LEP magnetic lattice\cite{JORGRF}.  
Such an approach allows the calculation
of the beam energy at any point in LEP, not just at the IPs.  
This feature permits the performance of the model to be studied 
through comparison with BPM data in the LEP arcs, which are 
sensitive to the effective $e^+e^-$ energy difference. 
The technique is illustrated in figure~\ref{fig:JorgRFfig}, 
where the energy offset between the electron and positron 
beam is clearly seen for two different RF configurations.
A comparison of the two sawtooths allows the parameters of the 
system to be determined, in particular the net RF phase
error at any LEP IP.  Two experiments performed late in the 2000
run, in which the RF at each IP was powered down and up in turn,
have been used to calibrate the method.

The average corrections for all of the LEP~2 running 
are shown in table~\ref{tab:rfcorrs}.
Comparisons made between the two models at selected 
energy points show agreement to within 1~MeV.


\begin{table}
\begin{center}
\begin{tabular}{|l|r|r|r|r|r|r|r|r|r|r|} \hline
Year          &\multicolumn{2}{|c|}{'96} & 
               \multicolumn{1}{|c|}{'97} &
               \multicolumn{1}{|c|}{'98} &
               \multicolumn{4}{|c|}{'99} & 
               \multicolumn{2}{|c|}{'00} \\ \hline
\Ecmn\ [GeV] &
\multicolumn{1}{|c|}{161} & 
\multicolumn{1}{|c|}{172} & 
\multicolumn{1}{|c|}{183} & 
\multicolumn{1}{|c|}{189} & 
\multicolumn{1}{|c|}{192} & 
\multicolumn{1}{|c|}{196} & 
\multicolumn{1}{|c|}{200} & 
\multicolumn{1}{|c|}{202} & 
\multicolumn{1}{|c|}{205} & 
\multicolumn{1}{|c|}{207} \\ \hline
IP 2 (L3)     & 19.8 & 19.4 &  8.2 &  6.0 &  8.8 &  8.2 &  8.0 &  8.0 &  3.4 &  3.0\\
IP 4 (ALEPH)  & -5.6 & -5.8 &-10.8 & -9.2 &-12.6 &-14.0 &-13.8 &-13.0 &-11.0 & -9.8\\
IP 6 (OPAL)   & 20.3 & 19.8 &  5.6 & -2.6 & -5.8 & -5.2 & -5.4 & -4.4 & -0.6 &  0.0\\
IP 8 (DELPHI) & -9.4 & -8.4 &-13.2 &-10.4 &-17.2 &-16.0 &-15.0 &-14.0 &-11.4 & -9.8\\
\hline
\end{tabular}
\end{center}
\caption{The luminosity-weighted RF corrections to \Ecmnmr\ in MeV at each
IP for each nominal energy point.}
\label{tab:rfcorrs}
\end{table}

\subsubsection{Error Assignment on the RF Corrections}

The error assignment for the RF \Ecm\ corrections 
arises from the following considerations:

\begin{itemize}
\item{Any discrepancy between calculated and measured control variables,
such as the $Q_s$ or the horizontal beam displacements,
indicates imperfections in the model.  For instance,
discrepancies in the $Q_s$ reveal a lack of knowledge of the
overall voltage scale or a phase error in the RF system;}
\item{From measurements made with a beam-based alignment 
technique~\cite{MIKEALIGN}, the locations 
of the cavities are only known with a precision of 1~mm;}
\item{A small  uncertainty comes from the 
unknown misalignments and non-uniformities
of all the magnetic components of LEP.   This contribution can be
estimated by simulating an ensemble of machines
with imperfections similar to those expected in LEP.}
\end{itemize}

\noindent In all cases the range of values of the energy corrections 
obtained when allowing the machine parameters to vary over their allowed 
values is taken as the systematic error. The procedure is discussed 
in detail in~\cite{LEP1PAPER}.
It should be noted that those energy-loss uncertainties 
important for the understanding of the $Q_s$ and detailed
in section~\ref{sec:qs}
have negligible impact on the \Ecm\ corrections at the IPs.

The total error on  \Ecmnmr\ from the RF correction is estimated
to be 8~MeV for the 183~GeV, 189~GeV and 192~GeV energy points,
and 10~MeV for all other running.
The conservative assumption is made that these uncertainties are fully
correlated between IPs and energy points.

The BPM data, such as those seen in figure~\ref{fig:JorgRFfig},
provide a very  powerful constraint on the MAD model of 
the individual beam sawtooth at the spectrometer, 
which is an important ingredient in the analysis presented in
section~\ref{sec:specana}.  
The error on this calculation for 
the dataset of spectrometer measurements is estimated to be 10~MeV.
This value is set by the uncertainty in applying the results
of the calibration measurements, made at the end of the 2000 run,
to the earlier spectrometer experiments.

\subsection{Possible Electron Positron Energy Differences}

The energy of the electron and positron beams are not expected to be 
exactly identical. Orbit differences lead to small
differences in the integrated bending field seen by each beam. The main 
cause for orbit differences at LEP is the energy sawtooth that 
separates the orbits at the highest beam energies by up to a few 
millimeters in the horizontal plane. Due to the strong energy dependence 
of the sawtooth, the expected energy difference, which is smaller than 
1~MeV at 50 GeV, can reach 3-4~MeV around 100 GeV, according to
simulation.   To cover this possibility an uncertainty of 4~MeV is assigned
on \Ecmnmr.

\subsection{Opposite-Sign Vertical Dispersion}

Beam offsets at the collision point can cause a shift in the
centre-of-mass energy due to opposite-sign vertical dispersion~\cite{LEP1PAPER}.
During operation beam offsets were controlled to within a few microns by
beam-beam deflection scans.   The dispersions were measured each year,
and also calculated within MAD.  The shifts in
centre-of-mass energy are estimated to be less than 2~MeV.
This value is taken as the systematic uncertainty,  with a 50~\% correlation
between years.

%% file: fl_ana.tex
\section{Constraining the Magnetic Model with the Flux-Loop}
\label{sec:fltest}

Each of the main dipoles had a copper loop embedded in the lower pole.
These were connected in series throughout each of the octants of LEP.
The flux variation in each octant was measured by a digital integrator.
This system constituted the  {\it flux-loop} (FL)~\cite{FLREF1}.
It is estimated that the FL sampled  96.5\% of the total bending field.
In the LEP~1 era,  prior to the routine use of RDP,  dedicated FL cycles were 
regularly performed. These included a polarity inversion of the dipoles
in order to determine the remanent field.   From these cycles, and through
expression~\ref{eq:ebdef}, the absolute energy
scale at the Z was determined with $\sim 10^{-4}$ precision~\cite{Z91}.

The need to extrapolate up to fields equivalent to $\Eb = 100~\rm{GeV}$
implies that it is impractical to use the FL as a tool of absolute 
energy calibration at LEP~2.   Instead ramps were made from fields
corresponding to RDP energies, up to fields equivalent to 100~GeV and beyond.
In the analysis the evolution of the (almost) total bending field,  
as measured 
by the FL, can be compared to that predicted by the NMR model, thereby
providing a constraint of the LEP~2 energy scale.

\subsection{Measurement Procedure and Datasets}

 Measurements using the FL system were carried out during 
dedicated experiments, without beam, in each of the years of LEP~2 running.
In each measurement the excitation current was ramped through a series 
of increasing values, which mostly corresponded to the physics energy 
settings, 
and the readings of the FL recorded in each of the eight LEP octants. 
The corresponding values of the 16 NMRs were also recorded. 
A summary of the experiments is 
given in table~\ref{tab:fldata}. Measurements were made in the region 
of 41 to 61 GeV, that is, in the region where there are also RDP data,
as well as at higher energies. Also given in table~\ref{tab:fldata} is 
the corresponding highest equivalent beam energy used for each year. 
In 1996 several FL measurements were also made, with equivalent beam 
energies up to 86 GeV. These were analysed 
on-line and are not part of the datasets considered here. 

The FL measurements used in the analysis 
are the averages over the individual measurements 
made in each of the eight octants of LEP. However, particularly in the
later years, not all of the octants were fully functioning
due to radiation damage. Also, as
discussed in section~\ref{sec:nmrpol}, the number of available 
NMR probes at any one time varied for the same reason.

 \begin{table}[htb]
\begin{center}
\begin{tabular}{|l|c|c|}
\hline
 Year      & Number of Ramps  & Highest Equivalent \Eb\ [GeV] \\
\hline
 1997      &  5                      & 101    \\
 1998      &  18                     & 101    \\
 1999      &  18                     & 103    \\
 2000      &  10                     & 106    \\
\hline
\end{tabular}
\end{center}
\caption{The number of FL ramps made in each year, together
with the corresponding highest equivalent beam energy measured in that year. }
\label{tab:fldata}
\end{table}

\subsection{Fitting Procedure}

Fits may be performed between the NMR probes and the FL
in the well-understood region of 41--61~GeV.   These fits can be used to
predict the average bending field as measured by the FL at the
settings corresponding to physics energies.
If the NMR probes can predict the FL field, and if the beam
energy is proportional to the total bending field, then it is a 
good assumption that the probes are also able to predict the 
beam energy in physics. The FL cannot be used to predict
the beam energy in physics directly, since neither the slope nor the offset
of the relationship between measured field and beam energy are known
with sufficient precision to make the extrapolation needed over the
$\sim 50$~GeV interval. 

Two methods are used to make an estimate of any possible 
non-linearity, with beam energy, in the procedure used in 
calculate \EOnmr\ at physics energies.

In {\bf method A}, for each FL excitation current 
the equivalent beam energy from the dipoles, \EOnmr, 
is determined from the NMR probe readings and expression~\ref{eq:epol}, 
using the values of $a^i$ and $b^i$  established from the RDP data.
In the 41-61~GeV interval of each FL ramp this is fitted against
\Efl, the equivalent energy as estimated by the FL, where

\begin{eqnarray}
\label{eq:flnew}
 \Efl  =  c + d \Bfl .
\end{eqnarray}

\noindent Here $c$ and $d$ are the fit coefficients, and \Bfl\ the FL
reading averaged over all available octants.   
The fit results are then used to find \Efl\ at high energy,
and this is compared with the value from the NMRs.

In {\bf method B} each NMR probe $i$ is used to make an
estimate of the FL
reading, \Bflpi\, through the linear relation

\begin{eqnarray}
\label{eq:correlcd}
 \Bflpi   =  e^i + f^i \Bnmri ,
\end{eqnarray}

\noindent where $e^i$ and $f^i$ are determined from a fit to \Bfl\
in the range 41--61~GeV.
These estimates,  averaged over 
all available probes irrespective of which octants they are in, 
give a mean NMR prediction of the FL reading, \Bflp.   
The difference between \Bflp\ and
\Bfl\ at high energy can be expressed as an energy through
multiplying by the ratio of average slopes
in expressions~\ref{eq:epol}
and~\ref{eq:correlcd} ($\langle b \rangle / \langle f \rangle $), 
to give a measure of $\Efl - \EOnmr$.

Both methods provide a comparison between the FL and the NMRs  
at high energy, and thus are sensitive to non-linearities in the magnetic
model.   The NMR data are however used in a different manner by the two
procedures, and this provides robustness against, for example, 
fluctuations caused by the varying number of probes available at each
measurement point.

\subsection{Comparison of FL Results Using Different NMRs}

\begin{figure}[htb]
\begin{center}
\epsfig{file=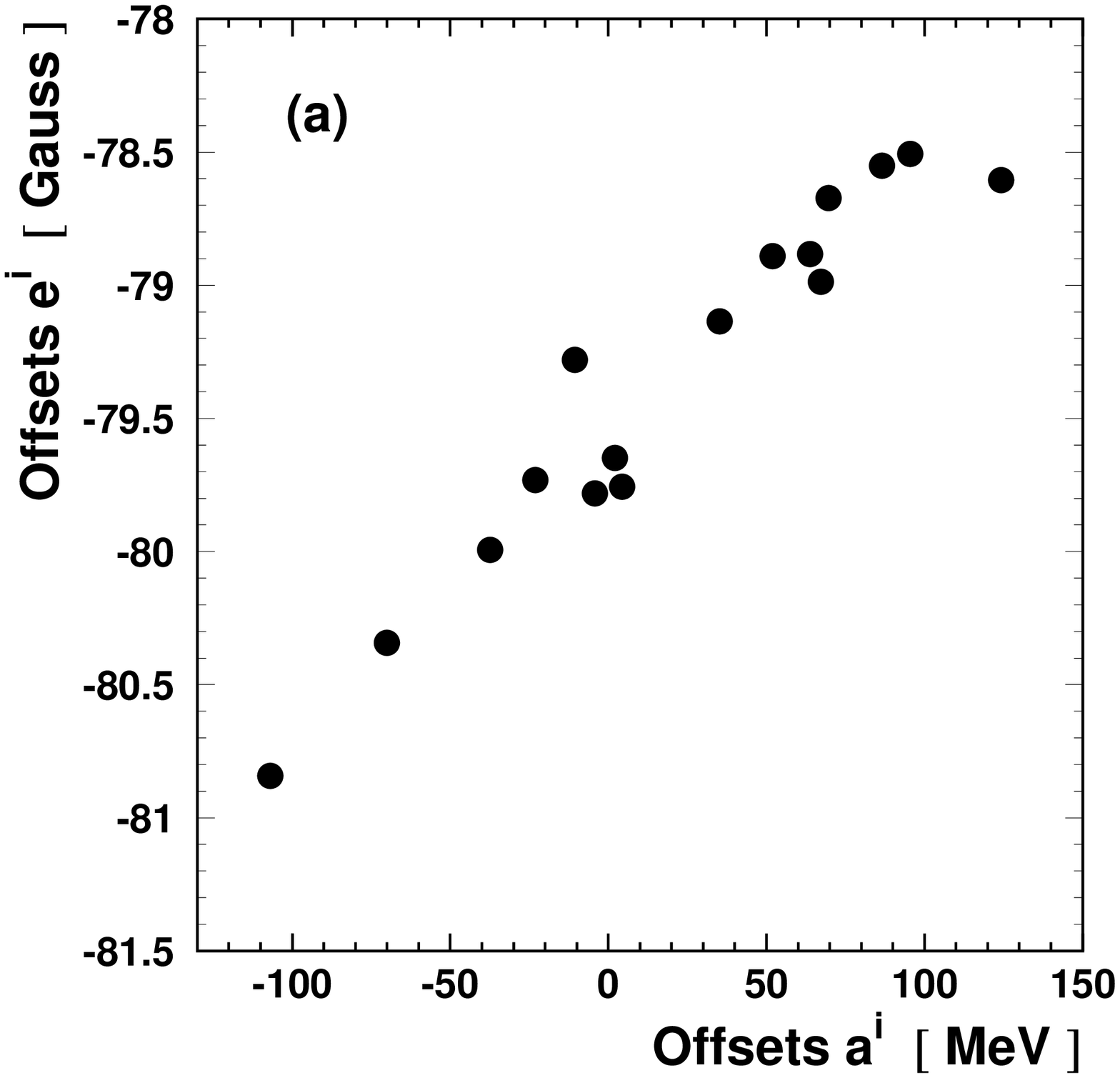,width=0.48\textwidth}
\epsfig{file=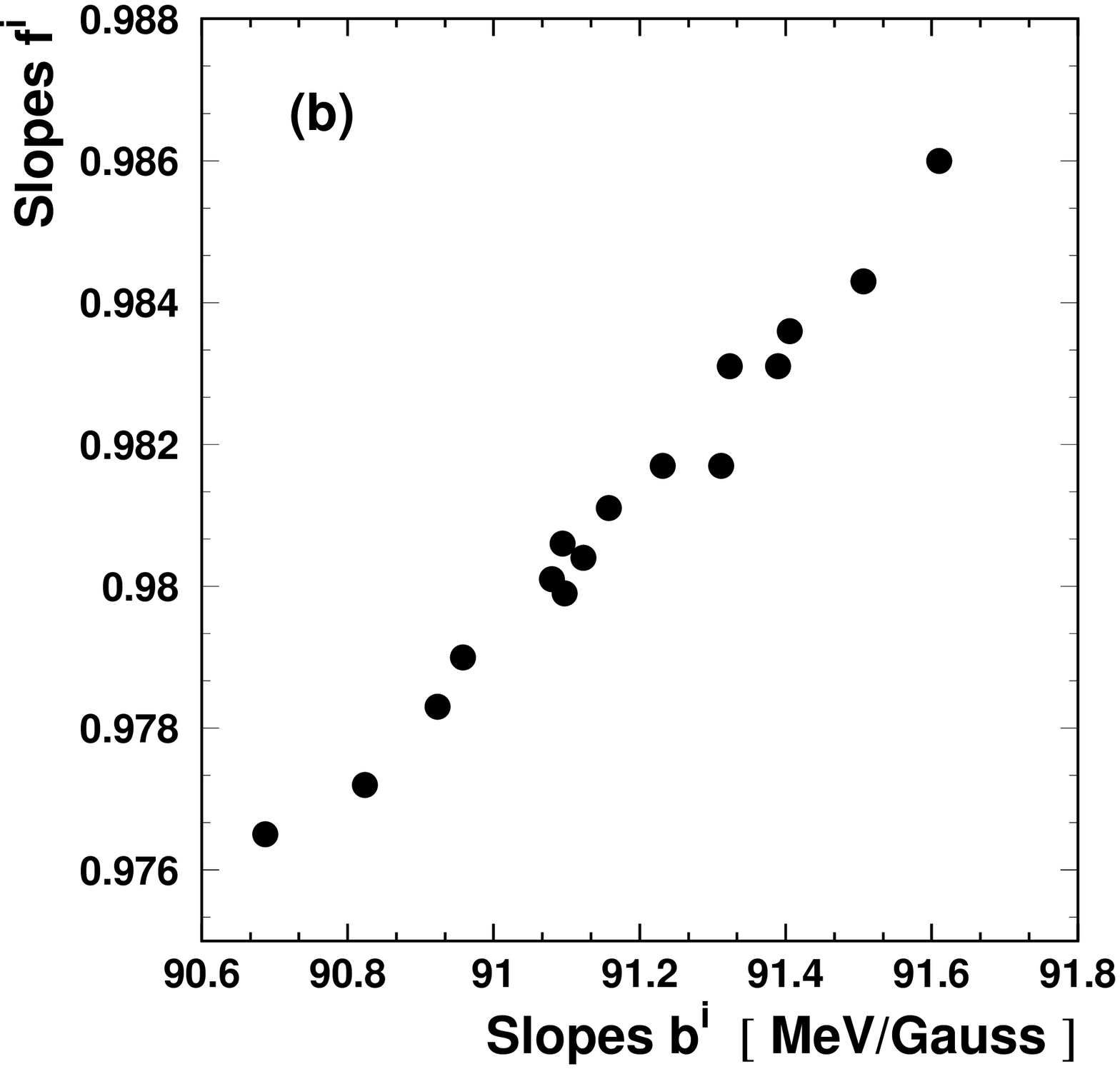,width=0.48\textwidth}
\caption{The offsets (a) and slopes (b) of 
equations~\ref{eq:epol} and \ref{eq:correlcd} 
comparing the field measured by each NMR probe
with the RDP and  FL measurements.
The values shown are averages over all the 
FL measurements. There is one entry per NMR probe 
in each plot.}
\label{fig:correl}
\end{center}
\end{figure}

A strong correlation is expected
between the offsets $a^i$ and slopes $b^i$ in equation \ref{eq:epol}
from the RDP calibration, and the offsets $e^i$ and slopes $f^i$ in
equation~\ref{eq:correlcd} from the FL.
The fitted parameters for each NMR 
are shown in figure~\ref{fig:correl}, and the expected
correlation is visible. The average offset, $\langle e \rangle$, 
is $-79.35$~Gauss, 
with an RMS spread over the 16 values of 0.69~Gauss. This
offset corresponds to the 7~GeV nominal beam energy setting
at the start of the FL ramp.
The average slope, $\langle f \rangle $, is 0.9810, with an RMS spread over
16 NMR probes of 0.0026. The field plates placed below the NMRs, in order
to improve the uniformity of the field, cause the
slope to be 2\% different from unity.

\begin{figure}[htb]
\begin{center}
\epsfig{file=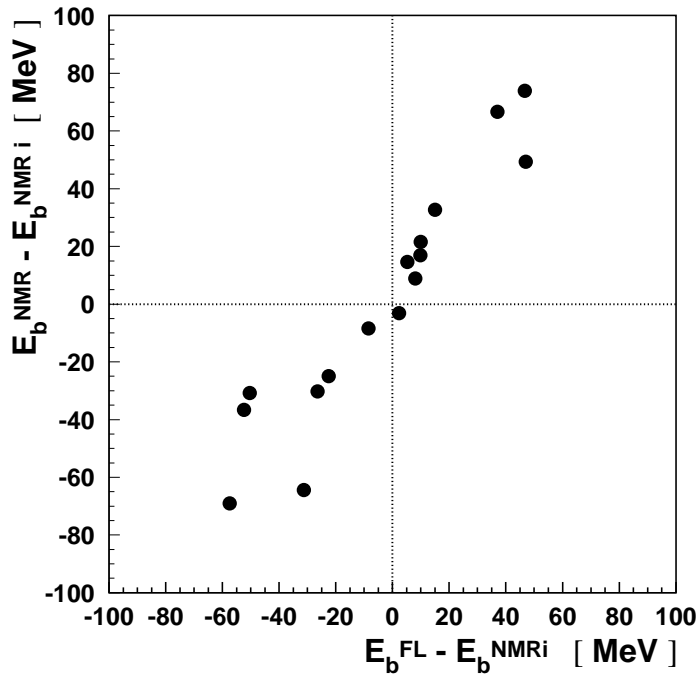,width=0.60\textwidth}
\caption{ For each NMR probe, the difference between the FL energy, from
method B, and that
estimated from the NMR probe is plotted against the difference between the
mean NMR energy, averaged over all probes, and that from the individual
probe in physics running. 
The differences are calculated at a beam energy of 100 GeV
} 
\label{fig:fl_bynmr}
\end{center}
\end{figure}

 The behaviour of the different NMRs can be seen in figure~\ref{fig:fl_bynmr}.
This shows the differences between the FL
estimate of the beam energy, calculated with method B, and the 
individual NMR estimates, \EOnmri, plotted against the corresponding 
probe residuals of figure~\ref{fig:nmrscatt}.
The comparison is made at a beam energy of 100 GeV.
Again, a strong correlation is observed. 

 These studies show that the FL measurements behave in a similar way to
the RDP measurements in terms of the results from individual NMRs and
give confidence that the FL data can be used to constrain the linearity
of the magnetic model.

\subsection{Variation of FL Results for Different Octants and Years}

\begin{figure}[htb]
\begin{center}
\epsfig{file=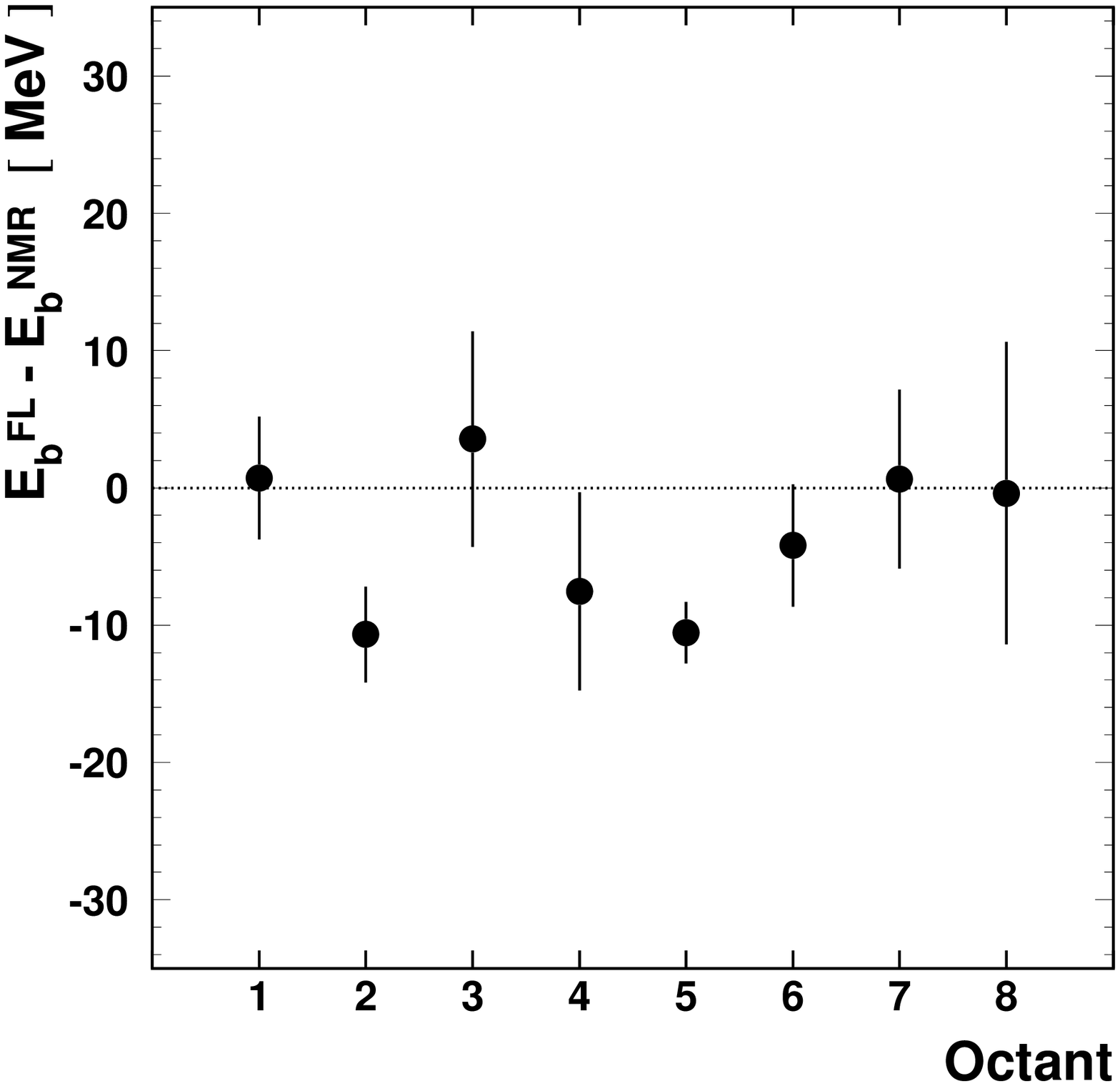,width=0.60\textwidth} 
\caption{The difference, in MeV,
between the magnetic field measured by the 
FL and predicted by the NMR probes
for each octant separately, using method A.
The differences are calculated at a beam energy of 100 GeV.
} 
\label{fig:fl_byoctant}
\end{center}
\end{figure}

The FL results used in the standard analyses 
are the averages of the available individual measurements for
each of the eight octants of LEP. The results for the individual 
octants using method A are shown in 
figure~\ref{fig:fl_byoctant}. The differences
between the FL value for each octant and the NMR values are computed
at a beam energy of 100 GeV.
The values for each octant have also been evaluated separately for each of
the four years in which there are data, and the 
errors shown in figure~\ref{fig:fl_byoctant} 
are half of the difference between the maximum and
minimum of these yearly values. 
The results are consistent between
octants, and exhibit year-to-year stability. The values from individual
octants span a range of approximately 10 MeV. The RMS of the mean
values from different octants is 5.5 MeV.

\begin{figure}
\begin{center}
\epsfig{file=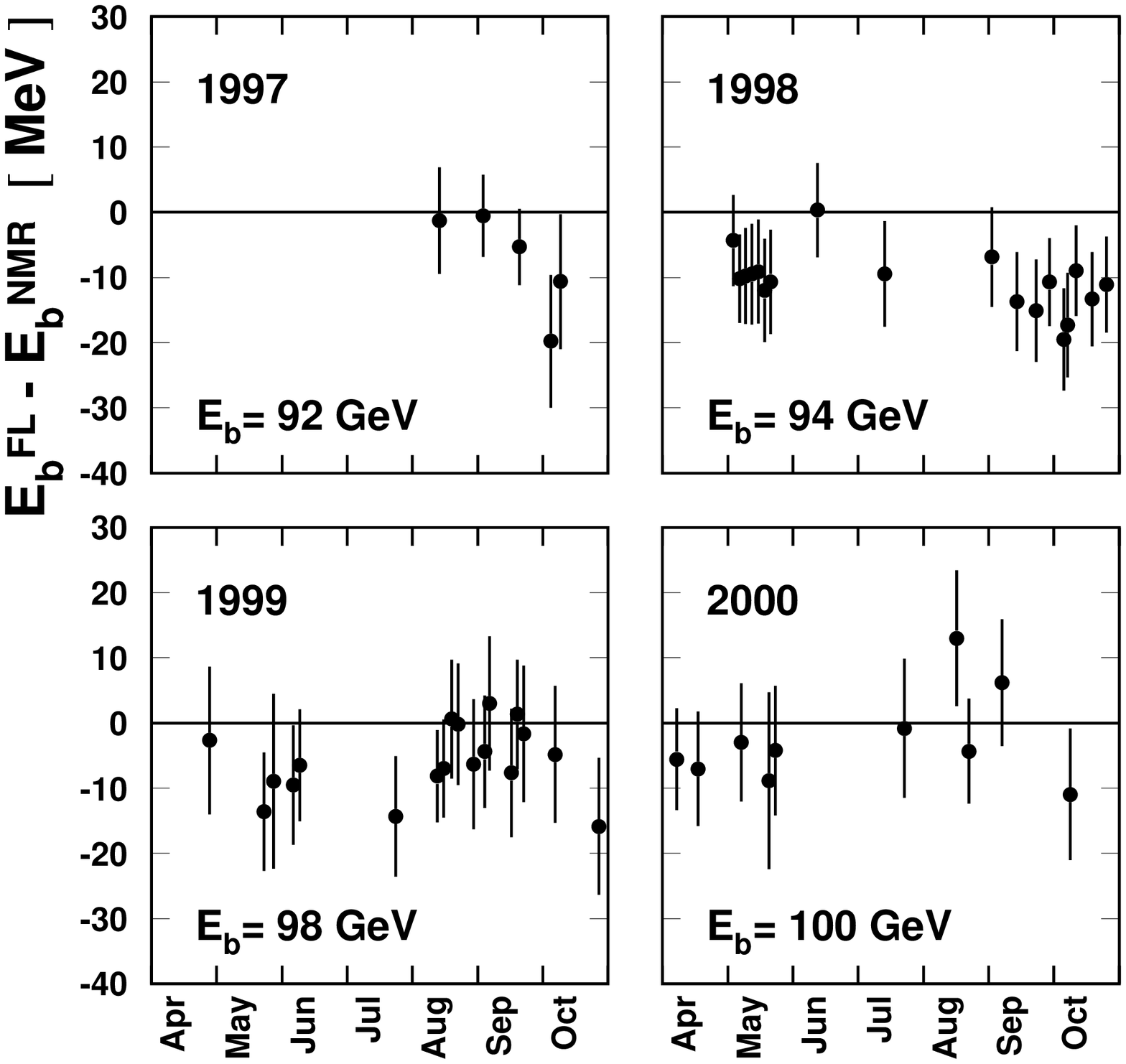,width=1.0\textwidth}
\caption{The difference, in MeV,
between the magnetic field measured by the 
FL and predicted by the NMR probes
for each of the FL measurements, using method B. The data are
shown separately for each year.
} 
\label{fig:fl_byfill}
\end{center}
\end{figure}

In figure~\ref{fig:fl_byfill} the values of $\Efl - \EOnmr$ from method B
are shown
as a function of time. Each entry corresponds to a single FL
ramp and the data from each year are separated. The beam energy
at which the differences are computed varies from year to year and is
indicated on the plot. It represents the main value at which physics data
were taken in the year. The error bars shown are the RMS values 
of the results from each of the NMR probes. 
It can be seen that there is no strong time dependence in the measurements.

\subsection{Comparison of FL and NMR Energy Model Results}

Table~\ref{tab:flbyyear1} lists the mean values
of the $\Efl - \EOnmr$ differences for each year, 
averaged over all the measurements in
that year, from method A, for a beam energy of 100~GeV. 
The mean value averaged over all data is also included. 

Table~\ref{tab:flbyyear2} presents the equivalent results
from method B.
A large part of the RMS scatter in the results comes from 
the different behaviour with energy of the NMR probes. Thus, if one or 
more of the NMR probes is not functioning for all, or part of, a 
particular measurement then this will increase the scatter. 

The two fitting methods A and B are very compatible
and the overall offset with respect to the energy model is small. 
However, the RMS values are smaller for method A,
since the values used in this method are already averaged  
over the NMR probes.

\begin{table}[htb]
\begin{center}
\begin{tabular}{|l|c|c|}
\hline
 Year             & \Efl\ $-$ \EOnmr\ [MeV]  & RMS  [MeV] \\
\hline
 1997             & 2.8                      & 4.4    \\
 1998             & -4.5                     & 6.1    \\
 1999             & -3.3                     & 6.3    \\
 2000             & -4.7                     & 12.2    \\
\hline
 All Years        & -3.3                     & 7.4    \\
\hline
\end{tabular}
\end{center}
\caption{Difference between the beam energy estimated by the FL and
that using the NMR model at 100 GeV for each year separately,
and also for all years together. The values given are from
method A.}
\label{tab:flbyyear1}
\end{table}

\begin{table}[ht]
\begin{center}
\begin{tabular}{|l|c|c|}
\hline
 Year             & \Efl\ $-$ \EOnmr\ [MeV]  & RMS  [MeV] \\
\hline
 1997             &  0.2                     & 10.3    \\
 1998             & -5.7                     & 12.6    \\
 1999             & -5.5                     & 14.6    \\
 2000             & -1.4                     & 18.2    \\
\hline
 All Years        & -4.2                     & 17.7    \\
\hline
\end{tabular}
\end{center}
\caption{Difference between the beam energy estimated by the FL and
that using the NMR model at 100 GeV for each year separately,
and also for all years together. The values given are from
method B. }
\label{tab:flbyyear2}
\end{table}

\subsection{Linearity in the High-Energy Region}

\begin{figure}[htb]
\begin{center}
\epsfig{file=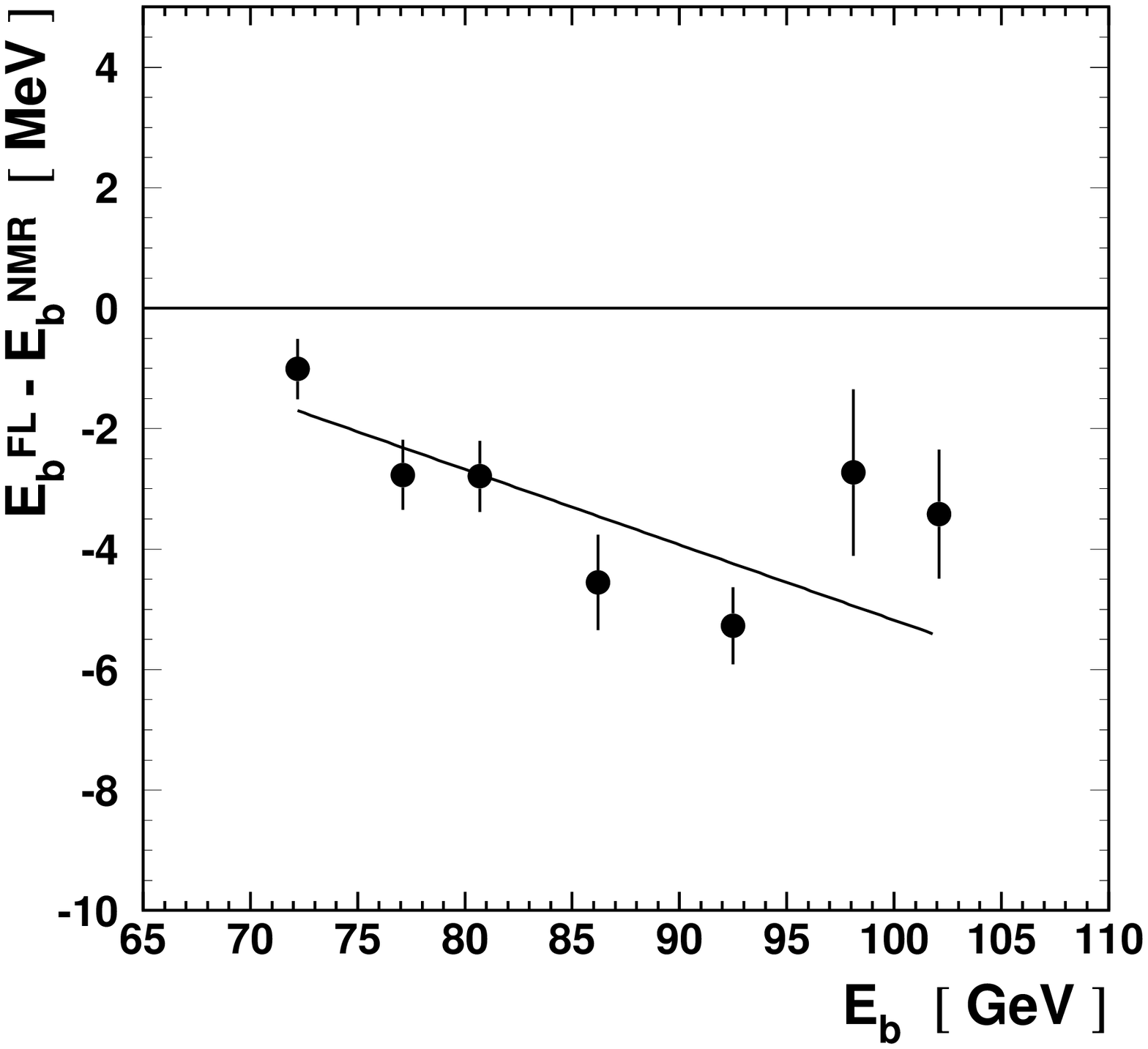,width=0.60\textwidth}
\caption{The difference, in MeV,
between the magnetic field measured by the 
FL and predicted by the NMR probes
as a function of the nominal beam energy, using
method A. Data
from all years are used. 
For plotting, certain energy points have been averaged together.
A linear fit to the differences is also
displayed.} 
\label{fig:flux}
\end{center}
\end{figure}

 The FL is the only device which allows a comparison with the NMR model
measurements over a wide range of effective beam energies.
The results of this comparison, averaged over all octants and all ramps,
are presented, as a function of \Eb,
in figure~\ref{fig:flux}. 
The error bars shown are calculated from the spread of the individual
FL measurements, over all years, at a given \Eb.
The estimate from the FL is slightly
lower than that from the NMR model and this difference grows
somewhat with increasing beam energy.
Also shown in figure~\ref{fig:flux} is a linear fit to the differences
over the range 72 to 106 GeV equivalent beam energy. This fit gives a slope 
of -0.125 $\pm$ 0.028 MeV/GeV and an offset, at a beam energy
of 100 GeV, of -5.2 $\pm$ 0.6 MeV. The $\chi^{2}$ for the fit is 13.2
for 5 degrees of freedom, giving a probability of 22$\%$.
The errors are computed from
the statistical spread of the FL measurements, and do not 
include any systematic effects.

\subsection{Robustness Tests and Systematic Uncertainties}

  Changes of the requirements in the fitting and extrapolation procedure
of method A have
been investigated. These include changing the minimum number of FL 
measurements in the range 41-61 GeV from 2 to 4, and changing the range of the
fit in the low-energy region from 41-61 GeV to either 41-57 GeV or
50-61 GeV. All these modifications to the procedure give changes 
in the difference between the FL and the NMR model at 100 GeV, and 
averaged over all data, of 3 MeV, or less. Especially in the later years
some of the octants did not always give FL data. Omitting each of the octants
in turn from the analysis changes the mean value of $\Efl - \EOnmr$ 
by less than 2 MeV.

 There is very little redundant information in the FL measurements which 
allows a rigorous study of the possible systematic uncertainties
to be performed. The accuracy of the device has previously
been estimated to be about $10^{-4}$~\cite{Z91}, which corresponds to an 
uncertainty of 10~MeV at $\Eb = 100~\rm{GeV}$.
This is compatible with the RMS values seen 
in the octant-to-octant variations, and the results of the various 
extrapolation methods used (although
part of this scatter is attributable to the behaviour of individual
NMR probes).

\begin{figure}
\begin{center}
\epsfig{file=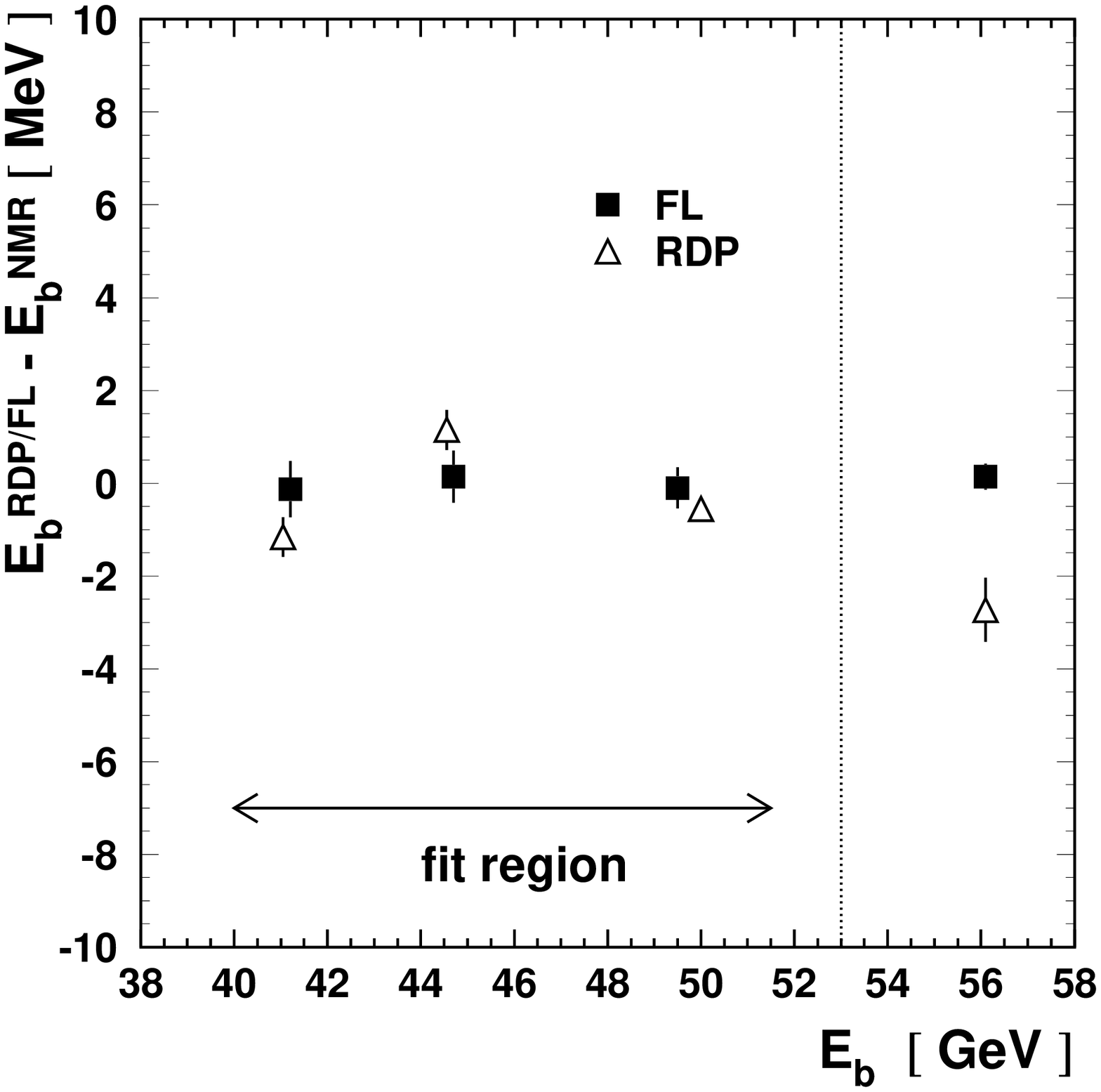,width=0.60\textwidth}
\caption{Residuals from the NMR fits to the RDP data and FL data
for all years, using method A. The error bars indicate the statistical 
scatter over the measurements. Only the data below 55 GeV are used in
the fits.}
\label{fig:residuals_epol_fl}
\end{center}
\end{figure}

 The main uncertainty in the results comes from the assumption that 
the measured FL values are linear with the excitation current,
 and thus the beam energy. 
This can only be tested where there are RDP
measurements, namely in the energy range 41-61 GeV. 
As a test of the linearity a special
fit has been made to the RDP data for all years, using 
equation \ref{eq:epol} as before, but excluding
the 55 and 61~GeV points. A similar procedure has been
carried out for the FL measurements using method A, and again not
using the 55 and 61~GeV points. 
In both cases the fits are compared with measurements at 56.1~GeV,
the mean value of the 55 and 61~GeV RDP data.
These residuals are plotted in figure~\ref{fig:residuals_epol_fl}.
The difference at 56.1~GeV between the RDP measurement and
the model, and the FL measurement and the model,
provides a linearity test over the sampled energy range.
Scaling up the observed difference in order to extrapolate to a beam energy of
100 GeV gives 15~MeV, and this is taken as an estimate of the 
uncertainty in the linearity of the FL device at high energy.
 
It is known that a small fraction of the
total bending field was not measured by the FL.   
This arose from three sources:

\begin{itemize}

\item{
The FL sampled only 98\% of the total bending field of each dipole. 
The effective area of the FL varied during the ramp 
because the fraction of 
the fringe fields overlapping neighbouring magnets changed.
The saturation of the dipoles, expressed as the change in
effective length, was measured before the LEP
startup on a test stand for different magnet cycles.
The correction between 50 and 100~GeV is of the order
of $10^{-4}$, corresponding to a 5~MeV uncertainty in the physics
energy at 100 GeV, and scaling linearly with energy for other
values.
}
\item{
 The weak dipoles matching the LEP arcs to the
straight sections contributed 0.2\% to the total bending field.
Assuming that their field was proportional to that of the standard 
dipoles between RDP and physics energies to better than 1\%, their
contribution to any non-linearity in the model is around 1~MeV.
}
\item{
The bending field of the double-strength dipoles in the injection
region contributed 1.4\% of the total. Their bending field has
been measured by additional NMR probes installed in the tunnel, 
and is found to be proportional to the bending field of the
main dipoles to rather better than $10^{-3}$, which gives a
negligible additional systematic uncertainty.
}
\end{itemize}

\noindent  The difference between FL and RDP residuals in 
figure~\ref{fig:residuals_epol_fl} may be partly caused by these unmeasured
contributions to the total bending field.  
To be conservative, however, they are 
considered as separate sources of uncertainty in the final error 
assignment.

\subsection{Summary of FL Results}

The central values of the FL analysis in the high-energy region
are taken from the fit to the data of figure~\ref{fig:flux}.

To determine the total systematic error to the FL measurement, 
it is assumed that the 15~MeV uncertainty arising from the non-linearity
comparison is independent from the estimated 5~MeV uncertainty 
associated with the bending field lying outside the FL.   Added in
quadrature these give a value of 15.8~MeV at $\Eb = 100~\rm{GeV}$.
This systematic uncertainty is taken to be fully correlated as
a function of beam energy and to increase linearly from a value
of zero at 47 GeV, where the FL measurements are normalised to
the RDP measurements. The range of FL measurements is from
72~GeV to 106~GeV, and this procedure gives an uncertainty which
grows from 7.5~MeV to 17.6~MeV over this span.

%% file: spec_setup.tex

\section{The LEP Spectrometer}
\label{sec:specsetup}

\begin{figure}[h]
 \begin{center}
  \includegraphics[width=\columnwidth]{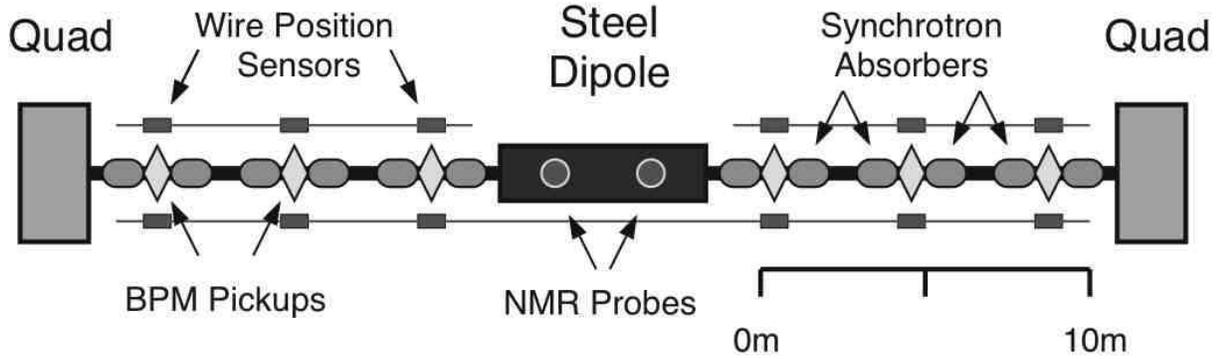}%
 \end{center}
 \caption{A schematic of the LEP spectrometer, situated between two
quadrupole magnets close to IP3.  The various components are discussed
in the text.}
 \label{fig:spectlayout}
\end{figure}

A project was initiated in 1997 to install an in-line energy spectrometer
into the LEP ring with the goal of measuring the beam energy 
to a precision
of $\sim 10^{-4}$ at $\Eb\ \sim 100~\rm{GeV}$.
By replacing two existing concrete LEP dipoles with a single 
precisely mapped steel dipole, 
and installing triplets of high-precision BPMs 
on either side, the local beam energy could be measured
as the ratio of the dipole bending field integral to the deflection angle.
The full apparatus was installed close to IP3  and commissioned in 1999,
and dedicated data taking took place throughout the 2000 run.  
A schematic of the spectrometer assembly is shown in 
figure~\ref{fig:spectlayout}.

While measuring the absolute deflection angle $\theta$ to the
required accuracy is too great a challenge, 
a high-precision relative
measurement can be performed by calibrating the spectrometer against
a low-energy reference point, \Ebr, well known through RDP,
and  measuring the change in bending angle, $\Delta \theta$,
as the beam is ramped to the high-energy point of interest.   
Then the relative difference between the energy determination from the
spectrometer,
\Espect, and that predicted by the energy model, \Enmr,
is given by:

\begin{equation}
\frac{\Espect \,-\, \Enmr}{\Enmr} \, = \,
\frac{\Ebr}{\Enmr} \, 
\frac{ \BL }{ \BLr }
\, \left( 1 \, + \, \frac{\Delta \theta}{\theta_0} \right ) \, - \, 1,
\label{eq:spec}
\end{equation}

\noindent  where \BLr\ and \BL\  are the integrated bending
fields at the reference point and measurement point respectively.
The spectrometer dipole is ramped with the LEP lattice, and so its bending
angle, $\theta_0$, remains approximately constant at a value of 3.77 mrad,
and $\Delta \theta \, <<  \, \theta_0$.

With a triplet lever-arm of roughly 10 meters, the spectrometer
BPMs must have a precision of $\sim 1$~\um\ in the bending plane and be 
stable against mechanical and electronic drifts at this same level.
This stability is only needed, however, for the few hours required
to span the data taking at the reference point and the measurement point.
How these problems were addressed is discussed in sections~\ref{sec:bpm}
and~\ref{sec:wps}.
The ratio $\BL \, /  \, \BLr$ must be known to better than
$10^{-4}$; the strategy pursued to achieve this is described
in sections~\ref{sec:mapping} and~\ref{sec:ambient}.

The beam energy at the spectrometer differs from the value of \Eb\ averaged
around the ring because of the RF sawtooth.   Correcting for the sawtooth
is an important ingredient in the spectrometer measurement.   The same
model was used as described in section~\ref{sec:rfcorr}.

\subsection{The Spectrometer Dipole}
\label{sec:mapping}

The spectrometer magnet was a custom-built 5.75 m steel dipole 
similar in design to those used in the LEP injection region.
It provided the same integrated bending field as the
two concrete core dipoles it replaced, but over a shorter length,
thereby maximising the space available for the BPM instrumentation.
As a steel cored magnet it was also less susceptible to aging and had better
stability under temperature variation.   Thermal
effects were further suppressed by water-cooling the excitation coils
through an industrial regulation circuit which limited the rise in coil
temperature, when ramping from \Ebr\ to high energy, 
to $3-4^\circ \rm{C}$.
Temperature changes were monitored by several 
probes installed at a variety of locations.

Mounted directly in the gap of the spectrometer magnet under the
beampipe were four NMR probes which continuously monitored the magnetic 
field strength.   Two of these probes were optimised for measurements
at fields equivalent to 60~GeV and below,  the other two for fields
corresponding to 40~GeV and above.
The probes were situated in precision mounts similar to those used
for the 16 NMRs of the magnetic model.
During LEP operation radiation damage
required that each probe had to be replaced two or three times during the
year.

\begin{figure}
 \begin{center}
 \epsfig{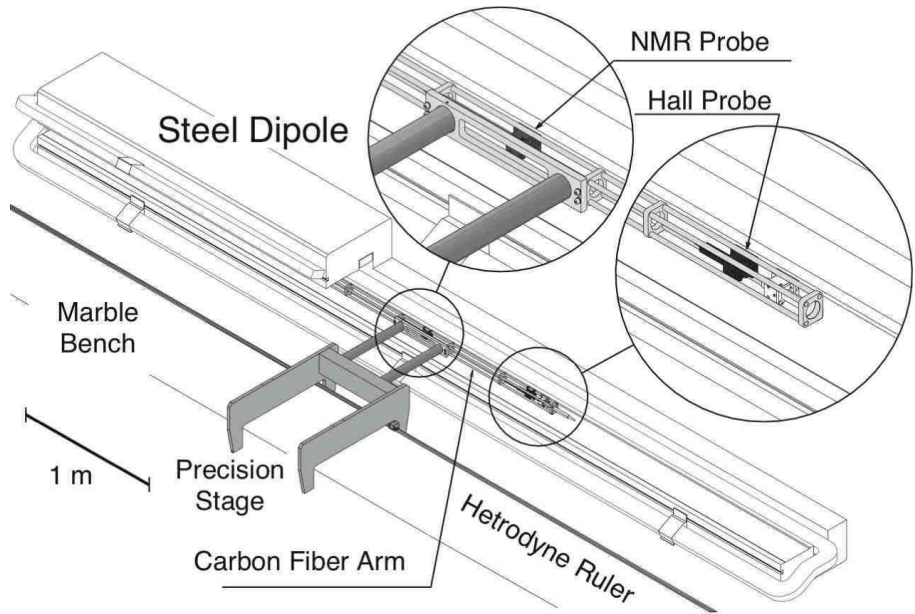}
 \end{center}
 \caption{The magnetic mapping test stand, showing inset
the components of the moving arm.}
 \label{fig:mapbench}
\end{figure}

Field maps of the total bending field were performed in the winter
of 1998-9 before the spectrometer magnet was installed in the LEP
tunnel (the `pre-installation measurement campaign'), 
and again in 2001-02 after the magnet was removed following
the LEP dismantling (the `post-LEP measurement campaign').
These maps were performed in a special mapping test stand as shown
in figure~\ref{fig:mapbench}.
Using a precision motor stage instrumented with an independent NMR 
probe mounted on a carbon fiber mapping arm, the core magnetic field 
of the dipole was sampled every 1 cm along the longitudinal axis
with an intrinsic relative precision of $10^{-6}$ for a variety of 
excitation currents and environmental conditions.
The length scale was determined to a relative precision of
$10^{-5}$ using a heterodyne ruler and 
verified with a laser interferometer.
In the end-field region where the mapping NMR probe no longer 
locked due to the high field gradient, temperature-stabilized Hall 
probes, also mounted to the movable arm, were used to complete the 
field mapping.
While these Hall probes had an intrinsic relative precision of 
$10^{-4}$, the end field represented only about 10\% of the total 
dipole bending field, and thus a relative precision per map 
of $10^{-5}$ was achieved.
With roughly 550 individual field readings taken per map, a single
dipole map required roughly 30 minutes to complete.

\begin{figure}
 \begin{center}
  \epsfig{file=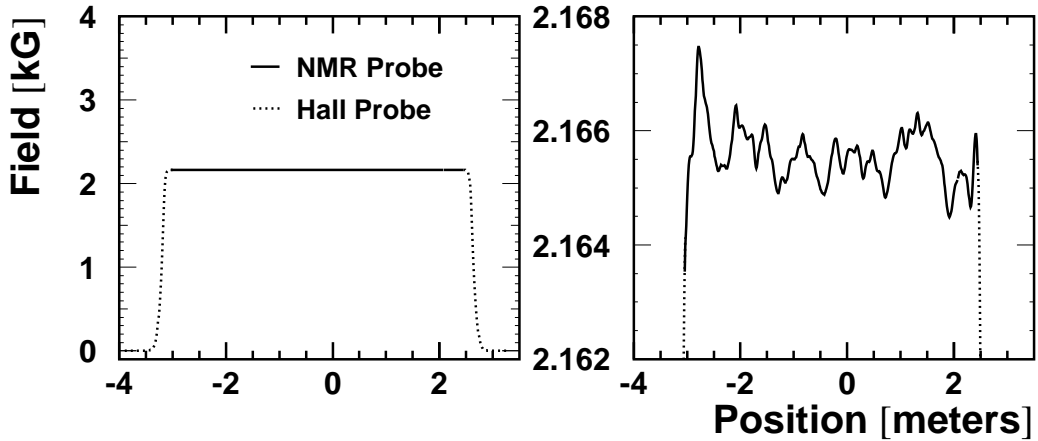,width=0.95\textwidth}
 \end{center}
 \caption{The longitudinal profile of the spectrometer dipole field at
an excitation current corresponding to 100~GeV as measured 
by the mapping campaigns.  
Also shown is a zoom of the core region for a single map.}
 \label{fig:mapprofile}
\end{figure}

The field profile at 100~GeV is shown in figure~\ref{fig:mapprofile},
indicating the extent of the end fields.   In both the mapping laboratory
and the tunnel these end fields were truncated 0.5~m away from the dipole 
with mu-metal shields.  Figure~\ref{fig:mapprofile} also includes a zoom 
into the core region for a single map, to illustrate the uniformity of
the field.

Using the results of the individual field maps, a model has been 
constructed to relate the total integral bending field of the
dipole to the local field value measured by
the four permanent reference NMR probes.
A two-parameter fit is performed between the probes and the integral
field for those excitation currents where each NMR was sensitive.
A $\sim 10^{-4} \, / \, ^\circ {\rm C}\,$ correction is included to 
account for the temperature dependence
of the end fields, which are not tracked by the NMR probes.
The model result is then taken to be 
the average of the individual predictions from all valid probe 
readings. 

\begin{figure}
 \begin{center}
\epsfig{file=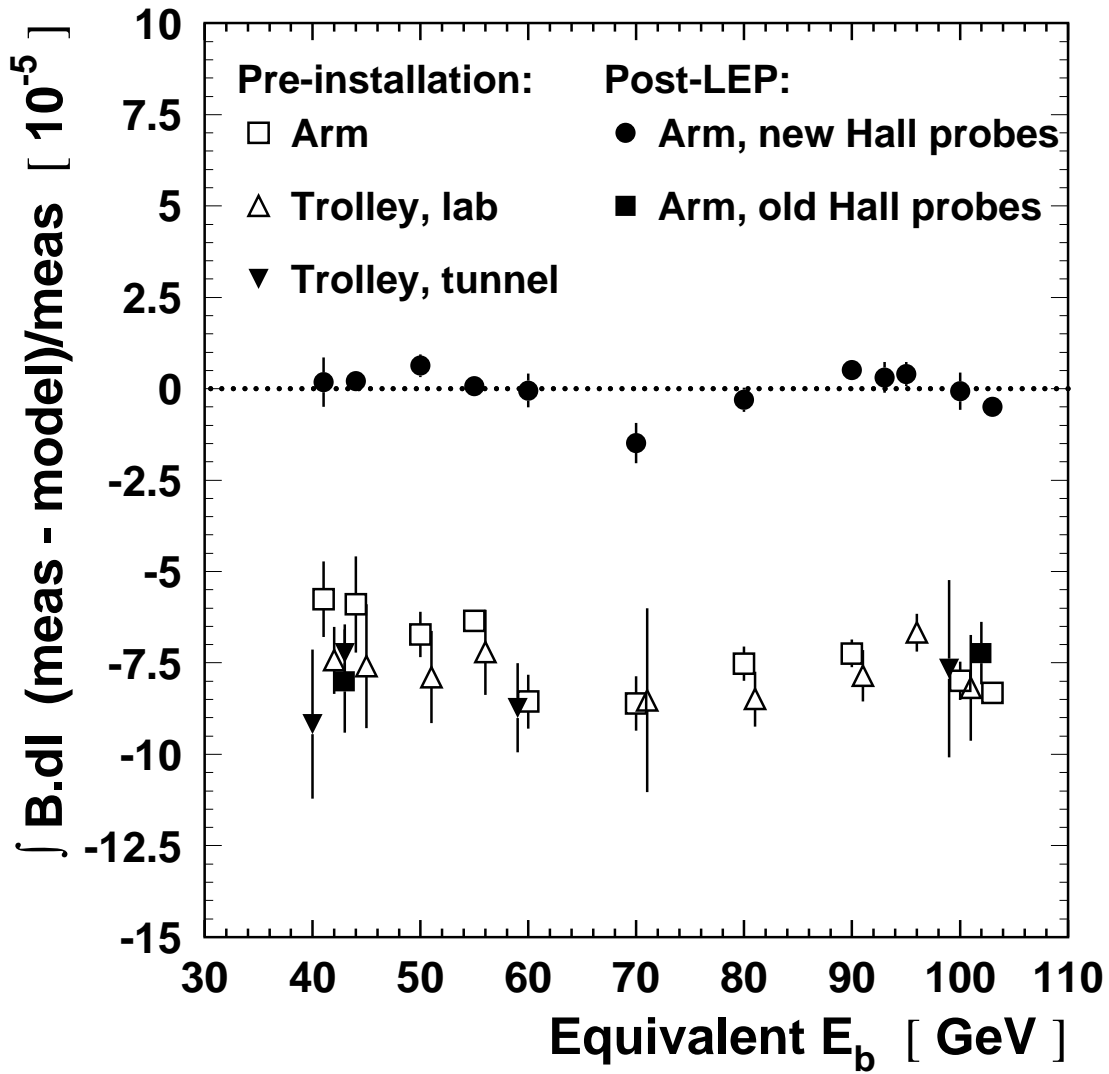,width=0.95\textwidth}
 \end{center}
 \caption{Relative residuals from the dipole magnet mapping campaigns, plotted
against energy. Results are shown for all the mapping configurations
discussed in the text. 
In all cases the model derives from a fit to the `Post-LEP: Arm, new Hall
probes' data. The horizontal positions of certain points have been slightly 
adjusted to aid clarity.}
 \label{fig:mapresid}
\end{figure}

The relative residual differences between the measured integrated dipole field,
for various datasets, and the model prediction after temperature correction, 
fitted to the post-LEP campaign data, are plotted in figure~\ref{fig:mapresid}.
Each point represents the mean value over all maps at an equivalent 
energy setting, and the error bar the RMS deviation over these maps.

The points corresponding to the post-LEP data (`Arm, new Hall probes') 
lie within $\pm 1.5 \times 10^{-5}$ of zero, and each have RMS deviations of
around $0.5 \times 10^{-5}$.  
When looking at the pre-installation data (`Arm'), however, 
an offset of 
$-8\times 10^{-5}$ can be seen. This is 
attributed to a bias associated with the Hall probe measurements
of the original campaign.  
Hall probes used in these maps had a sampling area which was
significant in size compared with the gradient of the end-field.
This deficiency was rectified in the second campaign.
 This explanation was confirmed by making a sub-set
of maps with the original instruments, the results of which are
included in the figure (`Arm, old Hall probes') and are seen to 
agree with the pre-installation data.

Additional maps were made with the arm displaced horizontally,  in
order to probe for any systematic effects which would arise from the finite
sagitta of the beam.  These show relative variations in the field integral 
of $10^{-6}$ for displacements of 1--2~cm, 
which is negligible for the energy calibration.
Excellent stability is also observed for maps made with small
vertical displacements.

Since the precision field mapping was performed in a magnetic 
test laboratory and not in the tunnel where the spectrometer operates, 
an additional {\em in situ} field-mapping technique was
developed using an NMR probe and miniature flux coil mounted
on a trolley which could be inserted directly into the LEP vacuum chamber.
A laser interferometer was used to monitor the position of the trolley.
The precision of this method is similar to that of the mapping-arm
approach.
Using this technique measurements were first made in the laboratory 
during the pre-installation campaign, with a section of vacuum chamber 
inserted into the dipole gap,  and then again in the tunnel prior to
the 1999 run.
The residuals of the field integrals measured with this method 
are also shown in figure~\ref{fig:mapresid}
(`Trolley').   These results are seen to be consistent with each other
and with the arm measurements of the pre-installation campaign,
indicating that the field the beam sensed in the tunnel was the same
as that measured in the laboratory.
Because the sampling area of the flux coil was 
significant relative to the end-field gradient,
these measurements shared a similar systematic
bias with the Hall probes of the first campaign.

In the spectrometer energy analysis, detailed in section~\ref{sec:specana}, 
it is the model fitted to the post-LEP data taken
with the new Hall probes which is used to calculate the integrated bending 
field.
As the mean values of the residuals are well determined at each magnet
setting, these are applied as corrections to the model.

More information on the spectrometer dipole and the mapping campaigns
and analysis can be found in~\cite{MAGNIM}.

\subsection{Environmental Magnetic Fields}
\label{sec:ambient}

In addition to the bending field provided by the spectrometer
dipole itself, in the LEP tunnel there were several other sources 
of magnetic fields which influenced the beam.
The single largest effect came from the earth's magnetic field,
which was measured to be $\simeq 400$~mG in the LEP tunnel.
Another contribution arose from the cables which 
provided current to drive both the 
main bending dipoles and the quadrupoles upstream from
the spectrometer, which were mounted on the tunnel wall about 1~m
from the beampipe.
The magnetic fields produced by these currents were non-negligible
and varied depending upon the nominal LEP beam energy and the specific
details of the machine optics.

\begin{figure}
 \begin{center}
 \epsfig{file=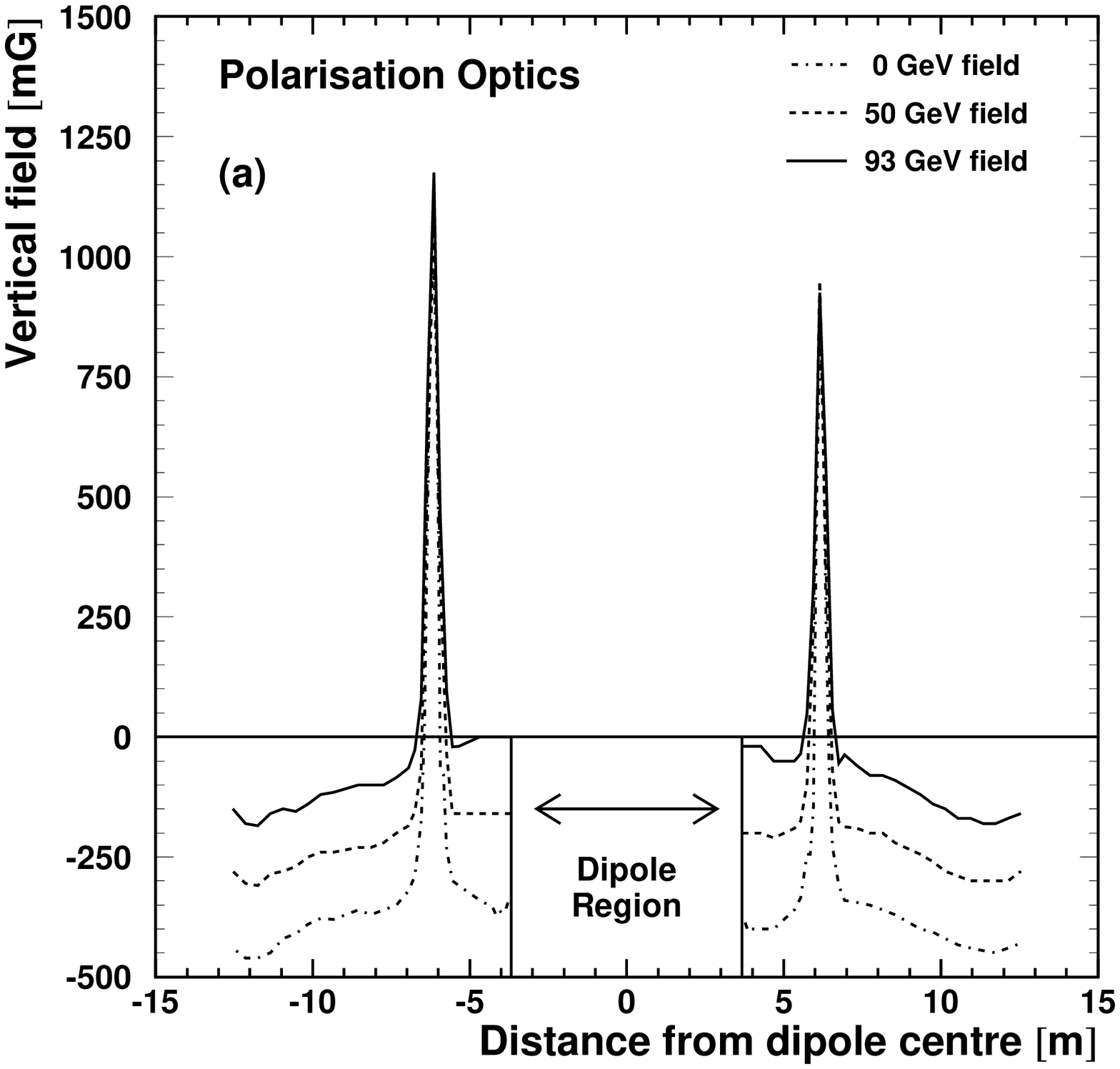,width=0.49\textwidth}
 \epsfig{file=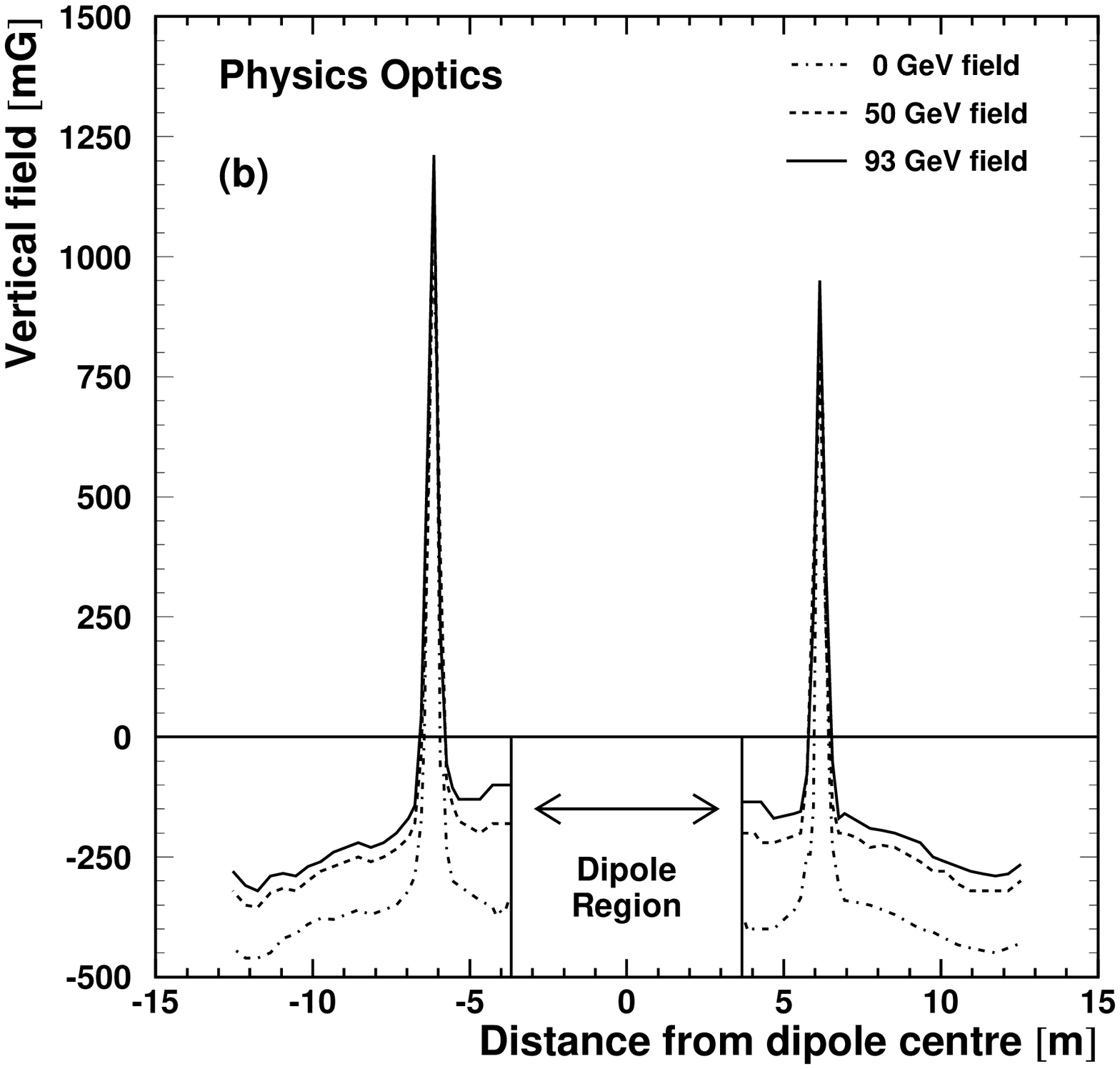,width=0.49\textwidth}
 \end{center}
 \caption{Environmental magnetic field readings in the vertical direction as
          a function of longitudinal position along the spectrometer
          for the polarisation (a) and physics (b) optics.   
          The dipole region is not shown.  The large spikes are caused by
          permanent magnets situated in vacuum pumps.}
 \label{fig:survey}
\end{figure}

The ambient field strength in the tunnel was explicitly
measured as a function of distance along the beamline 
while powering the main bending dipoles
at several nominal LEP energy settings for both physics and polarisation
optics.
The data from these vertical field surveys are shown in 
figure~\ref{fig:survey}.
The large spikes in the field, visible on either side of the
spectrometer magnet, correspond to the location of vacuum pumps
which contained permanent magnets.
Away from these spikes the absolute value of the field can be 
seen to decrease as the energy is raised, indicating that the
contribution from the magnet cables is in the opposite sense to
the earth's field.    The change in field has a stronger
energy dependence for the polarisation optics.

Each spectrometer arm was equipped with a fluxgate 
magnetometer capable of 3-axis field measurements,
situated immediately below the beampipe.
These instruments allowed any variations in the ambient 
magnetic field to be monitored with time.
Stable results are observed for all spectrometer data taking.

The effect of this ambient magnetic field was to 
bend further the beams while they traversed the BPM triplets.
Without correction, an error on the calculation of the
spectrometer bending angle of $\sim 10^{-4}$ is
made when ramping to high energy.
It is estimated that this field was monitored to a relative
accuracy of 10\%.

\subsection{Beam-Position Measurements}
\label{sec:bpm}

\begin{figure}
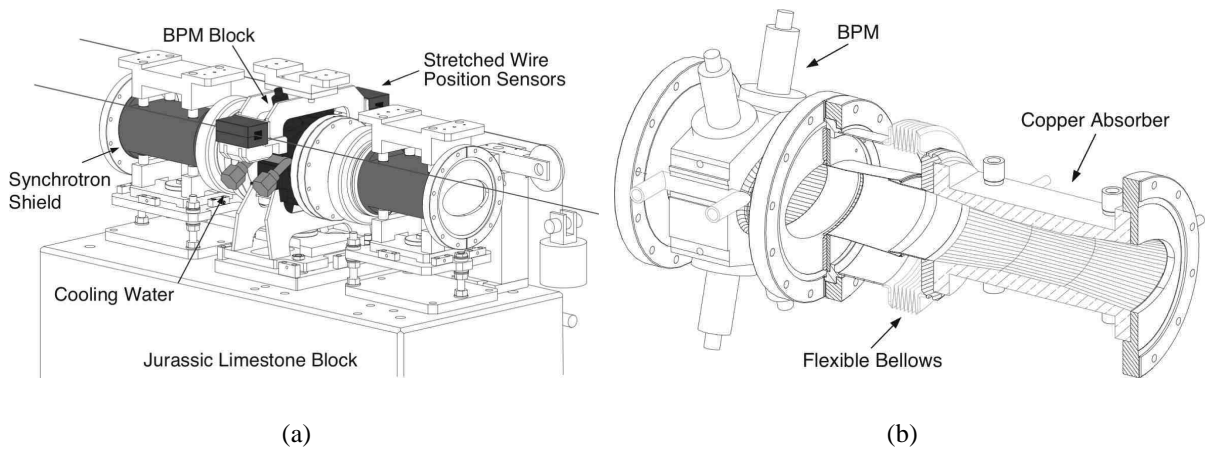

 \begin{center}
 \subfigure[]{
 \epsfig{file=BomStation.epsf,width=0.49\textwidth}}
 \subfigure[]{
 \epsfig{file=bpmstation_zoom.epsf,width=0.49\textwidth}}
 \end{center}
 \caption{A spectrometer BPM station (a), and a cut away view of the BPM
block and absorber (b). The various components are discussed
in the text.}
 \label{fig:bpmsta}
\end{figure}

Figure~\ref{fig:bpmsta}~(a) shows  one of the six BPM stations
of the spectrometer.   Each BPM-block was mounted on a stable limestone
base.  Surveys carried out after installation showed, that on average,
the blocks were well centred about their nominal positions with a 
RMS spread of 150~\microns\ in the transverse plane.    The horizontal
position of each block could be adjusted by a stepping motor with a 
reproducibility of $<$~100~\microns.

In order to ensure mechanical stability between low and high energy,
copper shielding absorbers, as shown in figure~\ref{fig:bpmsta}~(b),  
were designed to shadow the BPM pickup 
blocks from the intense synchrotron radiation present in the LEP 
environment.  During a ramp  from $E_\rm{ref}$ to high
energy the copper typically heated up by $15^\circ \rm{C}$,
whereas the presence of the shielding and 
independent temperature regulation suppressed the rise in 
the blocks themselves to $\sim 0.2 ^\circ \rm{C}$.

Any residual movement from temperature or other effects was tracked
by a stretched {\it wire-position sensor} system  (WPS).

\subsubsection{Geometry and Readout}

\begin{figure}
 \begin{center}
\epsfig{file=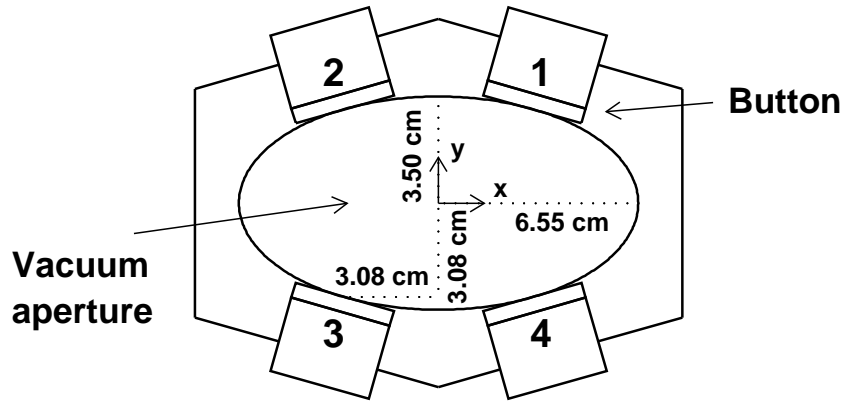,width=0.95\textwidth}
 \end{center}
 \caption{Schematic of a LEP BPM-block and its button sensors,
indicating the dimensions and the button numbering convention 
assumed in the text.}
 \label{fig:bpmblock}
\end{figure}

Standard LEP elliptical BPM-blocks were used, with four capacitive button
sensors.   The dimensions and button layout are illustrated in 
figure~\ref{fig:bpmblock}.  From the relative signal strengths of 
each button,  $S_i$ ($i=1,4$),  the BPM estimates of the beam position,
$x_\rm{BPM}$ and $y_\rm{BPM}$, are calculated according to the following
algorithm:

\begin{equation}
x_\rm{BPM} \: \propto \: \frac{ (S_1 - S_3) \, - \, (S_2 - S_4)}
{(S_1 + S_2 + S_3 + S_4)} ,
\label{eq:bpmx}
\end{equation}

\begin{equation}
y_\rm{BPM} \: \propto \: \frac{ (S_1 - S_3) \, + \, (S_2 - S_4)}
{(S_1 + S_2 + S_3 + S_4)}.
\label{eq:bpmy}
\end{equation}

To achieve the desired 1~\microns\ resolution and stability, 
customised BPM readout-cards were developed in collaboration with
industry, 
based on a design first used 
in synchrotron light source storage rings~\cite{BPMREAD}.
In the BPM electronics, the four analogue button signals from each 
BPM station were multiplexed into a common amplifier chain to
reduce the effects of gain drifts on the measured beam position.
The spectrometer BPM system, therefore, was not capable of 
turn-by-turn orbit measurements, but rather provided an integrated
mean beam-position with a frequency response of around 100 Hz.
Additional filtering was added to reduce noise and lower the overall
frequency response to below 1~Hz.
Gating allowed for the possibility of measuring both \elc\ and \pos\ positions
during two beam operation, but more stable results were obtained without
this feature enabled and with single beams.
The cards were housed in a barrack some distance from the spectrometer,
away from exposure to synchrotron radiation.  A cooling system kept 
their temperature stable during operation to  $0.1-0.2 \, ^\circ \rm{C}$.

Prior to installation, the response of the BPM readout-cards was
characterised in the laboratory using an electronic beam-pulse 
simulator.  The stability of the card response
was investigated against factors such as beam current and 
temperature.  No dependencies that would 
introduce significant systematic effects during LEP 
operation~\cite{BPMSTAB} were found.

\subsubsection{Relative-Gain Calibration}
\label{sec:ericgainbit}

\begin{figure}[htb]
 \begin{center}
 \epsfig{file=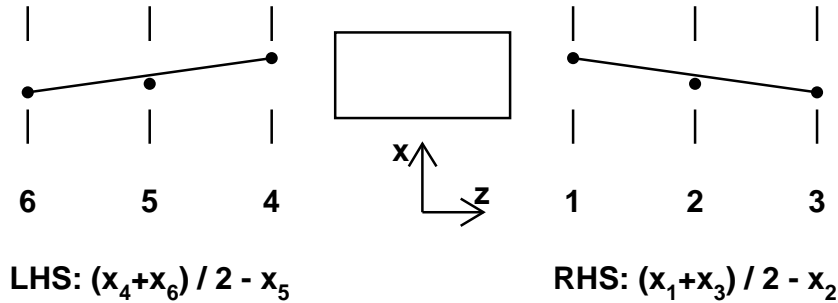,width=0.95\textwidth}
 \end{center}
 \caption{The definition of the BPM numbering scheme and bending plane 
triplet residuals.}
 \label{fig:tripdef}
\end{figure}

The response of the BPM readout differed between cards at the 
level of a few percent.   
In order to minimise errors on the measurement of the 
change in bending angle, online relative-gain calibrations were performed 
once or twice during almost all spectrometer experiments.
These calibrations consisted of using four local corrector magnets
to perform a series of beam translations and rotations, and minimising 
the {\it triplet residuals} in each arm separately, with the relative
gains of the inner and outermost BPMs left as free parameters in the fit.   
The definition of the triplet residuals is illustrated 
in figure~\ref{fig:tripdef} for the bending plane,  which also shows 
the BPM numbering definition.  Residuals can also be constructed
relating BPMs in different arms;  this was done in order to fix the
relative gain of the two triplets.   An analogous procedure was used
to determine the relative gains in the non-bending plane.

\begin{figure}
\begin{center}
\epsfig{file=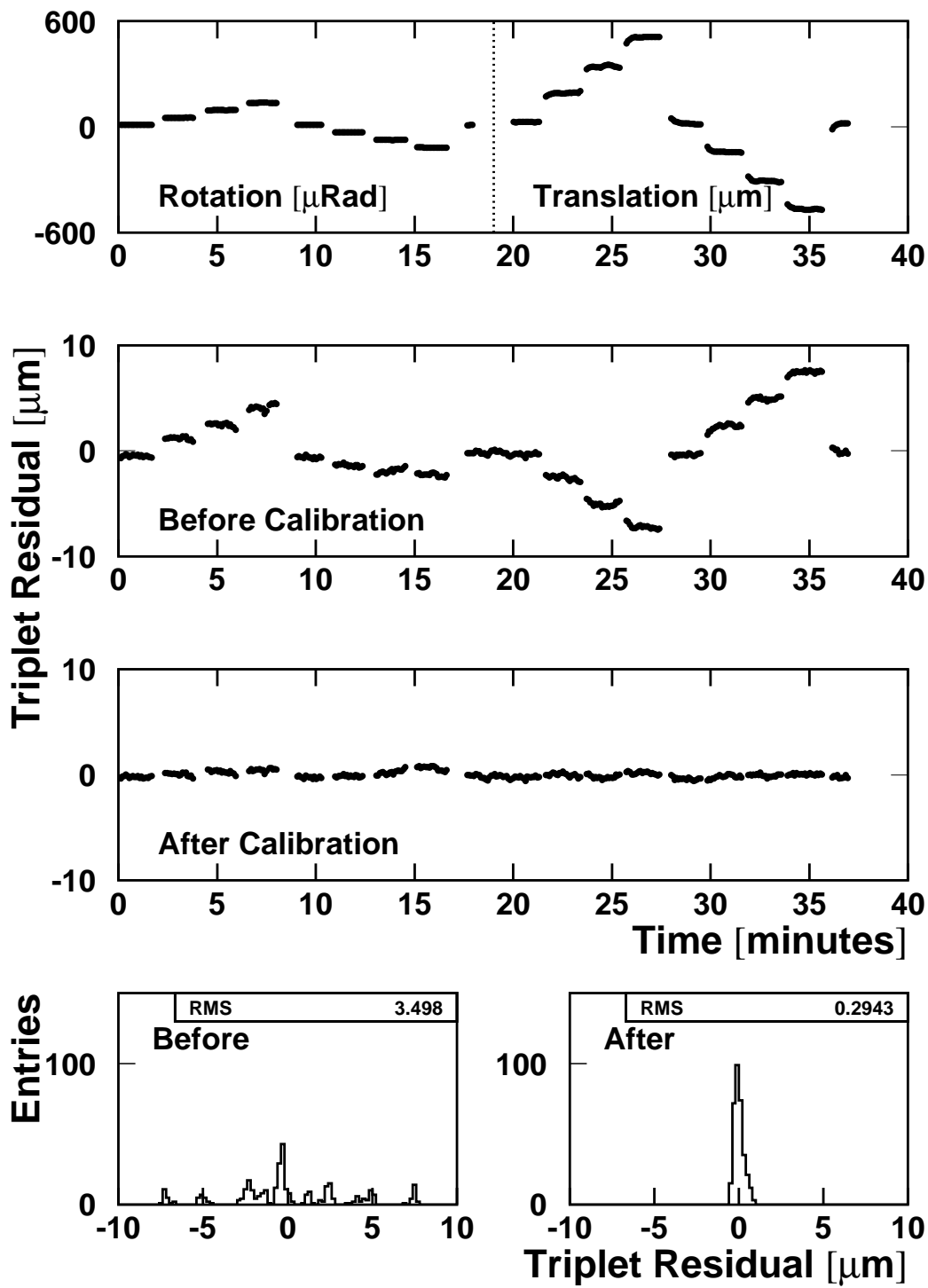,width=0.95\textwidth}
\end{center}
\caption{BPM relative-gain calibration.  Displayed, as a function of time,
are the beam movements in the calibration procedure, and the triplet
residual before and after calibration.  Also shown are the projections of
the triplet residuals.}
\label{fig:spectcal}
\end{figure}

Figure~\ref{fig:spectcal} shows a triplet residual for the same
data before and after relative-gain calibration.  The beam is 
undergoing rotations of up to 100~$\rm{\mu rad}$ and 
translations of up to 600~\um.   After calibration the
triplet residual has a width of 0.3~\microns.

Repeated calibrations during individual spectrometer experiments
indicate a relative-gain accuracy of $\simeq 0.2 \% $, and suggest
no dependence on beam energy or beam current. 
Larger variations are seen between experiments.

From calibrations performed close in time in both the horizontal
and vertical planes, cross-talk effects between the x and y
BPM readings can be studied.   
There are various possible sources of coupling between the x and y BPM
readings, including an unintentional rotation of the BPM
during installation, electrical cross-talk in the BPM readout
system and non-linear terms in the BPM response, as discussed in 
section~\ref{sec:bpmsyst}.
The data show no indication of geometrical
rotation, but do reveal  electrical cross-talk
of the order of 1\%  in some BPMs.  Coefficients have been determined
from these calibrations and then applied globally to all the experiments,
resulting in small corrections.

\subsubsection{Absolute-Gain Calibration}
\label{sec:absbpmgains}

\begin{figure}[htb]
 \begin{center}
 \epsfig{file=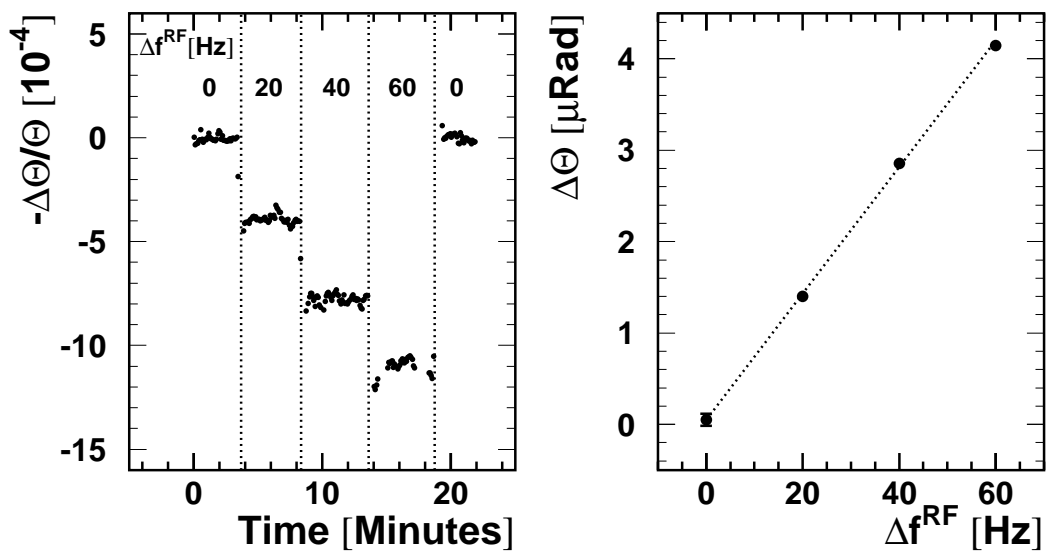,width=0.95\textwidth}
 \end{center}
 \caption{The change of bending angle measured in the spectrometer
as the energy is varied through a manipulation of the RF frequency.
The dependence is linear with a slope value consistent with expectations.}
 \label{fig:rfladder}
\end{figure}

While the {\em in situ} calibration procedure described in the
previous section can accurately determine the relative gain
of the spectrometer BPMs, the overall absolute gain is still
not constrained.
To verify that the absolute-gain scale of the BPM system
was sufficiently close to the assumed nominal value, spectrometer
data were taken while the LEP beam energy was varied through changes 
in the RF frequency.  An example of these measurements
is shown in figure~\ref{fig:rfladder}, in which the bending angle
is clearly seen to evolve linearly with the change in RF frequency, 
$\Delta f^\rm{RF}$.  From expression~\ref{eq:quadeffect},
and taking the spectrometer dipole field and local sawtooth correction
to be stable throughout the \fRF\ changes, 
the dependence is expected to be

\begin{equation}
\frac{\Delta \theta}{\Delta f^\rm{RF}} \, = \,
\frac{\theta_0}{\alpha_c \,f^\rm{RF}},
\end{equation}

\noindent in the case where the assumed absolute gain is correct.

Eight separate absolute-gain measurements were performed in 2000,
using both physics and polarisation optics.
All measurements show good linearity between the change in bending angle
and RF frequency,  and consistency amongst the BPMs in the bending angle 
measurement.
The ratio of the observed to the expected value of  
${\Delta \theta}/{\Delta f^\rm{RF}}$ for these experiments has
a mean value of $0.974 \pm 0.036$, which is consistent with unity.

An independent constraint on the absolute-gain scale was obtained 
using the stepping motors 
to move each BPM-block in turn during LEP operation.  The observed
change in triplet residual could then be cross-calibrated against
the physical movement measured by the wire sensors.  These measurements
also confirm the nominal gain to be correct with a precision of a few
percent.

For the energy calibration measurements the nominal value of the gain
scale is used with an uncertainty of $5\%$.

\subsubsection{Non-linearities and Beam-Size Effects}
\label{sec:bpmsyst}

Geometrical effects
introduce higher-order terms in the BPM response which
can be significant.
Consider an idealised circular BPM
with symmetrically distributed buttons at radius $a$,  and
a Gaussian beam of horizontal and vertical size $\sigma_x$ and
$\sigma_y$ respectively, positioned at coordinates $x,y$. 
It can be shown that, in this case, the algorithm expressed in 
equation~\ref{eq:bpmx} gives for the BPM
horizontal measurement~\cite{JOHNBPM}:

\begin{equation}
x_\rm{BPM} \: \propto \: x \, \left[ 1 \,-\, 
\left(3 \frac{\sigma_x^2 - \sigma_y^2}{a^2} \, + \, 
        \frac{x^2 - 3y^2}{a^2} \right) \right],
\label{eq:xbpmbias} 
\end{equation}

\noindent with a similar expression for the vertical coordinate.
Therefore both the beam size and quadratic position 
terms affect the measurement.

The energy calibration with the spectrometer relies on determining the change
in bending angle between \Ebr\ and high energy.
Therefore what is relevant in equation~\ref{eq:xbpmbias}
is how the higher-order terms change between
the two energy points.   The effect of the quadratic term can be 
suppressed by steering the beam at high energy as close
as possible to the position it was at \Ebr,
and ensuring that this position
is close to the centre of the BPM. 
This strategy also minimises any related errors arising from uncertainty
in the gains.

The beam-size term is more important, as $\sigma_x$ grows
with energy.  (As $\sigma_y \, << \,\sigma_x$ 
the change in the vertical beam size need not be considered.)
Furthermore, the beam size changes 
across the spectrometer, because of the evolution of the LEP 
betatron function. With the polarisation optics, the 
estimated horizontal beam sizes
at 50~GeV are 0.5~mm and 1.2~mm, for BPM~6 and BPM~3 respectively.  
At 90~GeV, these become 0.9~mm and 2.0~mm. 
Therefore, the bias to the position measurement over the energy step
is different across BPMs, and for a non-centred beam an 
apparent change in bending angle results.

To examine the problem in detail, a simulation program has been developed
to model the BPM response~\cite{JOHNBPM}.  In the case of a circular BPM this
gives results consistent with expression~\ref{eq:xbpmbias}.
For the elliptical BPMs of the LEP spectrometer, it is concluded that
the systematic effects introduced in the energy measurement are
small, provided that the beam passes within $\sim 1$~mm of the BPM centres
and is re-centred to better than a few 100~\um\ between the two energy points.

\subsection{The Wire-Position Sensor System}
\label{sec:wps}

Given the stringent 1~\um\ requirement on the stability of the
BPM system, additional instrumentation was installed to 
monitor independently  the BPM positions.
As shown in figure~\ref{fig:spectlayout}, the position of each BPM-block
was measured in both the horizontal and vertical plane by a pair of
stretched-wire capacitive-position monitors.
One of the two wires spanned the entire
30 meter length of the spectrometer apparatus to give an
independent reference line.
A pair of sensors mounted on either side of each block, around 30~cm apart,
allowed the effects of thermal expansion to be 
differentiated from relative transverse motion.
Six additional sensors (not shown), mounted on 
invar~\footnote{Invar is a 36\% nickel 64\% iron alloy with 
low thermal expansion
properties.}
supports, placed on the limestone bases,
provided reference measurements of the wire position,
independently of the BPM-blocks.

The intrinsic resolution of the sensors was found to be 
better than $0.2\um$.  The absolute value of the gains and
their stability with time were measured in the laboratory
with a moving stage and laser interferometer~\cite{JANTHESIS}.

\begin{figure}
\begin{center}
\epsfig{file=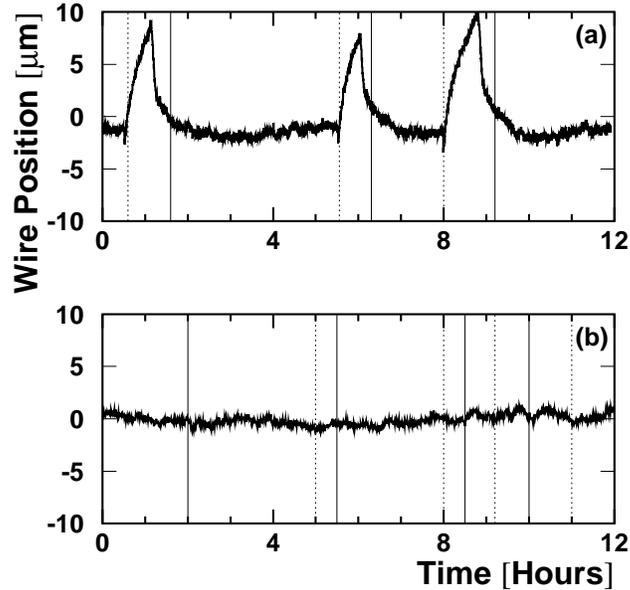,width=0.58\textwidth}
\end{center}
\caption{The response of a reference WPS over two 12-hour periods before (a) 
and after (b) additional synchrotron shielding was installed, and the 
wire centred within the sensor.
The vertical solid lines indicate the declared start-of-fill and the dotted
lines indicate a beam dump.  Note that (b) encompasses
fills in which actual spectrometer measurements were performed.}
\label{fig:wpsstab}
\end{figure}

During commissioning of the spectrometer, the WPS system was observed 
to be unexpectedly sensitive to the LEP environment.  
Figure~\ref{fig:wpsstab}~(a)
shows the response of a reference sensor against time, throughout
several successive LEP fills. Rapid positive changes in apparent position
are seen, coincident with injection and ramp, followed by rapid decreases
after beam adjustment. During the fills themselves apparent position
drifts of several microns sometimes occur.
Investigations showed this behaviour not to be physical; 
rather it was induced by a change in the dielectric constant
of air, brought about by the ionising effects of the synchrotron 
radiation~\cite{JANTHESIS}.
By installing additional synchrotron shielding, and taking care to centre
the wires in the sensors, these jumps were suppressed, as is displayed in
figure~\ref{fig:wpsstab}~(b), which shows the sensor reponse during
several fills in which actual spectrometer measurements were performed.

\begin{figure}
 \begin{center}
 \epsfig{file=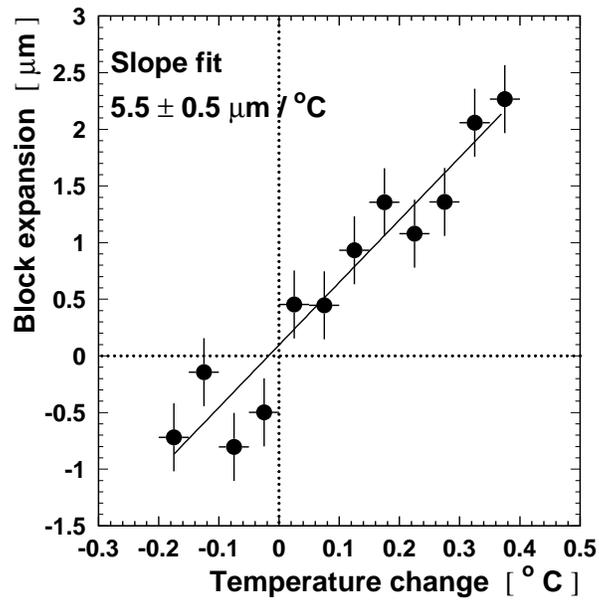,width=0.53\textwidth} 
\end{center}
 \caption{BPM-block expansion as measured by the WPS system
against temperature change during spectrometer measurements.
The data are averaged over all six stations.   A linear fit is
superimposed.}
 \label{fig:temp_vs_wps}
\end{figure}

In figure~\ref{fig:temp_vs_wps} is shown the BPM-block expansion,
as measured by the WPS system, plotted against the change in
block temperature for the spectrometer
ramps from low to high energy in 2000.  A clear 
linear dependence is seen, which agrees with the expected
expansion coefficient of aluminium to $25 \%$.

%% file: spec_ana.tex
\section{\Eb\ Measurement with the LEP Spectrometer}
\label{sec:specana}

\subsection{Datasets}

The dataset of experiments with usable spectrometer data at two
or more energy points ({\it multi-point}) 
consists of 18 single-beam fills distributed 
uniformly  throughout the 2000 LEP physics run.  
From these
experiments, two largely overlapping samples are defined.
The {\it high-energy} sample is made up of 17 fills in which 
spectrometer data were taken under stable conditions at
both 50~GeV and high energy, typically 93~GeV but
sometimes 90~GeV or 97~GeV depending on the available RF voltage.
In 5 of these fills, the 50~GeV point was calibrated by RDP.
The {\it low-energy} sample contains 8 fills with spectrometer
data at two or more energy points between 41~GeV and 61~GeV,
consisting of 21 such points in total. In this sample,
15 energy points in 6 fills were calibrated by RDP.
In the high-energy sample, some data were also recorded at 
intermediate energies of 70~GeV and 80~GeV.   
In total 10 (5 \elc, 5 \pos) of the fills were
taken with the physics optics, and 8 (5 \elc, 3 \pos)
with the polarisation optics.   
The important details of the multi-point fills are summarised
in table~\ref{tab:specfills}.  This table also lists those
fills used for the {\it bending-field spreading} (BFS) calibration,
described in section~\ref{sec:bfsmeas}.

\begin{table} 
\begin{center}
\begin{tabular}{|l|c|c|c|c|c|c|c|} \hline
Fill & Date   & Optics & Particle &  \Eb\ of measurements [GeV] 
& \multicolumn{3}{c|}{Interest of experiment} \\ 
\cline{6-8}
     &         &        &          &                                  &   
\hspace{0.1cm} HE \hspace{0.1cm}  & \hspace{0.1cm} LE \hspace{0.1cm}        &  BFS \\ \hline
7129 &  11 May &  Pol   &   \elc   & 41 (P), 45 (P), 50 (P), 70, 93   & $\bullet$  & $\bullet$   & \\
7251 &  25 May &  Pol   &   \elc   & 41 (P), 45 (P), 50 (P), 70, 93   & $\bullet$  & $\bullet$   & \\
7391 &  8 June &  Phy   &   \pos   &       50, 93                     & $\bullet$  &             & \\
7491 & 18 June &  Phy   &   \elc   &       50, 93                     & $\bullet$  &             & \\
7519 & 21 June &  Pol   &   \elc   &    41 (P), 50 (P), 93            & $\bullet$  & $\bullet$   & \\
7676 &  6 July &  Phy   &   \pos   &       50, 93                     & $\bullet$  &             & \\
7833 & 20 July &  Phy   &   \elc   &       50, 93                     & $\bullet$  &             & \\
7835 & 20 July &  Phy   &   \elc   &       50, 93                     & $\bullet$  &             & \\
7929 & 26 July &  Pol   &   \elc   &       41 (P), 50                 &            & $\bullet$   & \\
7931 & 26 July &  Phy   &   \pos   &       50, 93                     & $\bullet$  &             & $\bullet$ \\ 
8221 & 21 Aug  &  Phy   &   \pos   &       50, 90                     & $\bullet$  &             & \\ 
8224 & 21 Aug  &  Phy   &   \pos   &       50, 90                     & $\bullet$  &             & \\
8368 &  4 Sept &  Pol   &   \elc   &  41 (P), 50 (P), 55 (P), 61, 90  & $\bullet$  & $\bullet$   & \\
8443 & 10 Sept &  Pol   &   \pos   &  50, 60, 70, 80, 90              & $\bullet$  & $\bullet$   & \\
8444 & 10 Sept &  Pol   &   \pos   &  50, 60, 70, 80, 90              & $\bullet$  & $\bullet$   & \\
8556 & 25 Sept &  Pol   &   \elc   &  45 (P), 50 (P), 55 (P), 93      & $\bullet$  & $\bullet$   & \\ 
8559 & 25 Sept &  Pol   &   \pos   &       50, 90                     & $\bullet$  &             & \\ 
8566 & 26 Sept &  Phy   &   \elc   &       50, 97                     & $\bullet$  &             & $\bullet$ \\
\hline
\end{tabular}
\caption[]{
Fills from the 2000 run used in the spectrometer analysis, indicating date,
optics, particle type and energy points considered.
In the `\Eb\ of measurements' column `(P)' signifies that the energy 
was calibrated with RDP.  `Interest of experiment'
indicates which of the datasets the fill belongs in:
high-energy data set (`HE'), low-energy data set (`LE'),
or BFS calibration (`BFS').
}
\label{tab:specfills}
\end{center}
\end{table}

\subsection{Characteristics of the Multi-Point Data}

At each energy point in the multi-point fills, a period of data taking
where the beams were centred and stable is chosen, and
the spectrometer data analysed. 
Table~\ref{tab:specstab} shows the mean and RMS variation of 
certain important parameters between 
measurements for the dataset, such as beam position and BPM-block temperature.
It can be seen that for both the low and high-energy samples 
the beam was well re-positioned, and that the mechanical
stability of the apparatus remained good.
The change in bending angle when ramping between energies is
found to be small, with typical values $|\Delta \theta | < 1-2 \, {\mu \rm{rad}}$.
As a preliminary step to further analysis, corrections are applied to the BPM
readings to account for mechanical shifts,  as sensed by the
WPS system, and to compensate for the extra bending
in the spectrometer arms induced by the ambient magnetic field.

\begin{table}
\begin{center}
\begin{tabular}{|l|cc|cc|} \hline
Quantity   & \multicolumn{2}{c|}{Low energy} & 
             \multicolumn{2}{c|}{High energy} \\ \cline{2-5}
                                         &  Mean & RMS & Mean & RMS \\   \hline
Beam-position change in x [\microns]             & -30   & 190 & -57  & 145 \\
Beam-position change in y [\microns]             &  -1   & 230 & -43  & 184 \\
BPM-block temperature change [$^\circ \rm{C}$]   &-0.07  & 0.06& 0.16 & 0.15\\
BPM-block expansion [\microns]                   &-0.65  & 1.22& 1.37 & 1.33\\
Dipole-core temperature change [$^\circ \rm{C}$] & 0.71  & 0.58& 3.45 & 1.24\\   \hline
\end{tabular}
\caption[]{Stability of key parameters in the spectrometer
data set.  The values refer
to the change in parameter value between lowest and highest
energy point considered.  
The BPM quantities are calculated by taking the station with the maximum
excursion in each experiment.
}
\label{tab:specstab}
\end{center}
\end{table}

\begin{figure}[htb]
\begin{center}
\epsfig{file=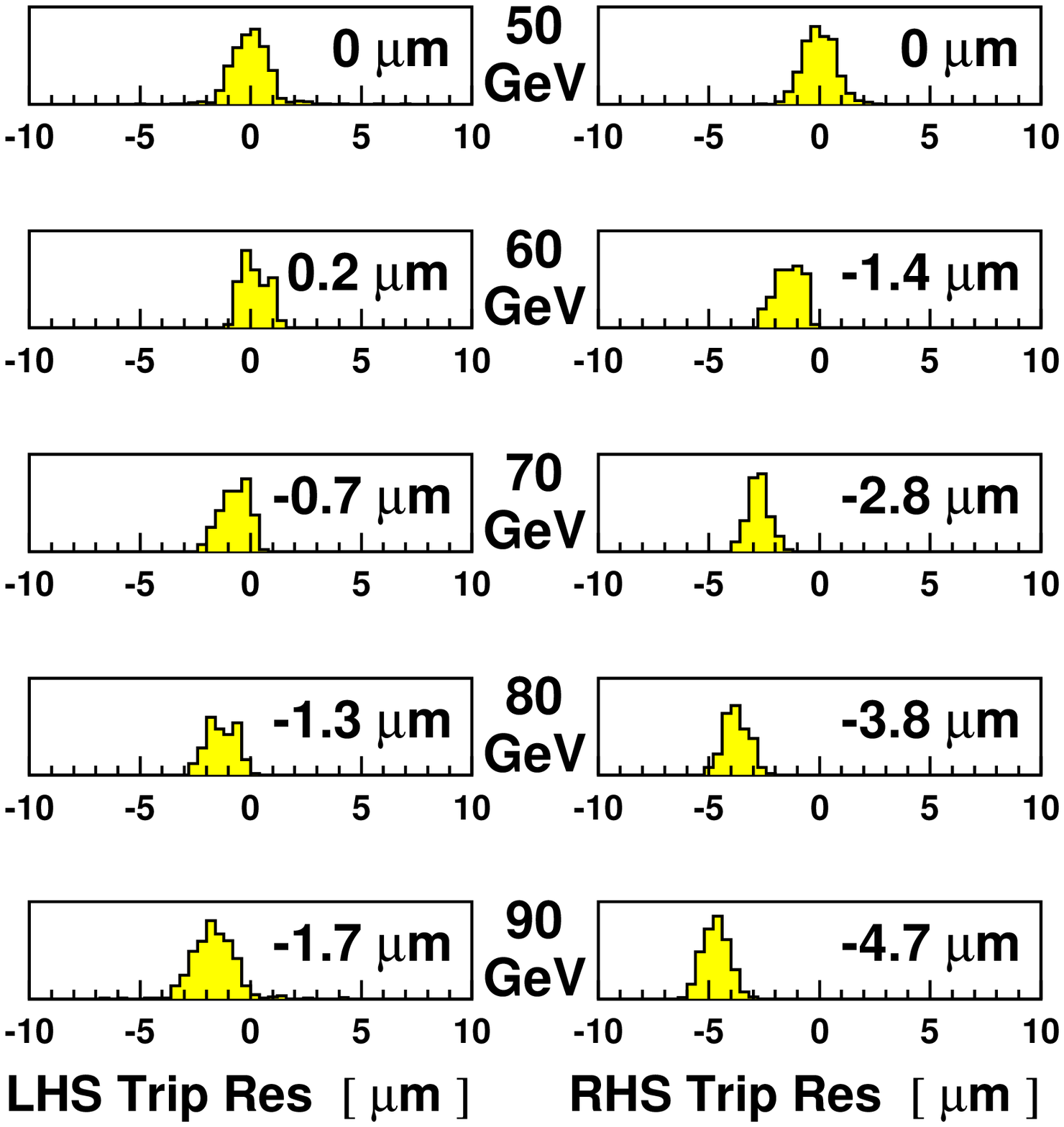,width=0.85\textwidth}
\caption[]{BPM triplet residuals for fill 8443.  The gains have
been calibrated and the triplet residuals centred at 50 GeV.
Good resolution is observed at the other energy points, but
accompanied by residual shifts.  The inset numbers indicate the
means of the distributions.}  
\label{fig:tripshiftex}
\end{center}
\end{figure}

The BPM triplet residuals are important figures-of-merit in
monitoring the integrity of the spectrometer data.   As
explained in section~\ref{sec:ericgainbit}, at a given energy
point after gain calibration, these residuals are stable with
a width of $< 1 \, \microns$.  Furthermore,
the calibration coefficients are equally 
applicable for other energy points within a given fill, giving
good resolutions throughout.   The central values of these 
residuals, however, are in general found to move between
energy points. 
Figure~\ref{fig:tripshiftex} shows a typical example from fill 8443,
where between 50~GeV and 90~GeV 
the triplet residuals are seen to move by
$-1.7 \, \microns$ in the left arm, and by $-4.7 \, \microns$ 
in the right arm.    Such {\it triplet-residual shifts}  (TRS)
indicate an effective relative movement amongst the BPMs,
when ramping between energy points.   These `movements'
cannot  be real, as they are not tracked in sign or 
magnitude by the WPS system. Rather they must 
arise in the response of the BPM themselves, or in the readout electronics.
 
\begin{figure}
\begin{center}
\epsfig{file=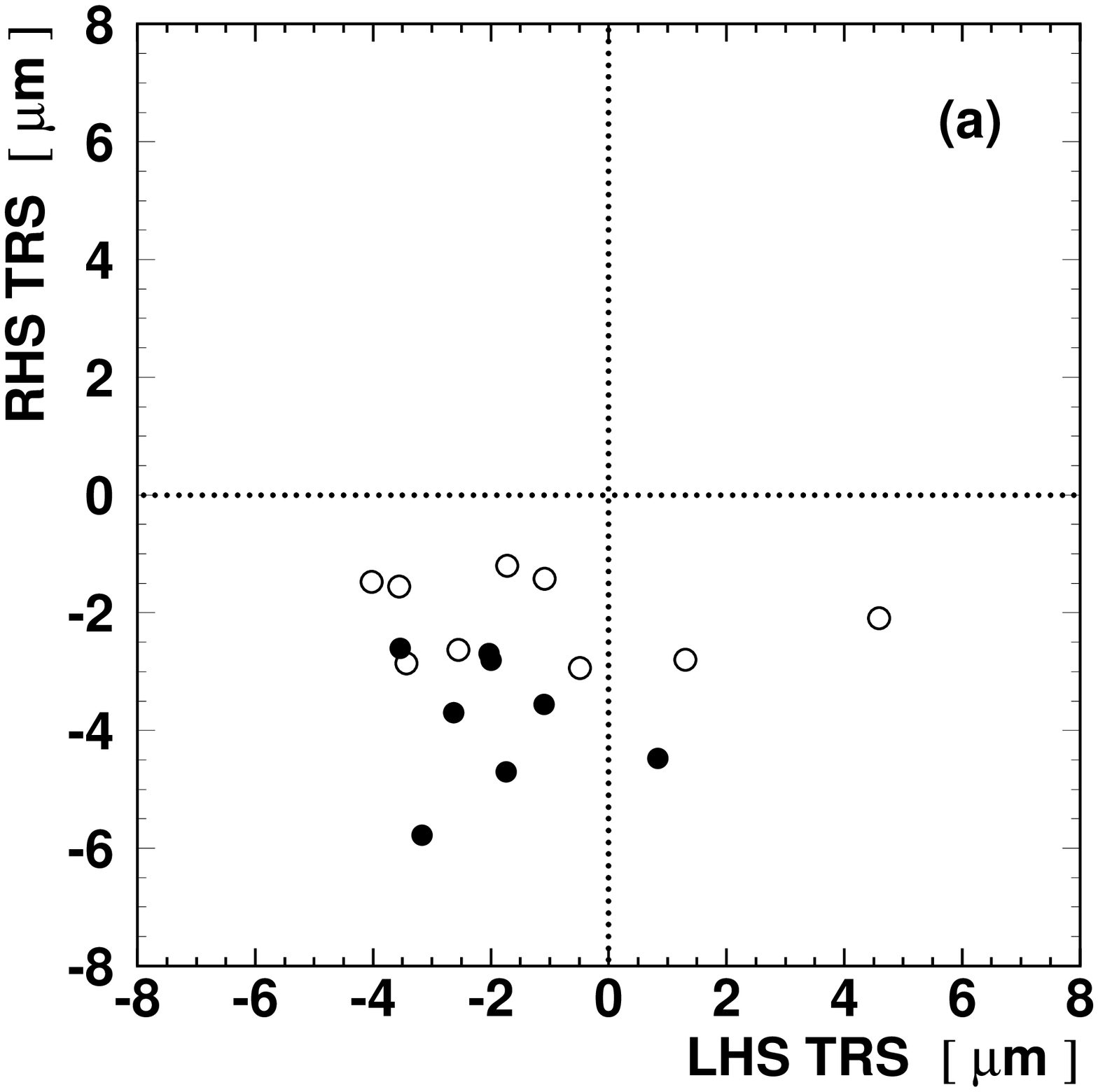,width=0.48\textwidth}
\epsfig{file=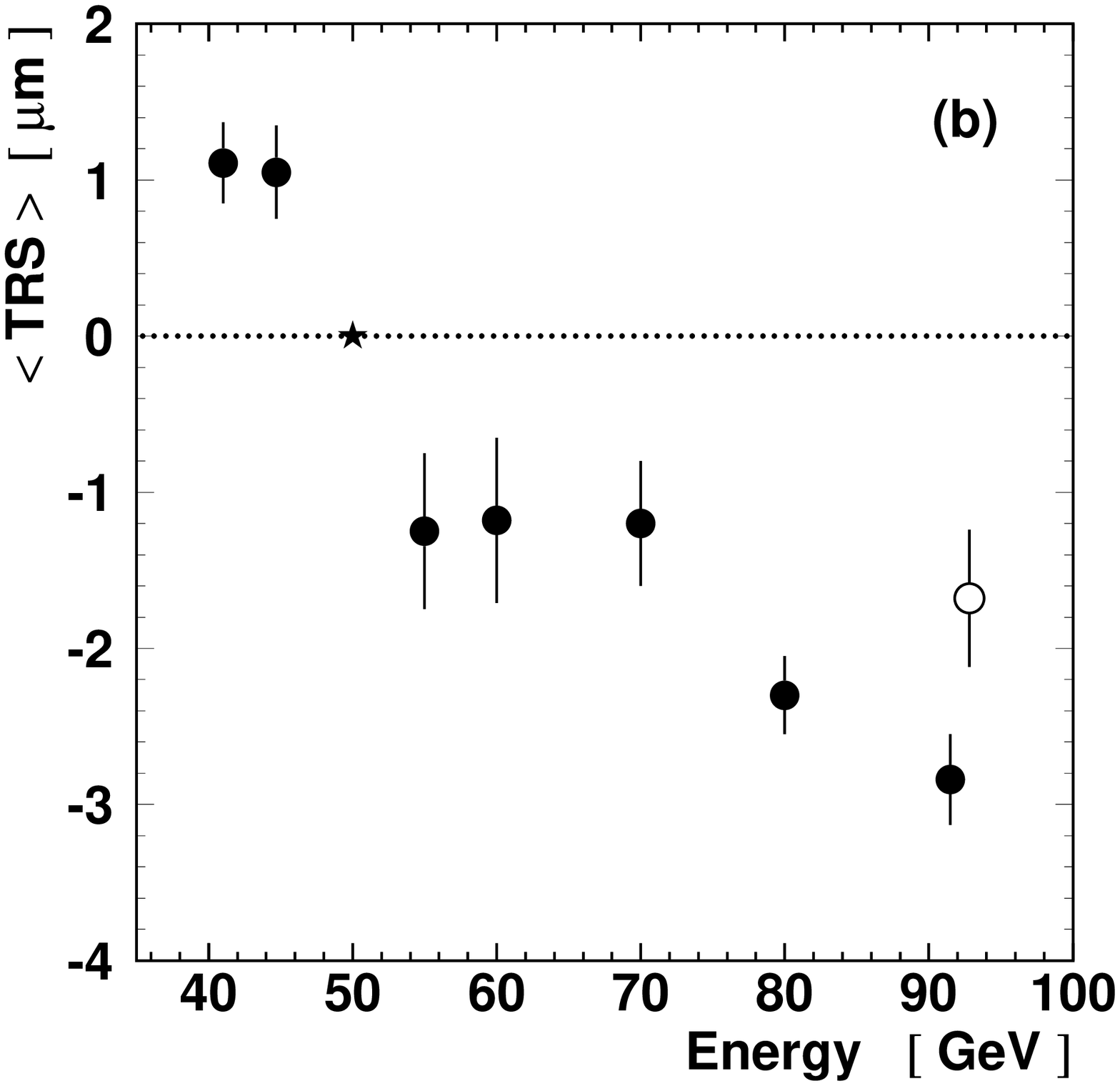,width=0.48\textwidth}
\caption{Characteristics of the triplet-residual shifts: (a) shows 
the TRS in the left and right arms for the high-energy sample; (b)
shows the mean TRS averaged over both arms (\mTRS), 
referenced to 50~GeV (asterix), as a function of energy, for
the full multi-point dataset.  The full points represent the
polarisation optics and the open points the physics optics.}
\label{fig:trschar}
\end{center}
\end{figure}

The characteristics of the TRS have been studied fill to fill.
Figure~\ref{fig:trschar}~(a) shows the values of the shifts in both
arms for ramps between 50~GeV and high energy.  They are 
predominantly negative, and vary in magnitude.
The means  are $-1.54 \pm 0.53 \,  \microns$ and  
 $-2.90 \pm 0.31 \,  \microns$ for the left and right arms respectively.
A similar behaviour is observed at lower energy.
Figure~\ref{fig:trschar}~(b) shows the mean value  at each energy point 
of the TRS averaged over both arms (\mTRS), referenced to 50~GeV, for the
full multi-point dataset.

The exact origin of the TRS is not well understood.
They are not correlated to temperature or to 
bunch current
and have no dependence on particle type.
The distribution in figure~\ref{fig:trschar}~(b) suggests a cause
which varies approximately linearly with energy,
thus disfavouring synchrotron radiation, and one which is more extreme
for the polarisation optics.
Variables which fulfil these criteria
are the bunch size and length, which for the
physics optics are similar between 50~GeV and high 
energy because of the routine
use of wigglers, but in the case of the polarisation optics steadily
increase.  In dedicated experiments at a fixed energy TRS were indeed
seen when wigglers were used to manipulate the beam parameters.
BPM misalignments and the beam-size dependence discussed
in section~\ref{sec:bpmsyst} might be one mechanism 
for the effect, but this is not proven.
In the following 
analysis, the redundancy provided by the triplet of BPMs
on either side of the spectrometer is exploited
to make an internal calibration of the dataset,
thereby minimising any biases brought 
about by the TRS in the energy determination.

\subsection{Analysis of the High-Energy Data}
\subsubsection{Survey of the Raw Spectrometer Estimates}

The energy model is used to calculate the mean beam energies at 50~GeV and
the high-energy point.  For each fill,
the spectrometer is referenced to the model
estimate at  50~GeV after applying a small correction to account for the known
difference to the true energy seen in figure~\ref{fig:banana}.
The change of bending angle and integrated 
magnetic field is then used to determine \Eb\ at high energy, and compared 
back to the model prediction, according to expression~\ref{eq:spec}.  
When relating  the mean beam energy to that determined at the spectrometer,
sawtooth corrections are applied.

The procedure of normalising the spectrometer measurement to a reference
energy means that in expression~\ref{eq:spec} the 
relative difference between the spectrometer
and the model estimate, \Enmr, is insensitive to any uncertainties 
which scale with 
energy.   This dependence is the case for all significant model contributions
detailed in section~\ref{sec:eb_model} which are relevant in these 
measurements.
To a very good approximation, therefore, it is non-linear systematics 
in \EOnmr\ alone which the spectrometer is constraining.  For this
reason, in the following the spectrometer results are compared with \EOnmr.

\begin{figure}[htb]
\begin{center}
\epsfig{file=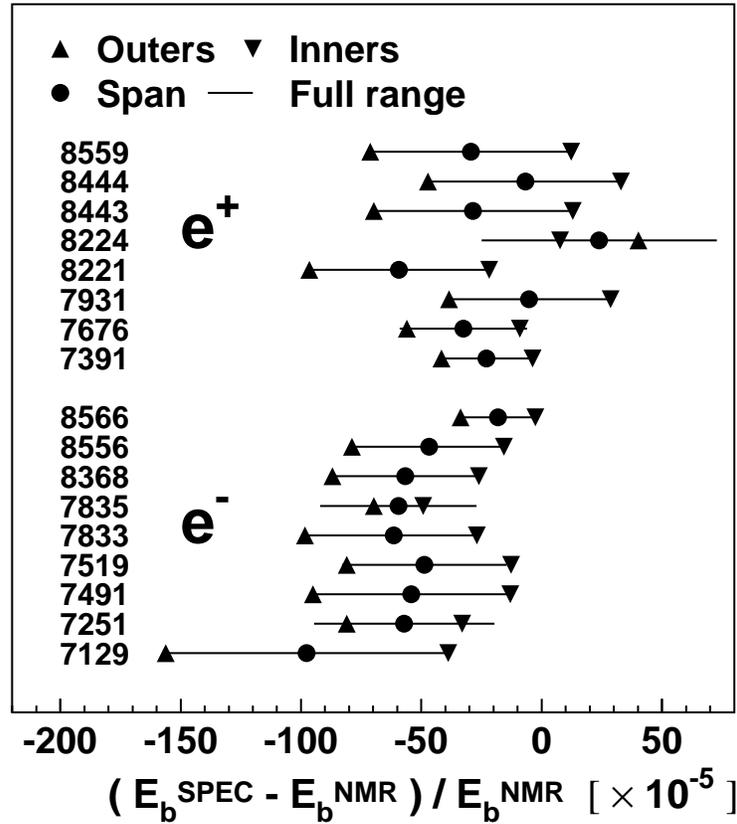,width=0.75\textwidth}
\caption[]{Spectrometer results compared with the NMR model at high 
energy.  For each measurement a spread is indicated showing the
variation coming from the choice of 9 possible BPM combinations (`Full range'),
of which 3 specific cases are indicated (`Outers', `Inners' and 
`Span', defined as in figure~\ref{fig:bpmcomb}).}
\label{fig:spec_unfitted}
\end{center}
\end{figure}

The bending angle and changes thereof can be constructed from any combination of 
two BPMs in one spectrometer arm, and two BPMs in the other, giving nine such
possibilities in total, each able to provide a separate determination of the
energy.  These determinations are not identical  because of the TRS .
Figure~\ref{fig:spec_unfitted} shows the difference between the spectrometer
and NMR model for all fills in the high-energy sample, where for each
a spread is indicated, which corresponds to the variation from the
different BPM combinations.  This spread takes values between $\pm 16$~MeV 
and $\pm 59$~MeV.
Of these nine combinations, three  can be defined of particular interest:

\begin{itemize}
\item{ {\bf Outers} -- formed from the two outermost BPMs
(6 \& 5 together with 2 \& 3); } 
\item{ {\bf Inners} -- formed from the two innermost BPMs
(5 \& 4 together with 1 \& 2); }
\item{ {\bf Span} -- formed by excluding the middlemost BPM
(6 \& 4 together with 1 \& 3).}
\end{itemize}

\begin{figure}
\begin{center}
\epsfig{file=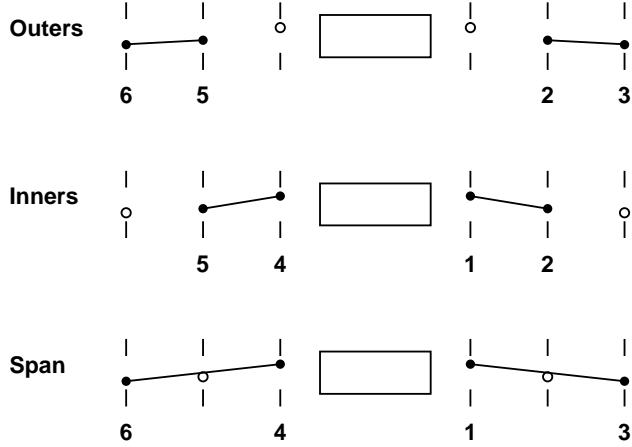,width=0.70\textwidth}
\caption[]{A schematic illustrating the three choices of BPM 
combinations used for determining the bending angle in
the spectrometer energy analysis.}
\label{fig:bpmcomb}
\end{center}
\end{figure}

\noindent  These combinations are illustrated in figure~\ref{fig:bpmcomb}.

The energy determinations with each of these combinations
are indicated in figure~\ref{fig:spec_unfitted}.  It can be seen that
in almost all cases the Outers give the lowest
energy estimate of the possible combinations, and the Inners the 
highest,  with the Span defining the
median value.  There are two exceptions: fill 8224 where the 
`Inner-Span-Outer'  hierarchy is inverted, and fill 7835 where 
other combinations give a much wider variation in result.
These two fills have a large positive TRS in one arm and are anomalous
within the sample. One other fill, 7251, exhibits a small positive
TRS in one arm, but one which is countered by a more significant negative
shift in the other arm.   Fills 7835 and 8224
are dropped from further consideration at this stage, 
leaving a sample of 15 measurements sharing a common systematic behaviour.
The nominal value of \Eb\ at high energy
for this sample is 92.3~GeV, averaged over the measurements.

\vspace*{-0.005cm}
The results in figure~\ref{fig:spec_unfitted} are divided into
electron and positron fills.  There is an indication that the
positron fills give a higher energy estimate than the
electron fills, with a difference in the raw means of 36~MeV.   
Although \Eb, when averaged around the ring,  
must be the same within a few MeV for electrons and positrons, 
it is unsurprising that larger differences are seen
in the spectrometer analysis.  The RF sawtooth  is anti-correlated
between the two particle types, and so any residual error in
calculating the correction will result in a separation between
electrons and positrons of approximately twice this amount.  
Conversely, the mean value of the two samples will give a result which is rather 
robust against imprecisions in the sawtooth modelling.

\subsubsection{Extracting \Eb\ from a Global Fit}
\label{sec:specheglobal}

A priori it is not known which combination of BPMs gives the most
reliable estimate of \Eb\, as the TRS only indicate relative 
effective motion between the blocks.    This question is best
answered by studying the full ensemble of measurements.
By considering the variation in results for the difference in
spectrometer and energy model 
as a function of \mTRS, the BPM combination which gives
the best stability can be identified.   In addition
an extrapolation can be attempted to the limit of
zero systematic effect.

\begin{figure}[htb]
\begin{center}
\epsfig{file=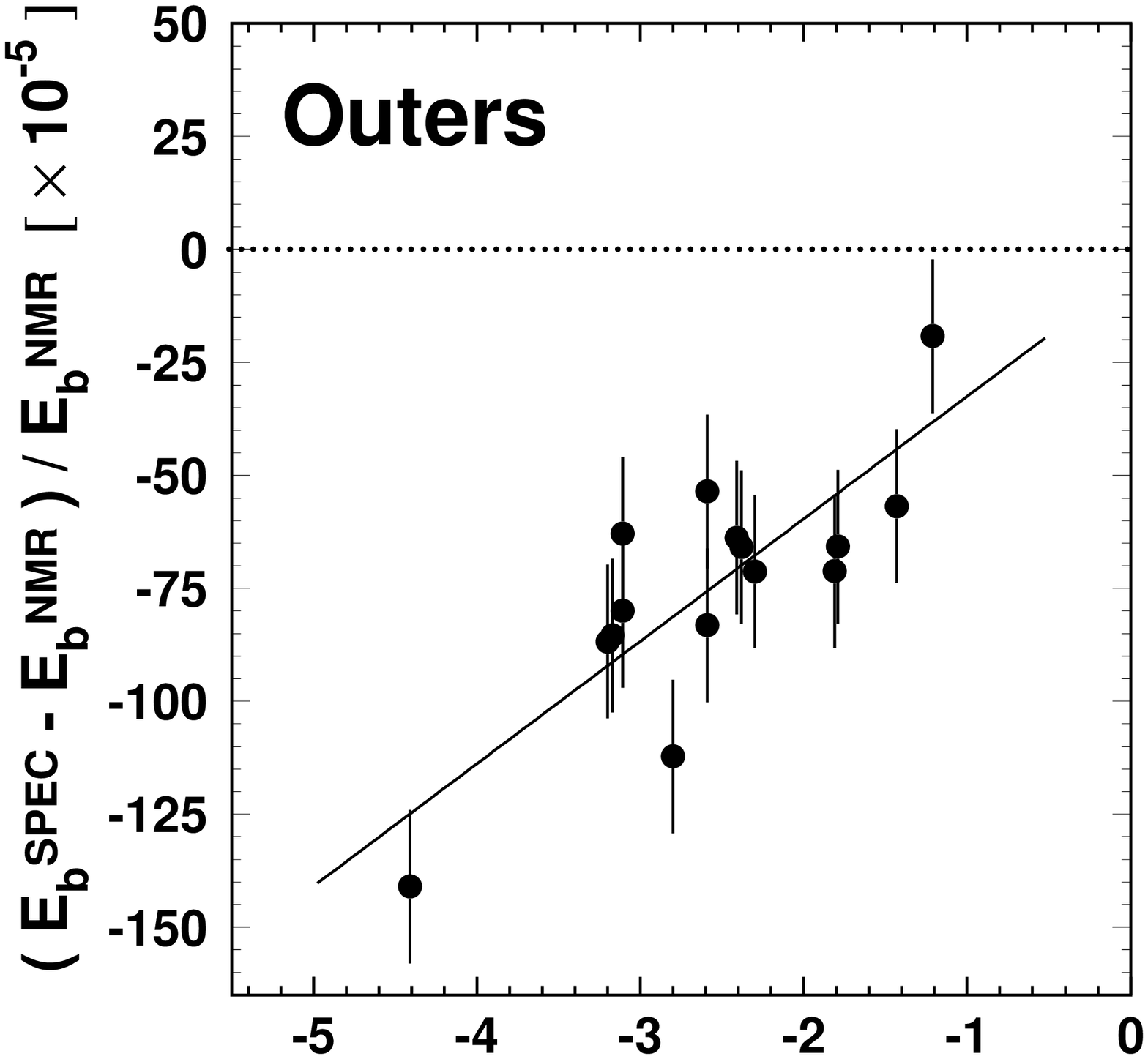,width=0.32\textwidth}
\epsfig{file=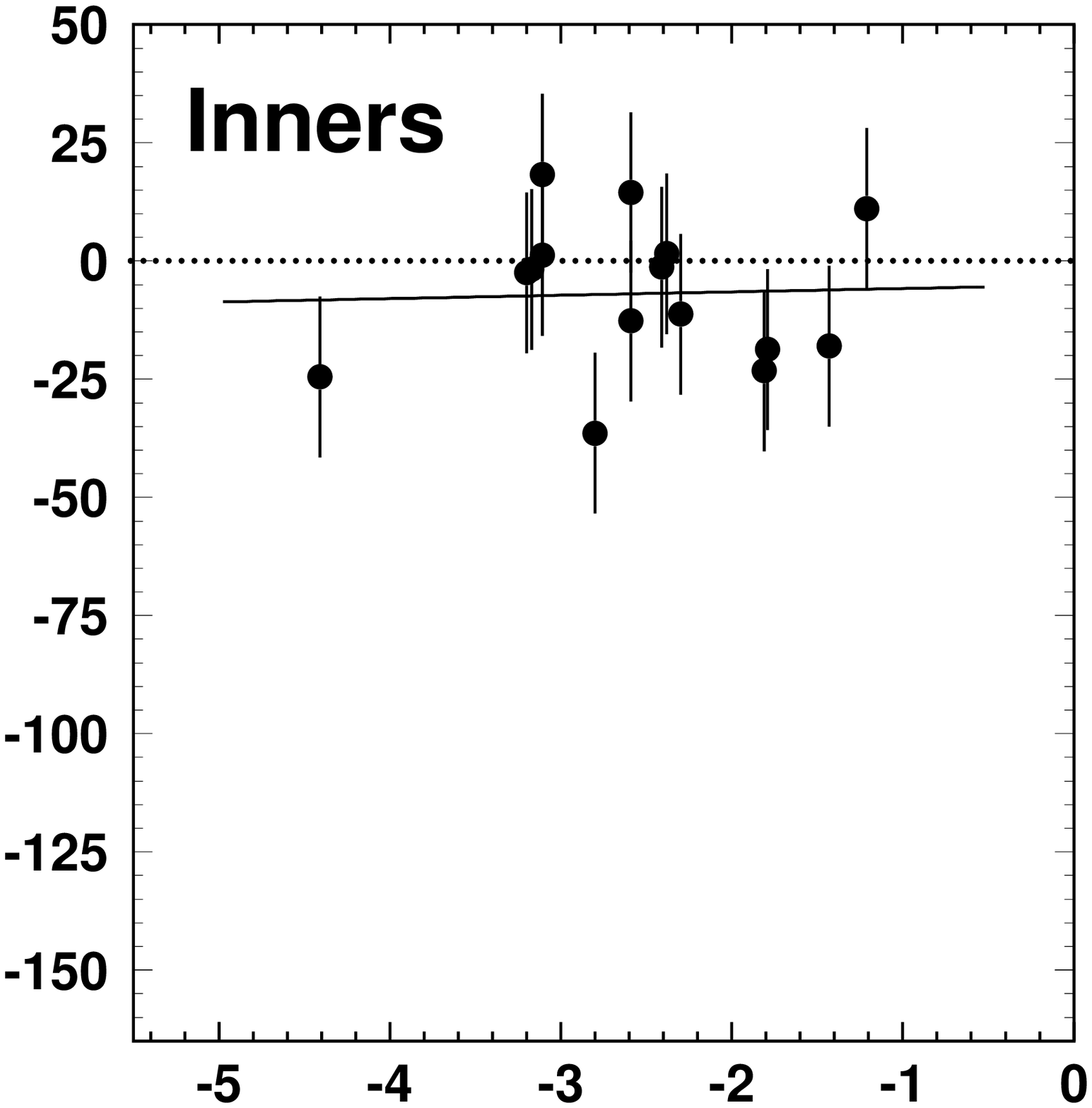,width=0.32\textwidth}
\epsfig{file=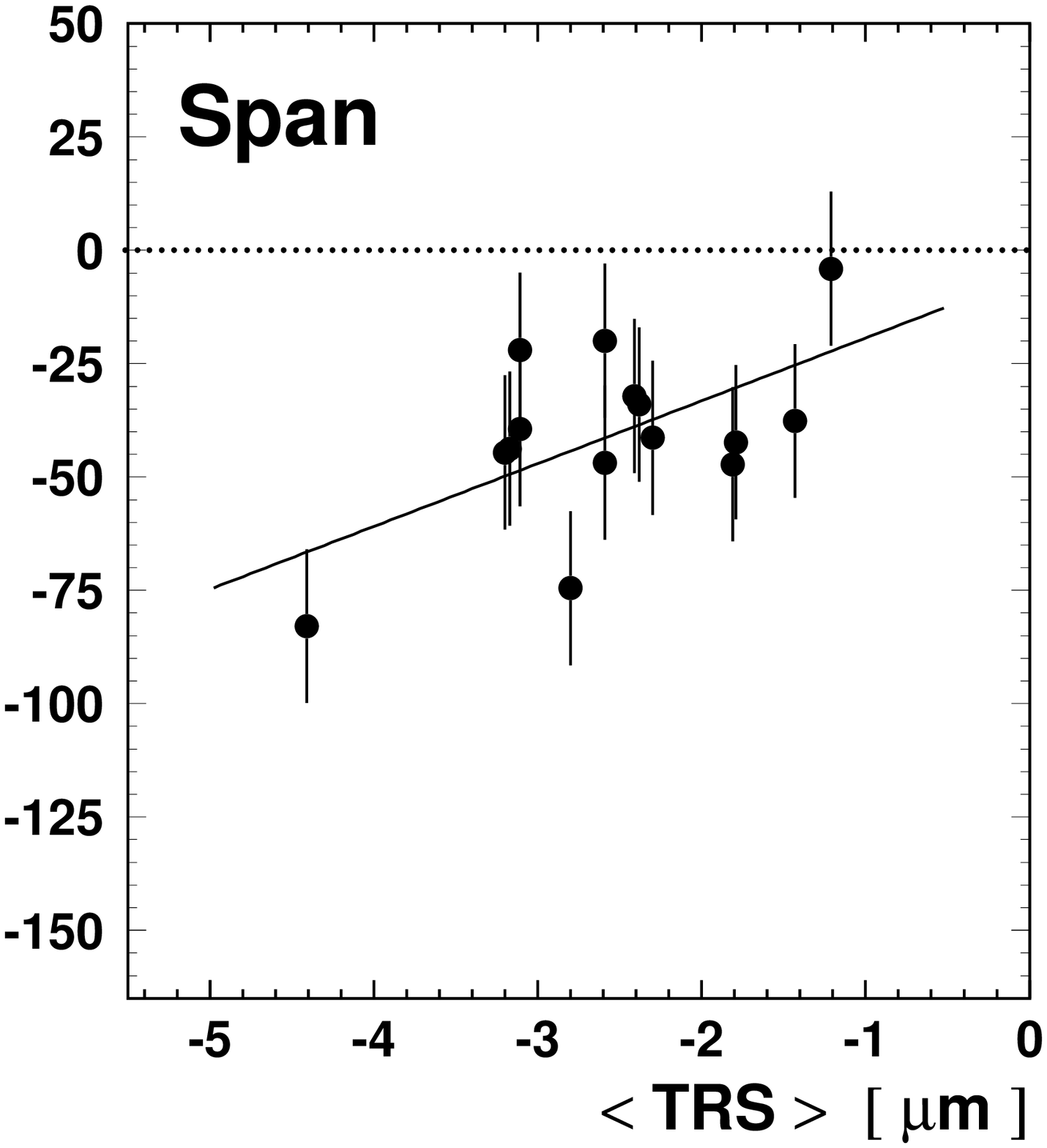,width=0.32\textwidth}
\caption[]{Spectrometer results at high energy as a function of \mTRS,
for different BPM combinations. 
The results of the fits described in the text are superimposed (solid line).}
\label{fig:he_result}
\end{center}
\end{figure}

Figure~\ref{fig:he_result} shows the spectrometer
result, as compared to the NMR model, 
plotted against \mTRS.
The plot is made separately for the Outers, Inners and Span results.
Prior to plotting a correction has been made to minimize the
difference between the electron and positron populations.
This correction is 
one of three parameters (`sawtooth') in a least-squared fit made between the 
spectrometer results and \mTRS:

\begin{enumerate}
\item{{\bf Offset} -- the extrapolated value of 
$(\Espect \, - \, \EOnmr)/\EOnmr$  at \mTRS=0;}
\item{{\bf Slope} -- the gradient of $(\Espect \, - \, \EOnmr)/\EOnmr$ 
with respect to \mTRS;}
\item{{\bf Sawtooth} -- the correction added to
the electron results, and subtracted from the positron results,
in order to compensate for residual errors in the sawtooth model.}
\end{enumerate}

\noindent The fit is made separately for each BPM combination.
In the fit each spectrometer measurement is assigned a relative error
of $17 \times 10^{-5}$, which gives a
$\chi^2/\rm{p.d.f.}$ of 1.06, 0.98 and 1.01 for the Outers,
Inners and Span fits respectively.  The fit results
are superimposed in figure~\ref{fig:he_result}
and listed in the `Standard' column of table~\ref{tab:he_result}.

\begin{table}
\begin{center}
\begin{tabular}{|l|l| c|c|c|c|c| } \hline
BPM        & Fit parameters           & \multicolumn{5}{c|}{Choice of input data} \\ \cline{3-7}
comb       &                          &  Standard      &  Pol optics &  Phy optics  & Early data  & Late data   \\ \hline
           & Offset                   &$-5.4 \pm 14.9$ &$13.7\pm25.1 $&$ 0.8\pm22.1$&$-15.6\pm19.9$&$4.7 \pm 42.2$\\
Outers     & Slope                    &$27.1 \pm 5.6$  &$32.2\pm8.4  $&$30.5\pm9.5 $&$ 24.3\pm 7.9$&$29.3 \pm16.6$\\       
           & Sawtooth                 &$14.1 \pm 4.1$  &$22.9\pm6.0  $&$6.1\pm6.1  $&$ 16.1\pm 6.7$&$12.7 \pm10.2$\\ \hline
           & Offset                   &$-5.1 \pm 14.9$ &$13.9\pm25.1 $&$-0.2\pm22.1$&$-13.1\pm19.9$&$4.4 \pm 42.2$\\
Inners     & Slope                    &$0.7  \pm 5.6 $ &$5.9 \pm8.4  $&$3.5\pm9.5  $&$ -1.4\pm 7.9$&$3.0 \pm 16.6$\\        
           & Sawtooth                 &$13.3 \pm 4.1 $ &$21.7\pm6.0  $&$5.7\pm6.1  $&$ 14.7\pm 6.7$&$12.5\pm 10.2$\\ \hline
           & Offset                   &$-5.5 \pm 14.9$ &$13.4\pm25.1 $&$ 0.3\pm21.6$&$-14.8\pm19.9$&$4.1 \pm 42.2$\\
Span       & Slope                    &$13.9 \pm 5.6 $ &$19.0\pm8.4  $&$16.9\pm9.4 $&$ 11.4\pm7.9 $&$15.9\pm16.6 $\\       
           & Sawtooth                 &$13.7 \pm 4.1 $ &$22.5\pm6.0  $&$5.6\pm6.0  $&$ 15.4\pm6.7 $&$12.5\pm10.2 $\\ \hline
\end{tabular}
\caption[]{Results of the global fit to the high-energy data.  Results are given
for the standard sample, and four example sub-sets.   The units are as follows:
Slope [$\rm \mu \rm{m^{-1}} \times 10^{-5}$], Offset [$\times 10^{-5}$] and
Sawtooth [MeV].}
\label{tab:he_result}
\end{center}
\end{table}

It can be seen that, within the assigned errors, the energy estimate coming from the 
Inners shows no evidence of a significant dependence on \mTRS.
The result 
from the Outers, on the other hand, shows a pronounced slope.  The result
from the Span fit lies between these two extremes.  These fits suggest
that there is little relative effective motion between the innermost pairs 
of BPMs, and it is the outermost BPMs in each arm, BPMs 6 and 3,
which exhibit instability with respect to the other four.
A calculation made under the hypothesis that all effective motion occurs
in BPMs~6 and~3 predicts slope values of 
$26.9, \, 0$ and $13.3 \times 10^{-5} \, \microns^{-1}$ for the 
Outers, Inners and Span respectively,  in very good agreement 
with the fit results to the data.



The fitted value of the offset in the global fit 
determines the spectrometer energy 
in the absence of TRS bias. It can be seen
that all three of the combinations considered
converge on the same value.  The value
corresponds to an offset of $-5 \, \pm 14$~MeV
with respect to the NMR model at a nominal
energy of 92.3~GeV.

The returned value for the correction to the sawtooth model is
14~MeV, which
is compatible with the 10~MeV uncertainty on the model 
estimated in section~\ref{sec:rfcorr}.

\subsubsection{Robustness Studies}
\label{sec:herobust}

In order to probe the homogeneity of the dataset,
and to cross-check the reliability of the
errors coming from the global fit, the fit
is repeated on various sub-samples of the data, namely:

\begin{itemize}
\item{Division between polarisation and physics optics;}
\item{Division between the first and second halves of the run;}
\item{Samples with each of the 15 fills dropped in turn;}
\item{Inclusion of the two anomalous fills 7835 and 8224;}
\item{Division according to whether the TRS were more significant 
in the left or the right arms;}
\item{Division into samples according to how well the beam was
re-centred between energies, and according to the position of the beam orbits
at the spectrometer;}
\item{Excluding those fills common to the low-energy sample analysed in 
section~\ref{sec:lowenergyres};}
\item{Excluding those fills with the largest fit residuals.}
\end{itemize}

\begin{figure}[htb]
\begin{center}
\epsfig{file=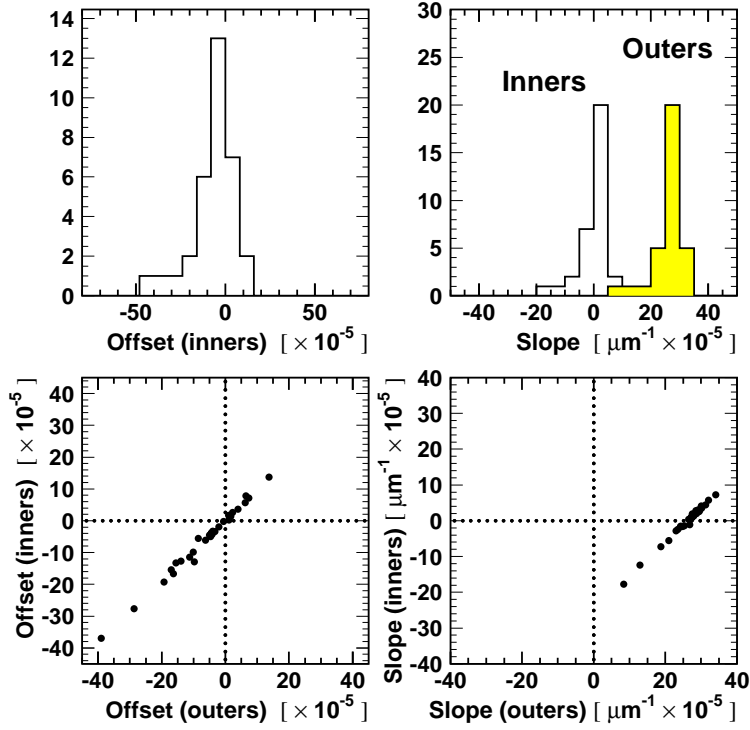,width=0.68\textwidth}
\caption[]{Variations in the fit result for the offset and the slopes
for the inner and outer BPM combinations calculated
from various combinations of fills 
in the high-energy sample.
Each entry corresponds to a distinct subset, varying in size between
7 and 17 fills.}
\label{fig:he_robust}
\end{center}
\end{figure}

The results from several of these studies are given in 
table~\ref{tab:he_result}.   The full variation of the
key parameters are histogrammed in 
figure~\ref{fig:he_robust}.   The observed 
fluctuations are well-behaved, with the RMS of the 
offset distribution found to be $12.6 \times 10^{-5}$. 
The largest deviations come from the smallest sub-samples,
and are always within
1--2 sigma in uncorrelated error.  The consistency between
results from different BPM combinations remains good in
all cases.

\subsection{Analysis of the Low-Energy Data}
\label{sec:lowenergyres}

The low-energy sample is important as it both allows the spectrometer's
performance to be evaluated in a regime where the energy is well
known, and also provides an independent dataset in which to study
the BPM behaviour and TRS characteristics.

Spectrometer energies are calculated as in the high-energy analysis
for all data in the 41--61~GeV range listed in table~\ref{tab:specfills}.
In fills with two low-energy points
\Ebr\ is set at 50~GeV; in fills with three or more points
the reference is chosen to be as close as possible to the mid-point 
of the full TRS excursion seen over the low-energy interval in that fill.
Where available, the true energy is defined by actual RDP 
measurements; otherwise the energy model is used.
As apparent in  figure~\ref{fig:banana}, there are
residuals of 2--3~MeV
between the fitted model and the energies as measured by RDP
in the 41--61~GeV regime.  Corrections are applied for these differences
so that a comparison can be made between the spectrometer and the
best possible estimate of the true energy, \Etrue.

The fractional differences between the energy estimate from the 
spectrometer and the 
true energy are fitted against \mTRS.   For this sample only two parameters
are considered:

\begin{enumerate}
\item{{\bf Offset} -- the value of
$(\Espect \, - \, \Etrue)/\Etrue$  at \mTRS=0;}
\item{{\bf Slope} -- the gradient of 
$(\Espect \, - \, \Etrue)/\Etrue$
with respect to \mTRS.}
\end{enumerate}

\noindent As the low-energy sample is dominated by electron fills it is not
possible to fit a correction to the sawtooth model.  
At low energy the sawtooth correction is significantly smaller, and
consequently the expected precision of the RF model is much better.
All data are included in the fit with their sawtooth correction fixed
to that of the model.
The mean residual of the two positron points is very close to
zero, indicating that indeed there is no significant problem
with the understanding of the sawtooth at these energies.

\begin{figure}[htb]
\begin{center}
\epsfig{file=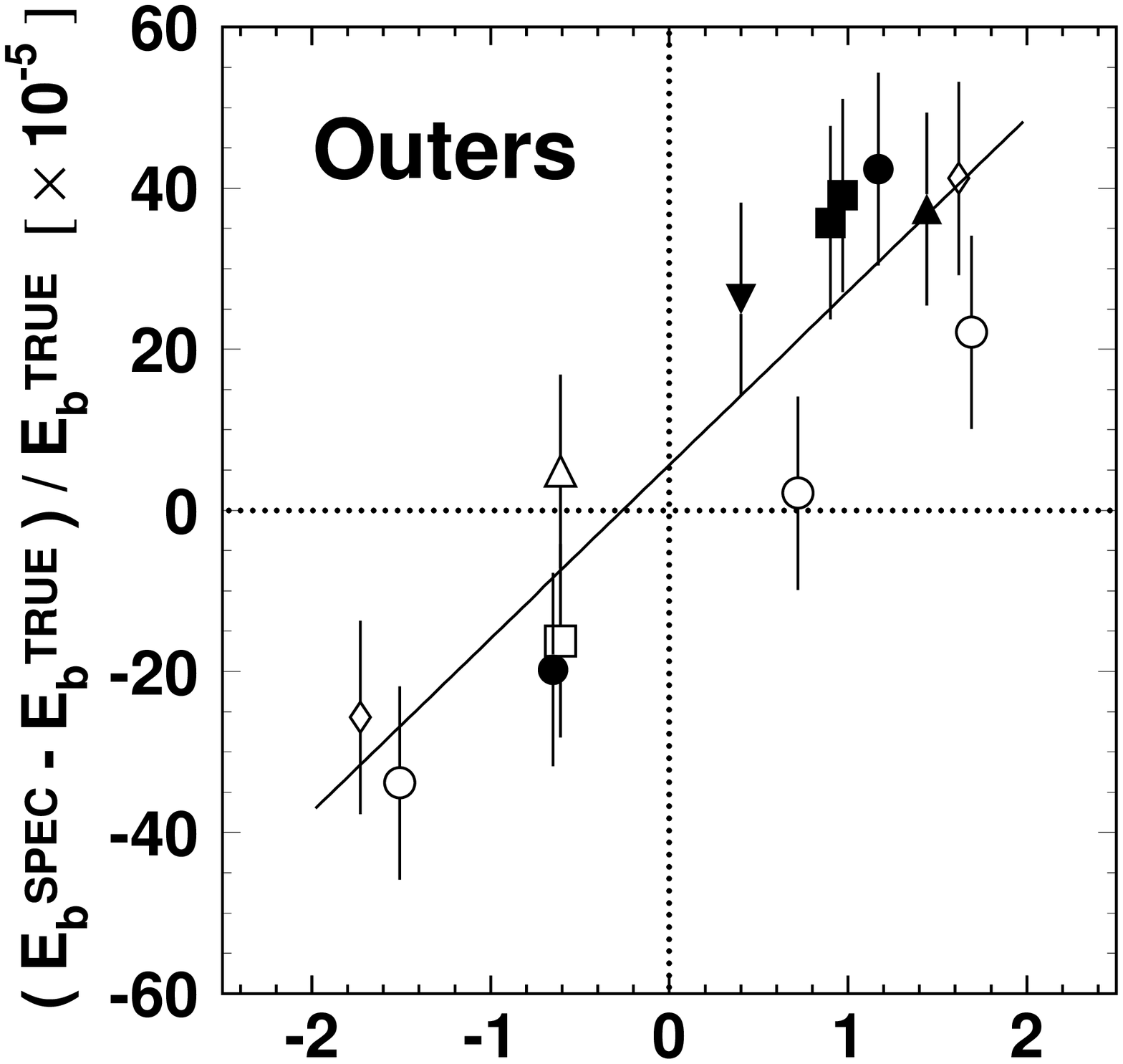,width=0.32\textwidth}
\epsfig{file=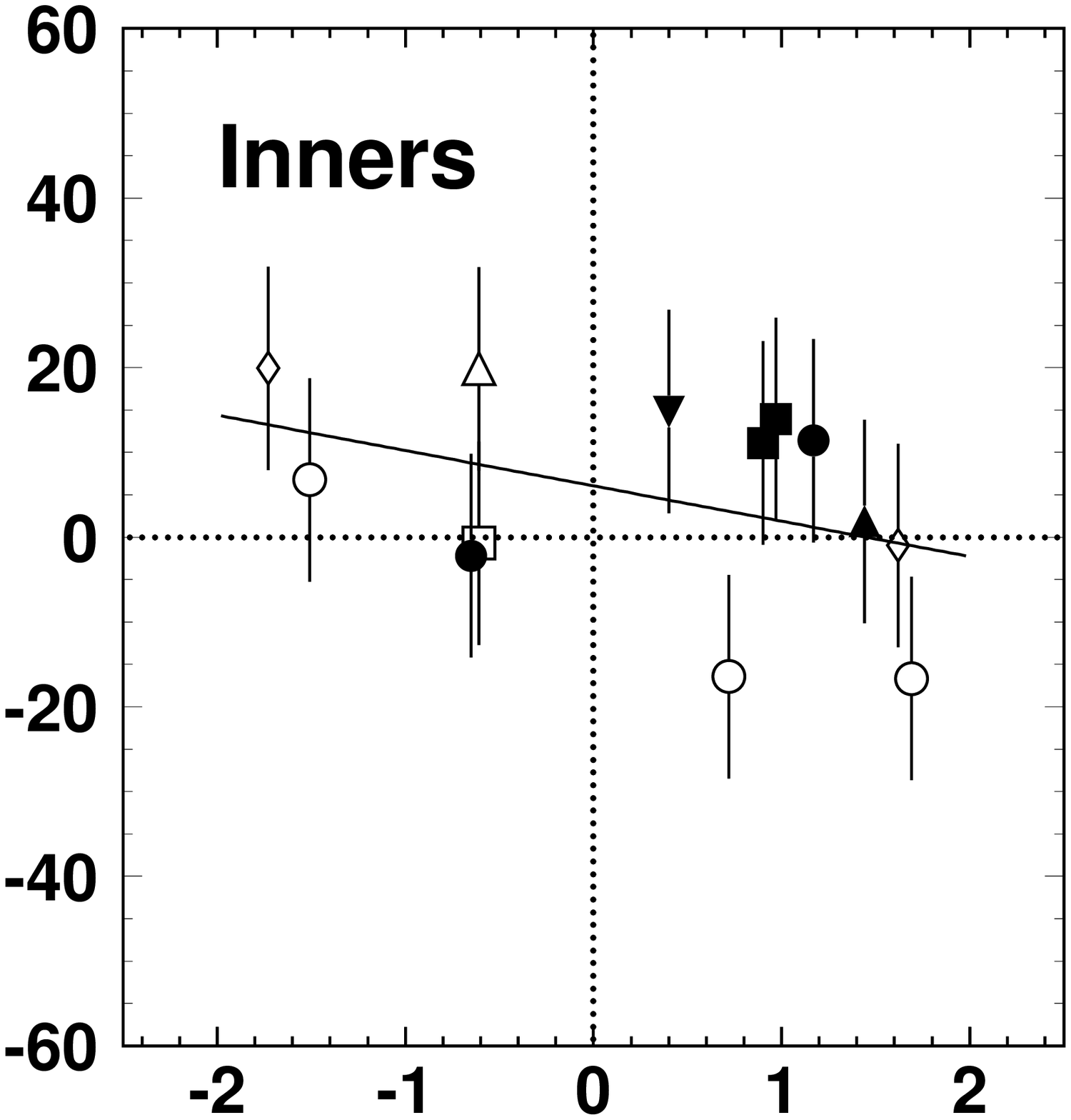,width=0.32\textwidth}
\epsfig{file=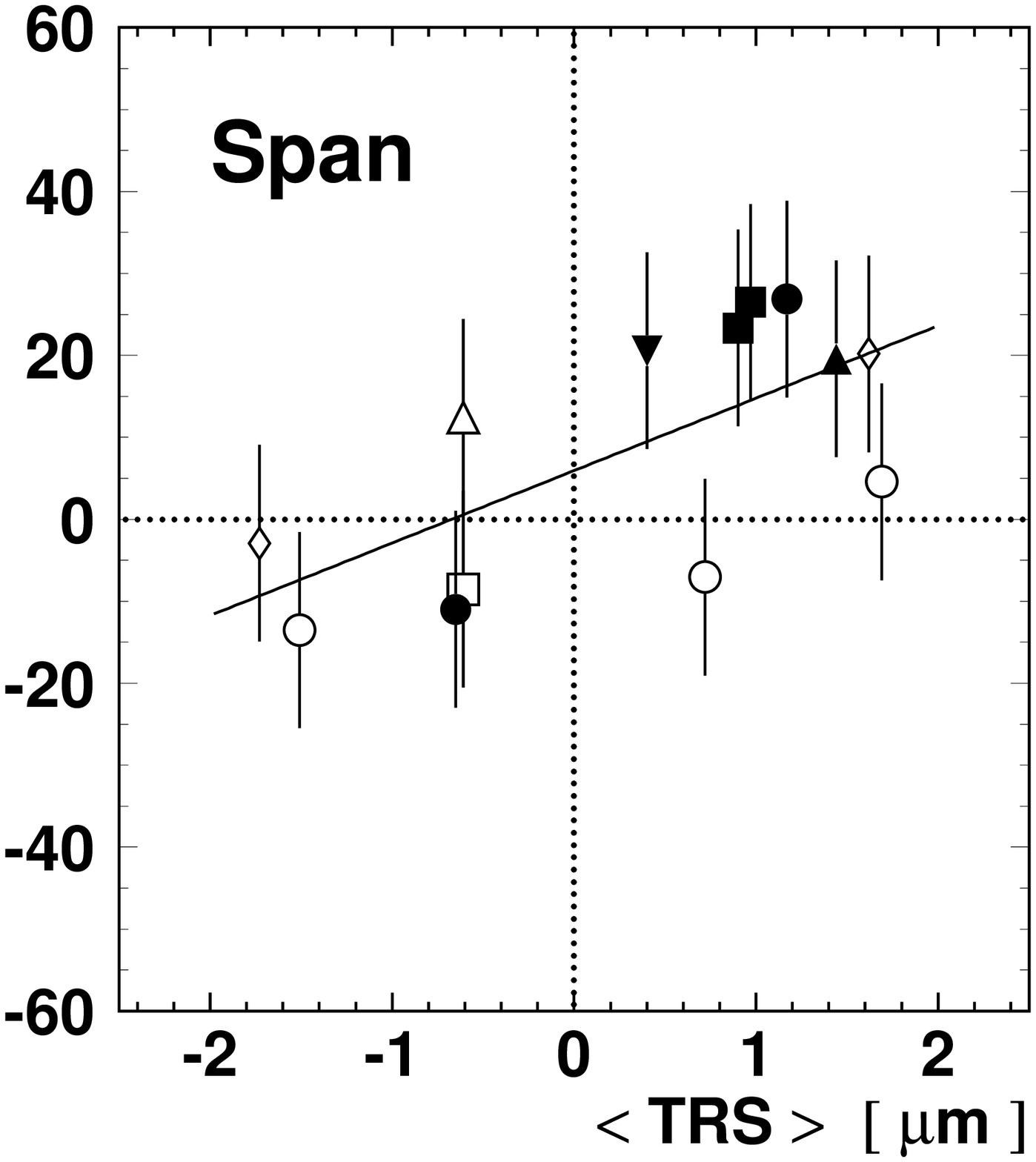,width=0.32\textwidth}
\caption[]{Spectrometer results at low energy as a function of \mTRS,
for different BPM combinations.  Common symbols are used to designate
measurements in the same fill.
The results of the fits described in the text are superimposed (solid line).}
\label{fig:le_result}
\end{center}
\end{figure}

The data points and superimposed fit are shown in figure~\ref{fig:le_result},
and the results listed in table~\ref{tab:le_result}.  The $\chi^2/\rm{p.d.f}$
of the fits are 1.10, 0.96 and 0.99 for Outers, Inners and Span
respectively. These have been obtained by arbitrarily assigning 
an error to each point of $12 \times 10^{-5}$, a smaller value than
was required in the fit of section~\ref{sec:specheglobal}. This difference
may be due to the increased significance of uncorrelated fill-to-fill
imperfections in the sawtooth model at high energy.

\begin{table} [htb]
\begin{center}
\begin{tabular}{|l|l| c| } \hline
BPM comb   & Fit parameters                          & Result \\ \hline
Outers     & Offset  [$\times 10^{-5}$]                 &$5.6  \pm 3.4$  \\ 
           & Slope   [$\rm \mu \rm{m^{-1}} \times 10^{-5}$]  &$21.5 \pm 2.9$  \\ \hline
Inners     & Offset  [$\times 10^{-5}$]                 &$6.1  \pm 3.4 $ \\ 
           & Slope   [$\rm \mu \rm{m^{-1}} \times 10^{-5}$]  &$-4.2 \pm 2.9 $ \\ \hline
Span       & Offset   [$\times 10^{-5}$]                &$5.9  \pm 3.4 $ \\ 
           & Slope    [$\rm \mu \rm{m^{-1}} \times 10^{-5}$] &$8.8  \pm 2.9 $ \\ \hline
\end{tabular}
\caption[]{Results of the global fit to the low-energy data.  }
\label{tab:le_result}
\end{center}
\end{table}

The slope values returned by the fit agree with those obtained at
high energy.  This shows that the same systematic behaviour is
present in the two regimes, and that the low-energy sample can
be used to assess the performance of the spectrometer at high energy.

The offset quantifies the agreement between the spectrometer estimate 
and the true energy when the TRS have been accounted for.
This is found to be non-zero at a significance
of almost two statistical sigma, suggesting a possible
bias in the energy reconstruction of 3-4~MeV.
No evidence is observed for an energy dependence in this offset.
The bias, if real, is not understood; therefore a conservative 
error assignment is favoured over using the
result to correct the high-energy fits.

\begin{figure}[htb]
\begin{center}
\epsfig{file=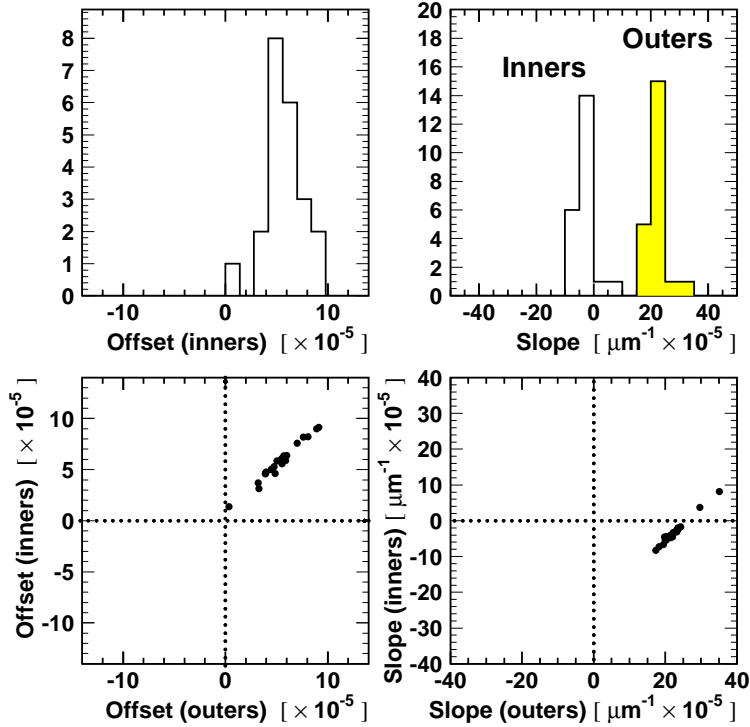,width=0.68\textwidth}
\caption[]{Variations in the fit result for the offset and the slopes
for the inner and outer BPM combinations 
from various combinations of fills
in the low-energy sample.
Each entry corresponds to a distinct subset, varying in size between
4 and 13 energy points.}
\label{fig:le_robust}
\end{center}
\end{figure}

The robustness of the result has been explored by isolating distinct sub-samples
of points within the dataset, applying similar criteria to those used in 
section~\ref{sec:herobust}, and repeating the global fit.   Distributions
of the spread of results are shown in figure~\ref{fig:le_robust}.  Further investigations
have been made varying the choice of reference point, the gains assumed in the BPM analysis,
the RF sawtooth model and the value of the integrated dipole fields.
From all of these studies it is concluded that the spectrometer correctly measures the
relative energy change in the low-energy sample within a tolerance of $10 \times 10^{-5}$.
This uncertainty also encompasses the value of the fitted offset, plus one sigma.

\subsection{Systematic Error Assignment}
\subsubsection{Global Fit Results} 

The global fit offset result of $-5 \times 10^{-5}$
at high energy provides the central value of
the relative spectrometer energy determination with respect
to the NMR  model, a result  
which is the same for all BPM combinations considered, as
is seen in table~\ref{tab:he_result}.
The accompanying error from the fit of $15 \times 10^{-5}$ 
is taken as the uncertainty 
associated with the scatter of the measurements.

The studies at low energy, presented in section~\ref{sec:lowenergyres},
show a systematic behaviour
fully compatible with that observed at high energy.
The low-energy fit, however, suggests a possible small offset 
of the spectrometer 
measurements with respect to the true energy.  
On consideration of this, and the variations in this offset
under different fit strategies, an error of 
$10 \times 10^{-5}$ is assigned
to represent the validity of the spectrometer 
performance as cross-checked in the 41--61~GeV
regime.

\subsubsection{Beam Size and BPM Non-Linearities}

As discussed in section~\ref{sec:bpmsyst}, the measured beam position,  as
conventionally calculated from the electrode signals of a BPM, 
ignores higher-order effects which introduce a 
potential systematic uncertainty into the spectrometer measurement.
Contributions come from both the finite transverse beam size and from 
non-linear dependencies.

The variation in beam size has been calculated as a function of optics, longitudinal
position in the spectrometer, and energy.     The mean transverse offsets of 
the beam from the BPM centres, as estimated from consideration of the beam orbits
of the measurements, and knowledge of the BPM alignments within the spectrometer,
are found to be less than 1~mm.
A simulation of the BPM response is then used to determine the change in
apparent positions, and thus beam angle, with the variation in beam size
for the measurements.   From this study a relative error of $4 \times 10^{-5}$
is assigned to the global energy determination.

As is seen in table~\ref{tab:specstab}, the beam was reliably 
re-positioned between
low and high-energy measurements, with a spread of less than
200~\um.   This is sufficient to render non-linear  
effects negligible.

\subsubsection{Knowledge of the BPM Gains}
\label{sec:beamsizesyst}

The dataset has been reanalysed changing the assumed scale of
the BPM gains by $\pm 5 \%$, and the global fit 
repeated.  Variations of $0.5 \times 10^{-5}$
are seen in the result.   This weak dependence on the knowledge of
the gains is a consequence of the care taken  at high energy to re-steer the
beam close to its low-energy position.

\subsubsection{Knowledge of the Dipole Bending Field}

The standard analysis is based on values of 
the integrated bending field derived from
a model fitted to the data of the
post-LEP mapping campaign.
The analysis has been repeated using the
model based on the pre-installation campaign data.
This results in a change of the offset of
$+1.5 \times 10^{-5}$.   Other models
considered, relying on different 
fit strategies, and alternative magnet temperature
corrections, give smaller
variations.  The difference between
the pre-installation and post-LEP based results
is taken as the error arising
from bending field uncertainties.

\subsubsection{Knowledge of the RF Sawtooth}

At high energy the sawtooth correction is an important
input in relating spectrometer measurements to the
NMR model.  The division of the dataset into nearly equal
numbers of $e^-$ and $e^+$ fills, and the high
degree of anti-correlation of the correction between
the two particle types, enables the validity of the
sawtooth model to be assessed, and in turn ensures that the 
energy determination from the global fit
is largely insensitive to model imperfections.

Results have been obtained using both sawtooth models
described in section~\ref{sec:rfcorr},  and using alternative tunings
of each model.  From these studies
an uncertainty of $5 \times 10^{-5}$ is assigned.

\subsubsection{Uncertainty in Corrections to the Bending Angle Calculation}

Prior to the calculation of the bending angle, corrections were applied 
from the WPS system to account for movements in the BPM-blocks.
A further correction was applied for the effect of 
the ambient magnetic field on the beam trajectory.
As part of the systematic error analysis
these corrections are removed in turn, and the global fit is
repeated.  The results are shown in table~\ref{tab:fitwithoutcor}.

\begin{table}[htb]
\begin{center}
\begin{tabular}{|l|l| c|c| } \hline
BPM        & Fit parameters           & \multicolumn{2}{c|}{BPM correction dropped} \\ \cline{3-4}
comb       &                          &  WPS movements    &  Ambient field  \\ \hline
           & Offset                   &$-14.5\pm 10.5$ &$-10.1\pm 17.6$ \\
Outers     & Slope                    &$23.6 \pm 4.0$  &$32.2\pm5.4  $ \\
           & Sawtooth                 &$14.6 \pm 4.1$  &$14.1\pm4.1  $ \\ \hline
           & Offset                   &$-14.0\pm 10.5$ &$-13.2\pm 17.6 $ \\
Inners     & Slope                    &$-2.7 \pm 4.0 $ &$1.8 \pm5.4  $ \\
           & Sawtooth                 &$14.2 \pm 4.1 $ &$13.6 \pm4.1  $ \\ \hline
           & Offset                   &$-14.1\pm 10.5$ &$-13.4\pm17.6 $ \\
Span       & Slope                    &$10.5 \pm 4.0 $ &$13.7\pm5.4  $ \\
           & Sawtooth                 &$14.3 \pm 4.1 $ &$13.7\pm4.1  $ \\ \hline
\end{tabular}
\caption[]{Results of the global fit to the high-energy data with each of the
two corrections to the BPM readings removed in turn.  The units are as follows:
Offset [$\times 10^{-5}$],
Slope [${\mathrm \mu \rm{m^{-1}}} \times 10^{-5}$] and
Sawtooth [MeV].  The results should be compared to the `Standard' column in 
table~\ref{tab:he_result}.}
\label{tab:fitwithoutcor}
\end{center}
\end{table}

The fit error in both cases is different from that coming from the
standard treatment.  This is because the corrections move both the energies
and the values of \mTRS\ for each fill.    For individual measurements
the effect of the ambient field correction is larger than is apparent from this
table.  For results determined with the Outers and the polarisation optics,
for instance, the correction is $\sim 25 \times 10^{-5}$. However, a correlated 
correction is made at the same time to \mTRS, in such a manner that the
resulting variation in the global fit results is much smaller.  
With no ambient-field correction,
the consistency in the offset result for the different BPM combinations 
is degraded.

Following the discussion in sections~\ref{sec:ambient} and~\ref{sec:wps},
systematic errors corresponding to 10\% and 25\% of the full
shift are assigned for the ambient-field and
WPS corrections respectively.

\subsection{Spectrometer Result at High Energy}

The component uncertainties in the energy determination using the spectrometer
are summarised in table~\ref{tab:specerrsum}, together
with the total, under the assumption that the contributions
are uncorrelated.

\begin{table}[htb]
\begin{center}
\begin{tabular}{|l|c|} \hline
Contribution &   Value \\
             & [ $\times 10^{-5}$ ] \\ \hline
High-energy scatter     & 15.0 \\
Validity at low energy  & 10.0 \\
Beam size               &  4.0 \\
BPM gains               &  0.5 \\
Integrated dipole field &  1.5 \\
Sawtooth model          &  5.0 \\
WPS correction          &  2.2 \\
Ambient bending field   &  0.7 \\ \hline
Total                   & 19.3 \\ \hline
\end{tabular}
\caption[]{Summary of the error contributions to the spectrometer determination
of the relative difference of \Eb\ with respect to the energy model.
}
\label{tab:specerrsum}
\end{center}
\end{table}

The spectrometer measures the following
offset with respect to the NMR model at a nominal \Eb\ of 92.3~GeV:
\begin{eqnarray}
( \, \Espect - \EOnmr \, )_{\, 92 \, \rm{GeV}} & = & -4.9 \, \pm \, 17.8 \: \rm{MeV}.
\nonumber
\end{eqnarray}

\subsection{Spectrometer Data at Intermediate Energies and in 1999}

Table~\ref{tab:specfills} lists four fills
with spectrometer data taken at a nominal 
energy of 70~GeV.  These may be analysed to provide a spectrometer result
in this intermediate energy regime.

The sample of 70~GeV fills is too small to allow an independent study of
the TRS behaviour.  Instead, any systematic bias is corrected for using
the slope results of the standard fit to the high-energy data.
In fact, as can be observed from figure~\ref{fig:trschar}~(b), the 
\mTRS\ evolution between 50~GeV and 70~GeV is rather small.  Furthermore,
in two of the fills, data at 60~GeV are used to define the reference point,
as this choice further suppresses the TRS systematic.

The four fills give consistent results and are therefore combined to give
a mean energy determination from the spectrometer at 70~GeV.   The accompanying
error from the sawtooth model is assigned to be half of the value of the
applied correction.   The other components in the uncertainty are estimated
as for the high-energy data.    The resulting offset between the spectrometer
and the NMR model is found to be:
\begin{eqnarray}
( \, \Espect - \EOnmr \, )_{\, 70 \, \rm{GeV}} & = & -0.6 \, \pm \, 9.7 \: \rm{MeV}.
\nonumber
\end{eqnarray}

\noindent The correlation with the measurement at 92.3~GeV is dominated by the
common uncertainty arising from the verification of the spectrometer performance 
in the low-energy sample, and is estimated to be 75\%.   

This 70~GeV result is 
used in section~\ref{sec:ebcom}, together with the measurement at 92.3~GeV,
to constrain any evolution of non-linearity of the NMR model with energy.
The 80~GeV points in fills 8443 and 8444 
have not been analysed because of large TRS systematics
and a very high correlation with the measurements at the other energy points.

\vspace{0.5cm}

\begin{figure}
\begin{center}
\epsfig{file=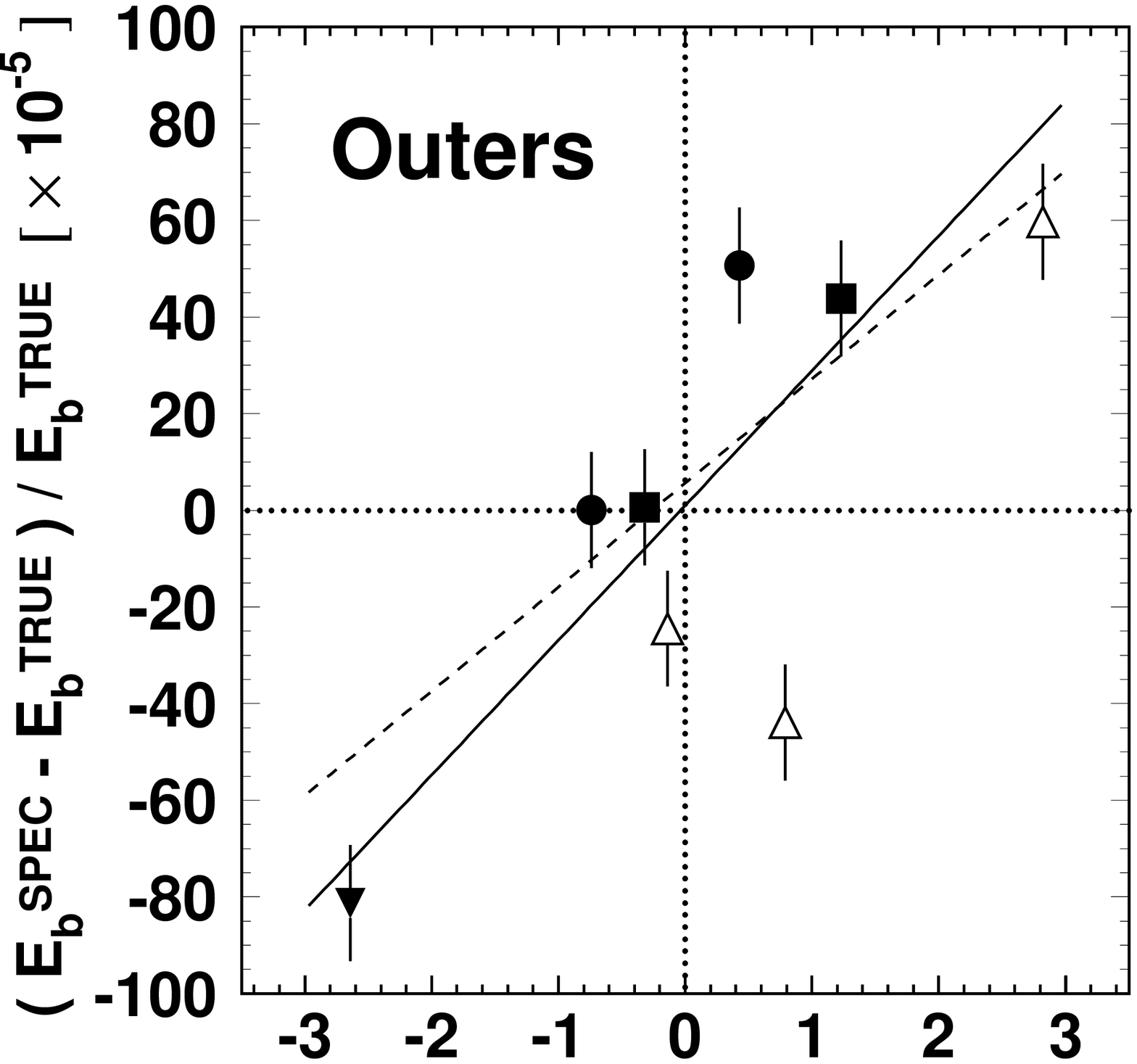,width=0.32\textwidth}
\epsfig{file=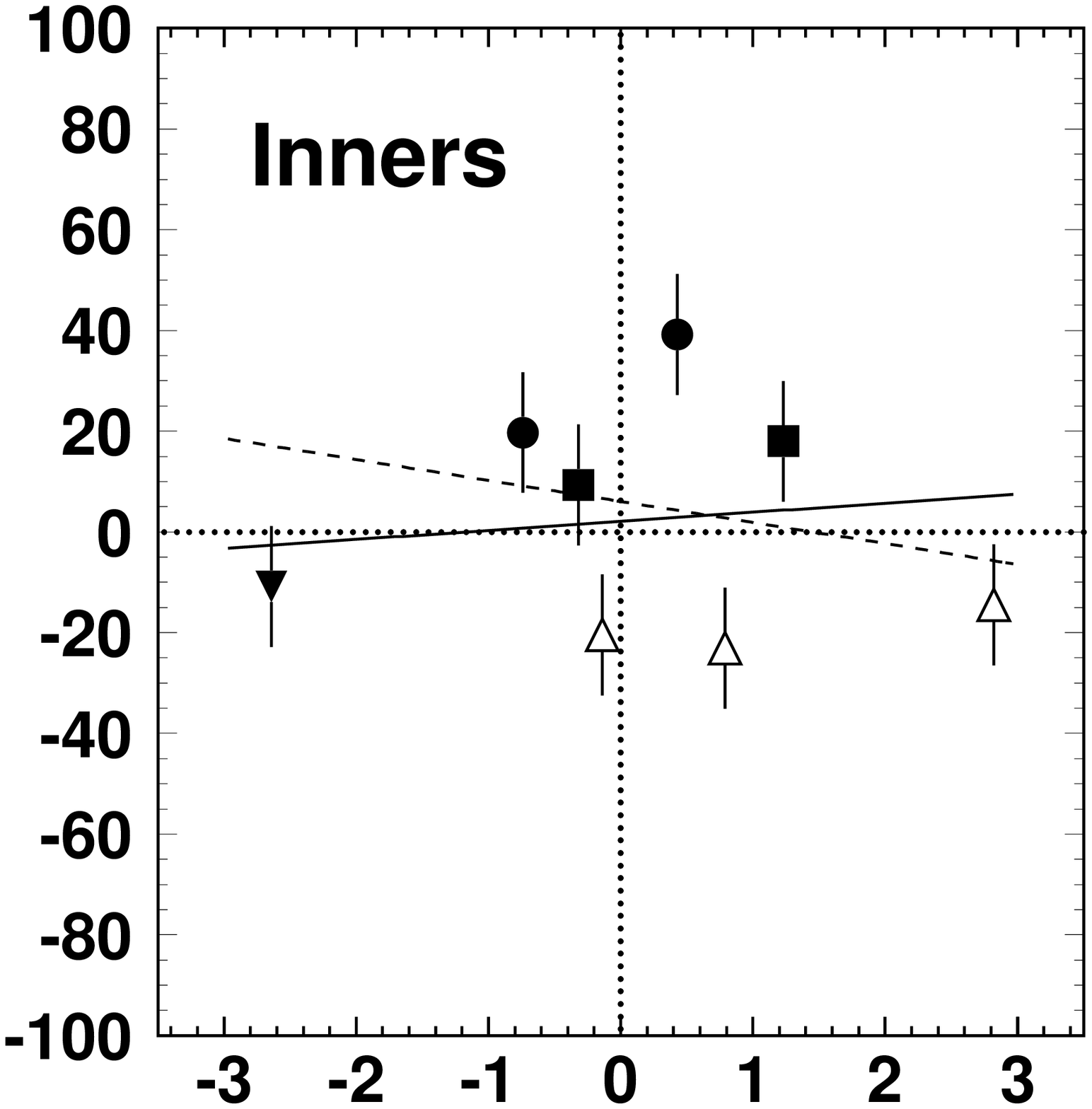,width=0.32\textwidth}
\epsfig{file=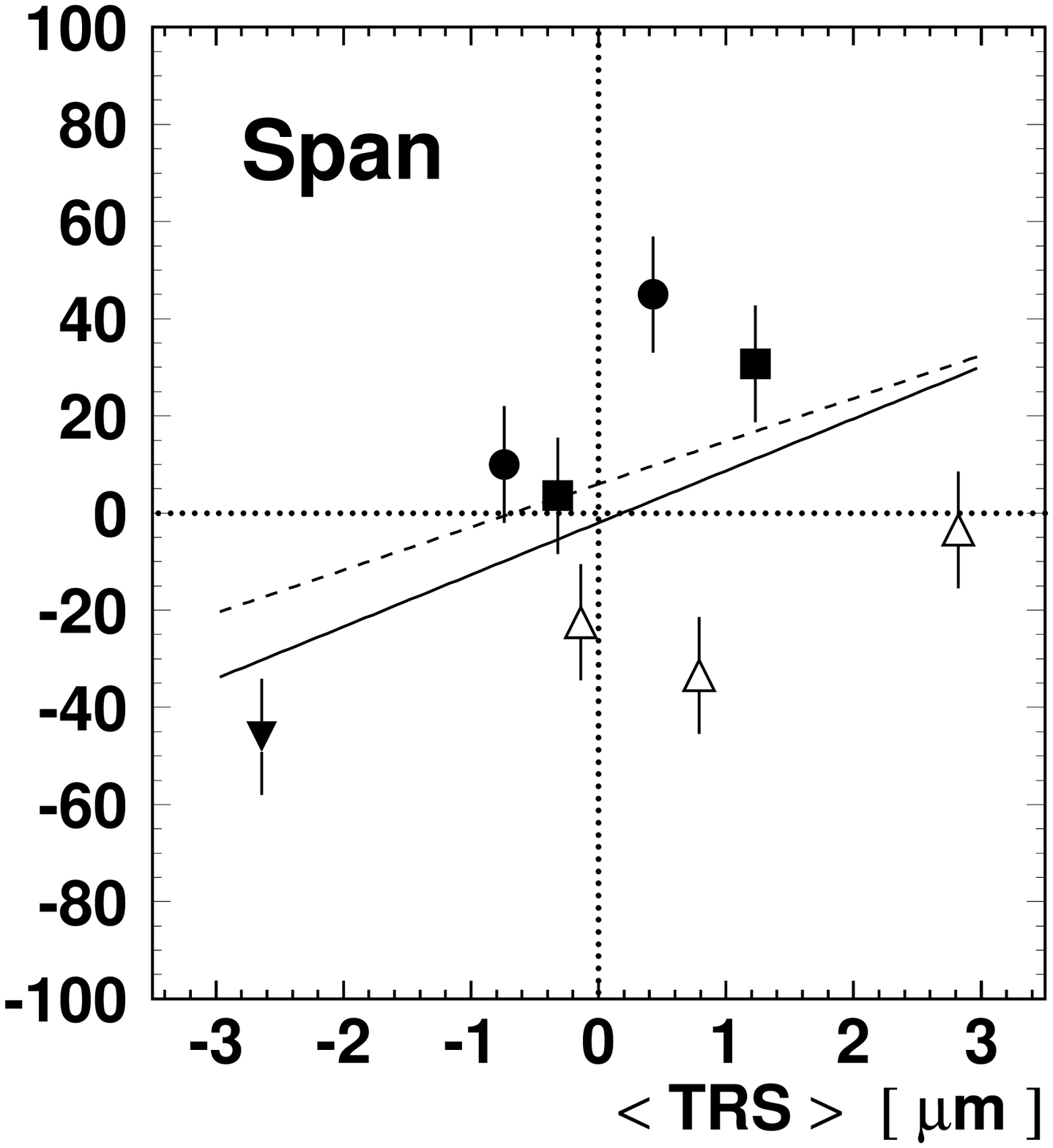,width=0.32\textwidth}
\caption[]{Spectrometer results in 1999 at low energy as a function of \mTRS,
for different BPM combinations.  Common symbols are used to designate
measurements in the same fill.
The bold lines show the results of linear fits made to the data. 
Superimposed 
as dashed lines are fits to the 2000 data set.}
\label{fig:le_result_99}
\end{center}
\end{figure}

\begin{figure}
\begin{center}
\epsfig{file=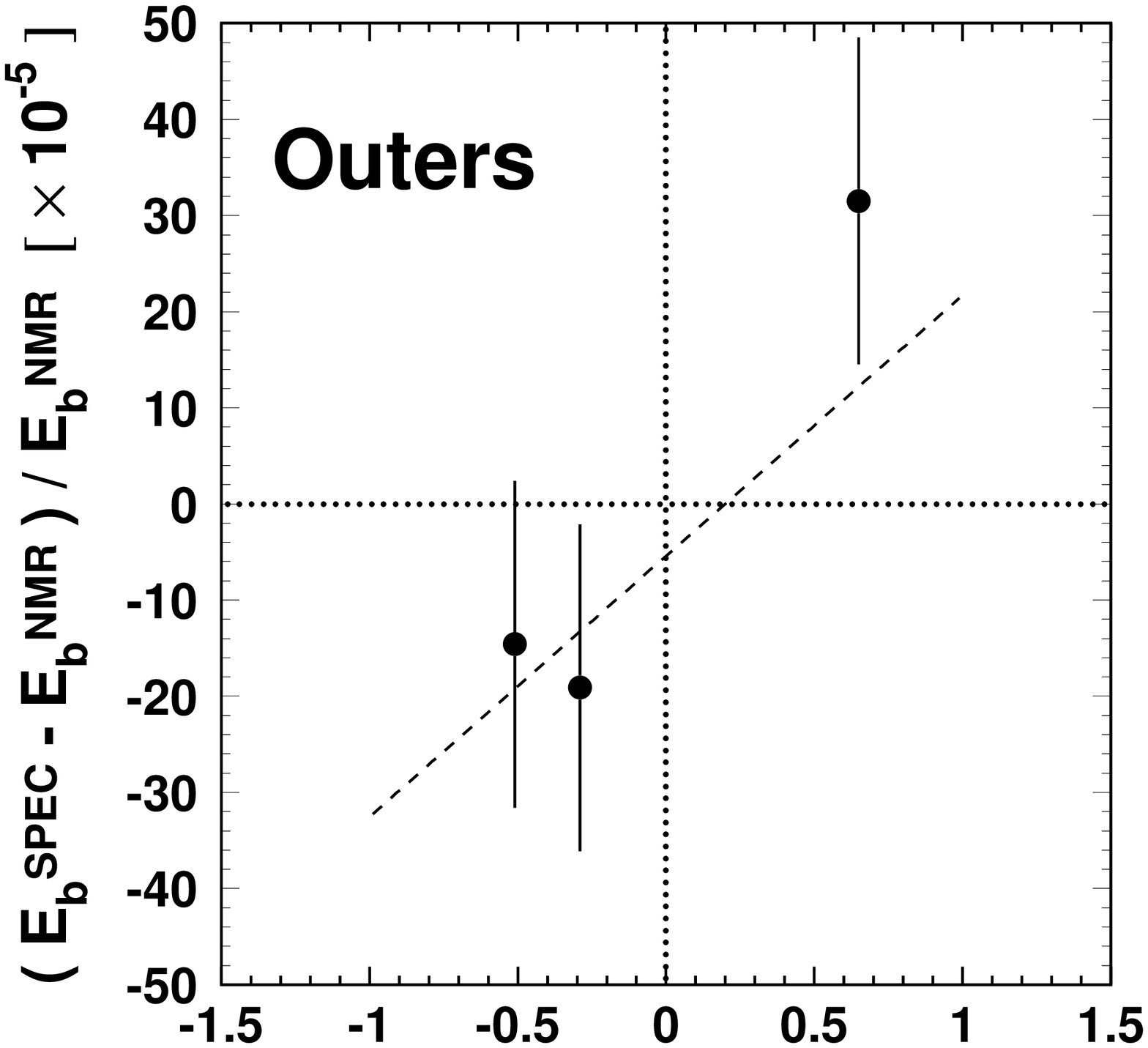,width=0.32\textwidth}
\epsfig{file=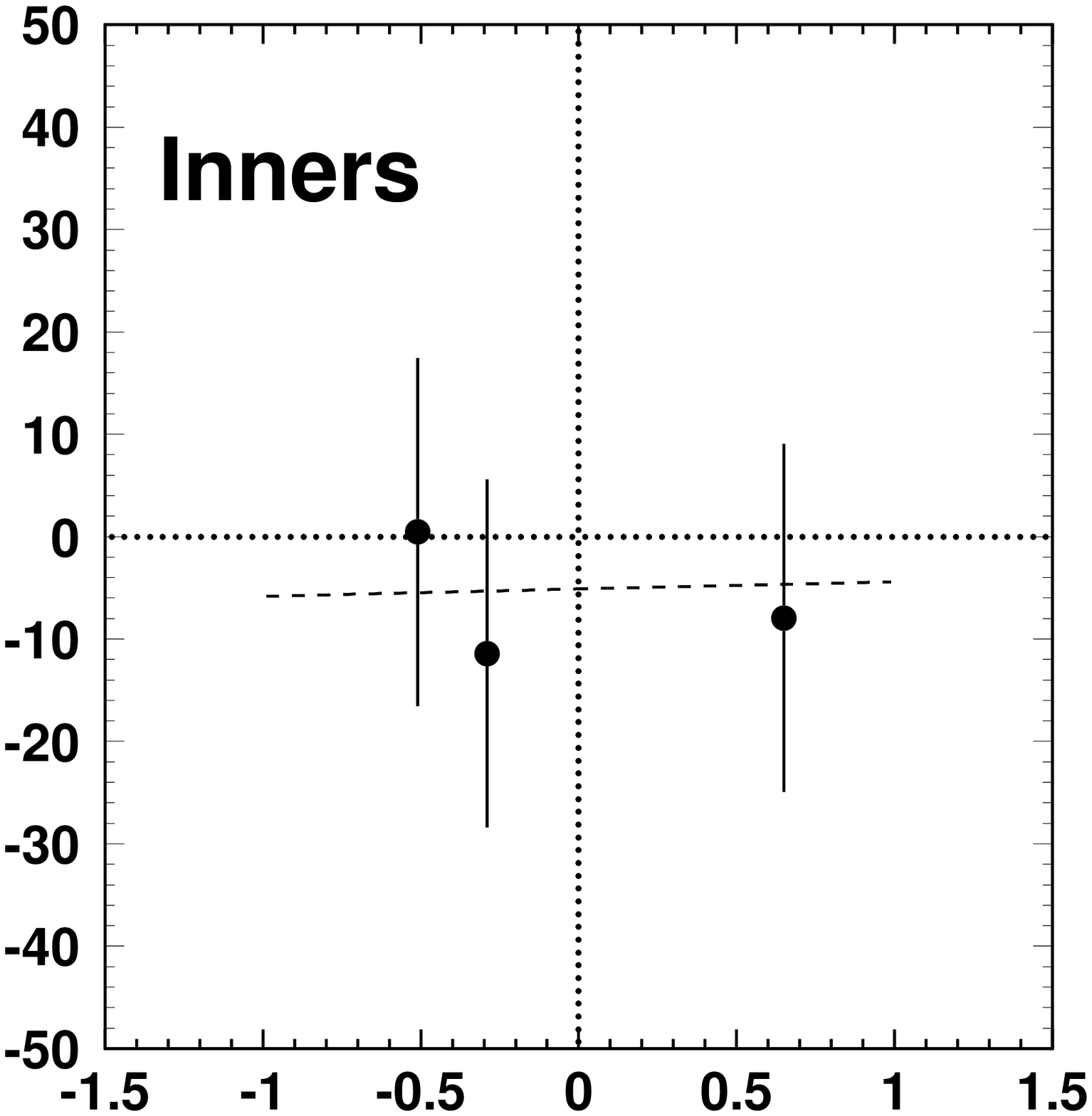,width=0.32\textwidth}
\epsfig{file=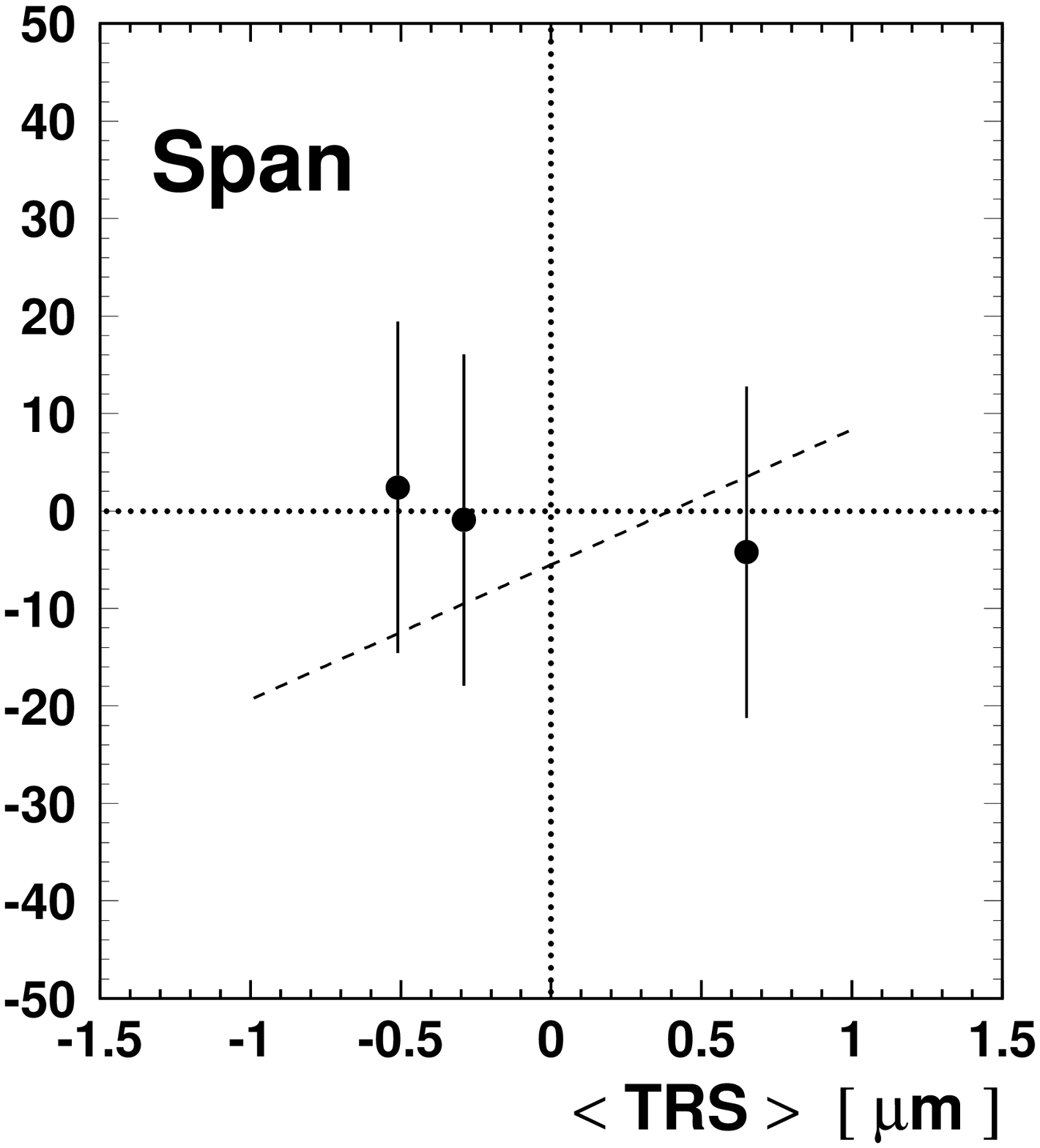,width=0.32\textwidth}
\caption[]{Spectrometer results in 1999 at high energy as a function of \mTRS,
for different BPM combinations. 
Superimposed 
as dashed lines are fits to the 2000 data set.}
\label{fig:he_result_99}
\end{center}
\end{figure}

During the latter period of the 1999 run, data were taken at low and high energy
in order to commission the spectrometer.   The stability of the operating 
conditions were significantly inferior to 2000.  Furthermore, all the experiments
were made with an electron beam, not allowing constraints to be placed 
on the sawtooth model for this year.   For these reasons, no quantitative
results are presented.  The comparison of the spectrometer measurements with the
true energy as a function of \mTRS\ in figure~\ref{fig:le_result_99}  shows, however,  
that the behaviour at low energy for each BPM combination appears to be
very compatible with that observed in 2000,  although the points
exhibit a larger scatter.  Figure~\ref{fig:he_result_99} 
indicates that the results at high
energy are also consistent.   (Some of these data have positive TRS in one
arm, which when averaged with the negative values in the other arm, 
lead to smaller values of \mTRS\ than in 2000.)   This adds confidence for the 2000
analysis.


\subsection{Measurement of the BFS Boost with the Spectrometer}
\label{sec:bfsmeas}

In addition to constraining the magnetic extrapolation,
the spectrometer is also used to calibrate the {\it bending-field spreading} 
(BFS) boost.
In both fills 7931 and 8566, after the usual spectrometer
measurements had been made at high energy, a BFS boost was
then applied.
This action induced a noticeable change 
of bending angle in the spectrometer.   In order to minimize
BPM-related systematics, the RF frequency was increased
so as to introduce a  known energy change of opposite sign
to the BFS, and thereby return the bending angle to close to
its original value.  Care was also taken to re-steer the beam
back to its position prior to the boost.   
The residual change
in bending angle is measured, and from this and the 
change in \fRF\, the effect of the BFS boost is determined.

The results of the experiments are shown in table~\ref{tab:bfsres},
giving both the nominal and measured values of the boost applied.
In calculating the systematic error, contributions
are considered from the variation in result with BPM combinations;
from a $5\%$ uncertainty in the absolute-gain scale;
through any variation in relative gains seen in the online calibrations
within the fill; and from a $1\%$ error in the momentum compaction factor.
Because of time constraints, the beam was significantly 
less well re-centred in fill 8566 than in 7931, and this explains the 
difference in precision between the experiments.  Both measurements,
however, show the value of the  BFS 
boost to be consistent with expectations.  

\begin{table}
\begin{center}
\begin{tabular}{|c|c|r|} \hline
Fill   & Nominal BFS & \multicolumn{1}{|c|}{Measured BFS}   \\ \hline
7931   &  219 MeV    &  213.5 $\pm$ 7.8 MeV \\
8566   &  297 MeV    &  304.1 $\pm$ 33.5 MeV \\ \hline
\end{tabular}
\caption[]{Results for the BFS calibration experiments.}
\label{tab:bfsres} 
\end{center}
\end{table}

%% file: qs_ana.tex
\section{\Eb\ Measurement with the \Qs\ Fit}
\label{sec:qs}

The combined effects of synchrotron radiation loss, and the boost from
the RF system, leads to particles undergoing longitudinal oscillations.
The frequency of these oscillations is dependent on the particle energy.
An analysis based on measurements of 
the oscillation frequency, therefore,  offers
an alternative way to determine \Eb\ and constrain the energy model.

\subsection{Energy Loss and Synchrotron Oscillations}

\begin{figure}[htb]
\begin{center}
\epsfig{file=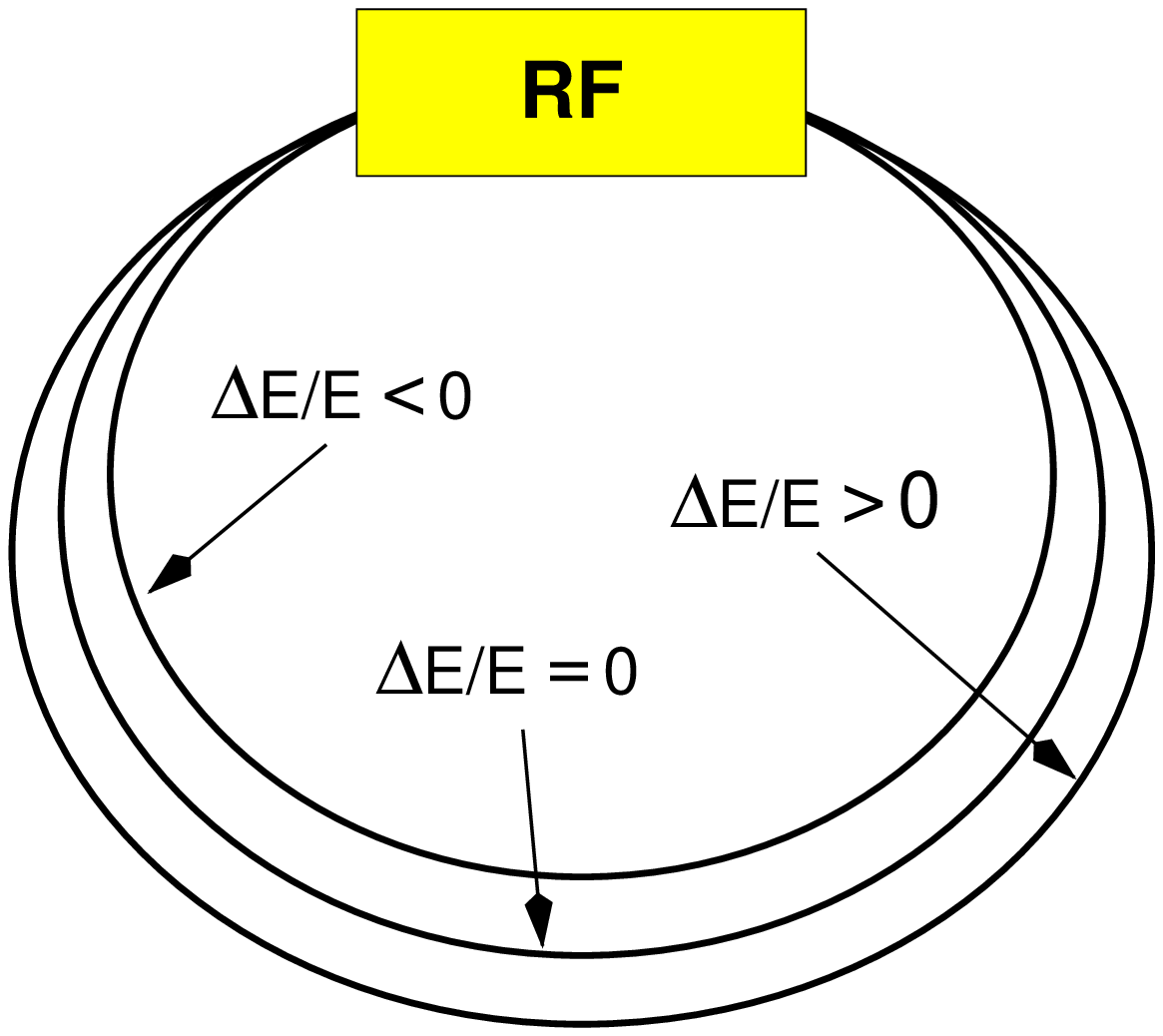,width=0.38\textwidth}
\epsfig{file=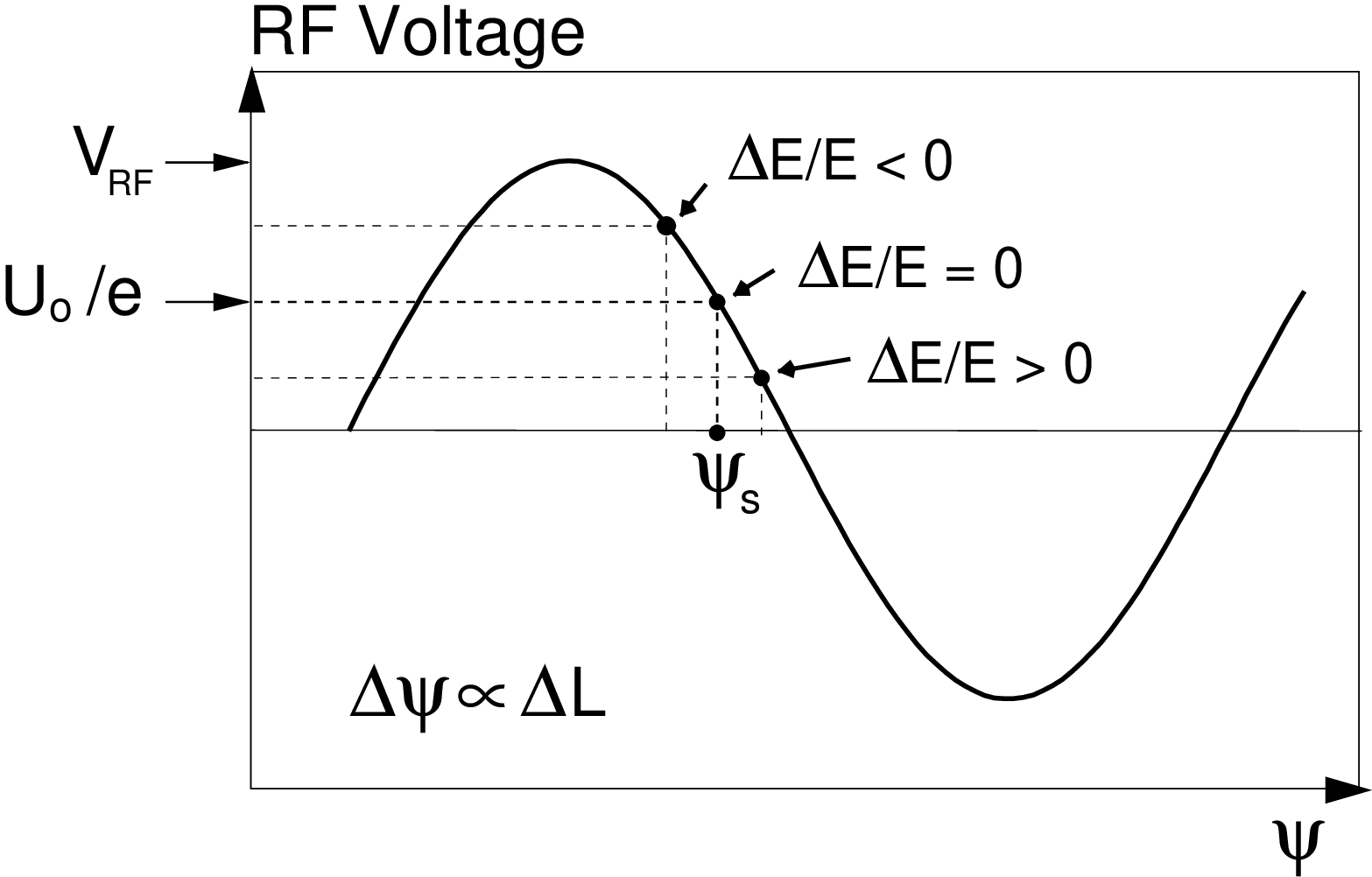,width=0.53\textwidth}
\caption[]{Change of orbit length ($\Delta L$) for particles with 
energy deviation ($\Delta E$) in 
an accelerator with a single RF cavity (left).  Accelerating voltage as a
function of phase; particles with different energies arrive at different
phases, thus seeing a voltage different from that needed to compensate the
nominal energy loss per turn (right).}
\label{fig:rf_schem}
\end{center}
\end{figure}

Consider the case of a beam of energy \Eb, 
experiencing an energy loss per 
turn in the dipole magnets of $U_0$, 
as given by expression~\ref{eq:u0def2}.
This energy loss is restored by the RF system, which,
for the purposes of discussion, is taken to be a single  cavity as shown in 
figure~\ref{fig:rf_schem}.~\footnote{
The results quoted, however, are derived 
under the assumption that the RF voltage is distributed homogeneously around
the ring.} 
The voltage provided by the cavity to the arriving beam can be expressed as
\begin{equation}
V(\psi) = V_\rm{RF} \sin \psi ,
\end{equation}

\noindent where $V_\rm{RF}$ is the peak voltage provided by the 
system and $\psi$ the phase.   
The stable phase
angle, $\psi_s$, of particles with the nominal energy,
is defined by the condition $U_0 = e \, V_\rm{RF} \sin \psi_s$.
Particles with lower-than-nominal energy follow a shorter path length 
and, in the ultra-relativistic regime, arrive at an earlier time in
the RF cycle, therefore experiencing a larger energy boost than
particles at $\psi_s$. The converse is
true for particles with higher than nominal energy.  These effects
lead to {\it synchrotron oscillations} of angular  frequency $\Omega$.
Assuming that the amplitude of oscillations is small, and the damping
due to synchrotron radiation is negligible, it can be shown~\cite{ACCELBOOK1}
that:
\begin{equation}
  \Omega^2 ~=~ \omega_{\mathrm{rev}}^2 
              \left( \frac{\alpha_c h}{2 \pi \Eb} \right) 
               ~e~ \frac{dV}{d\psi} (\psi_s) ,
\label{eq:omega2}
\end{equation}  

\noindent
where $\omega_{\mathrm{rev}}$ is the angular revolution frequency,
$\alpha_c$ is the momentum compaction factor and $h$ the harmonic
number of the accelerator, that is the ratio between the RF frequency
and the revolution frequency  (31320 in the case of LEP).

The {\it synchrotron tune}, $Q_s$, is defined as the ratio of the oscillation
frequency to the revolution frequency. Expression~\ref{eq:omega2},
together with the definition of the stable phase condition, gives
the following relation:
\begin{equation}
  Q_{s}^{2} = \left( \frac{\alpha_c h}{2 \pi \Eb} \right) 
                        \sqrt{e^{2}~V^{2}_\rm{RF} - U_{0}^{2}}.
\label{eq:simmod}
\end{equation}

In principle, therefore, fitting expression~\ref{eq:simmod} to
measurements of the synchrotron
tune at different RF voltages enables
the beam energy to be determined.
In practice, however, this expression is inadequate for energy
calibrations of the required precision.  It neglects 
energy losses in the quadrupoles, correctors and from other sources.
Further corrections are necessary to account for the particular 
distribution of RF cavities at LEP and the possibility of large-amplitude 
oscillations.   These refinements are discussed in
section~\ref{sec:improvedqs}.

\subsection{Measurement Procedure and Datasets}

\begin{figure}[htb]
\begin{center}
\epsfig{file=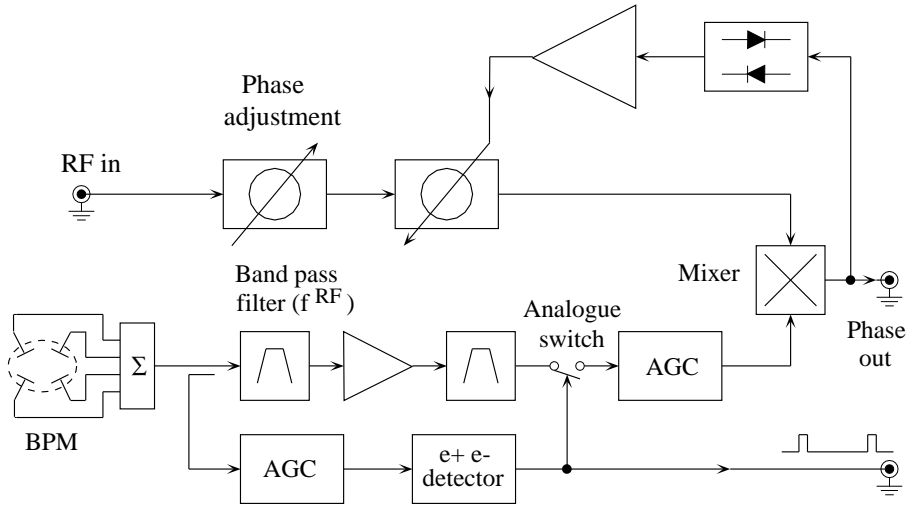,width=0.75\textwidth}
\caption[]{A schematic block diagram of the LEP synchrotron oscillation 
detector.}
\label{fig:qsschem}
\end{center}
\end{figure}

The determination of the $Q_s$ was based on a measurement of the phase 
between a bunch and the RF frequency.  
Figure~\ref{fig:qsschem} shows
a block diagram of the LEP bunch phase monitoring system.  
The summed 
signal from a four-button BPM was processed with band pass filters centred
at the RF frequency, amplifiers and an automatic gain control (AGC)
loop.  The phase of the resulting signal
was compared to the RF frequency (mixer), and the output fed into
a spectrum analyser.   The $Q_s$ peak in the resulting 
Fourier-analysed spectrum was located 
manually~\footnote{During LEP 1 operation an automatic peak finder yielded
the $Q_s$ data used to help understand the modelling of the
RF system~\cite{LEP1PAPER}.
This proved unreliable during high-energy running.}
with a typical accuracy of 0.0003.   In general, the signal was averaged
over several turns and over all bunches.  

\begin{figure}[htb]
\begin{center}
\epsfig{file=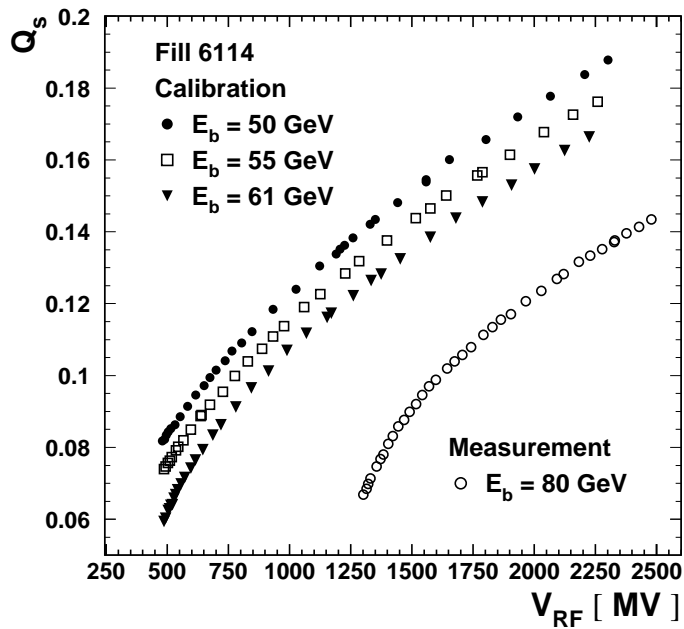,width=0.61\textwidth}
\caption{Measured $Q_s$ is shown against
$V_{\rm{RF}}$ for different beam energies. The calibration set for the 
voltage calibration factor is indicated, together with the high-energy
data. }
\label{fig:ecalqs}
\end{center}
\end{figure}

In a $Q_s$ energy calibration experiment, measurements were first
made at one or more low-energy points, before
ramping to high energy. The purpose of the
low-energy measurements was to enable the absolute scale of the
RF voltage to be fixed through cross-calibration against the energy model
in a regime where the model is known to be reliable.
At each energy point the total RF voltage, $V_{\rm{RF}}$, was 
varied over the same range, stepping between
the lowest value compatible with stable operation at high energy, 
to the maximum available.  
The need to span a significant range in $V_{\rm{RF}}$
dictated that the choice of high-energy point, most usually 80~GeV,
was typically somewhat lower than that attainable
by the full RF system during physics operation.  
At each value of $V_{\rm{RF}}$ the synchrotron tune was measured.
Data from a typical  $Q_s$ experiment are shown in figure~\ref{fig:ecalqs}.

Energy calibration experiments using $Q_s$  were made in 1998, 1999 and 2000
and are listed in table~\ref{tab:qsmd}.  In total six fills were used
to measure \Eb\ at high energy.   Other fills were used to constrain
uncertainties in the higher order corrections to the 
model.
All but one of these
were made with single beams of positrons.
(Fill 5128 was made with positrons and electrons simultaneously.)  
The choice of optics was 102/90 for all
experiments apart from 8445, which was performed with 101/45.
The bunch currents were set low, with typical values of
$50 \, {\rm \mu \rm A}$,  so as to minimise the parasitic mode energy
loss discussed in section~\ref{sec:addeloss}.

\begin{table}
\begin{center}
\begin{tabular}{|l|c|c|c|} \hline
Fill    & Date        &  \Eb\ of measurements [GeV] & Interest of experiment \\ \hline
5128    & 4 Sept '98  &  66, 91       & Energy calibration  \\
 &  &  & \\
5981    & 24 July '99 &        61       & Parasitic mode loss \\
6114    & 13 Aug '99  &  50, 55, 61, 81  & Energy calibration  \\
6338    & 15 Sept '99 &  50, 55, 61, 80  & Energy calibration  \\
 &  &  & \\
7456    & 14 June '00 &  42, 45, 48, 50, 55, 61 & Bending radius constraint \\
7832    & 20 July '00 &        61       & Parasitic mode loss \\
8315    & 29 Aug '00  &  50, 55, 61, 80       &  Energy calibration  \\
8445    & 10 Sept '00 &  50, 55, 61, 65, 80 &  Energy calibration  \\
8809    & 18 Oct '00  &  50, 55, 61, 65, 80 &  Energy calibration  \\ \hline
\end{tabular} 
\end{center}
\caption[]{List of $Q_s$ fills, giving the energy points analysed and
the main purpose of the experiment.}
\label{tab:qsmd}
\end{table}

\subsection{The Improved Synchrotron Oscillation Model}
\label{sec:improvedqs}
\subsubsection{RF Calibration and Distribution}
\label{sec:qsimprovedrf}

The effective voltage seen by the beam can be significantly different from
the sum of all individual nominal cavity voltages due
to uncertainties in the voltage calibration, 
phasing errors, 
and longitudinal alignment errors. 
A crucial correction to expression~\ref{eq:simmod} is therefore 
to replace $V_{\rm{RF}}$, the nominal total RF voltage, by $g \, V_{\rm{RF}}$,
where $g$ is a correction factor to account for these effects.

\begin{figure}[htb]
\begin{center}
\epsfig{file=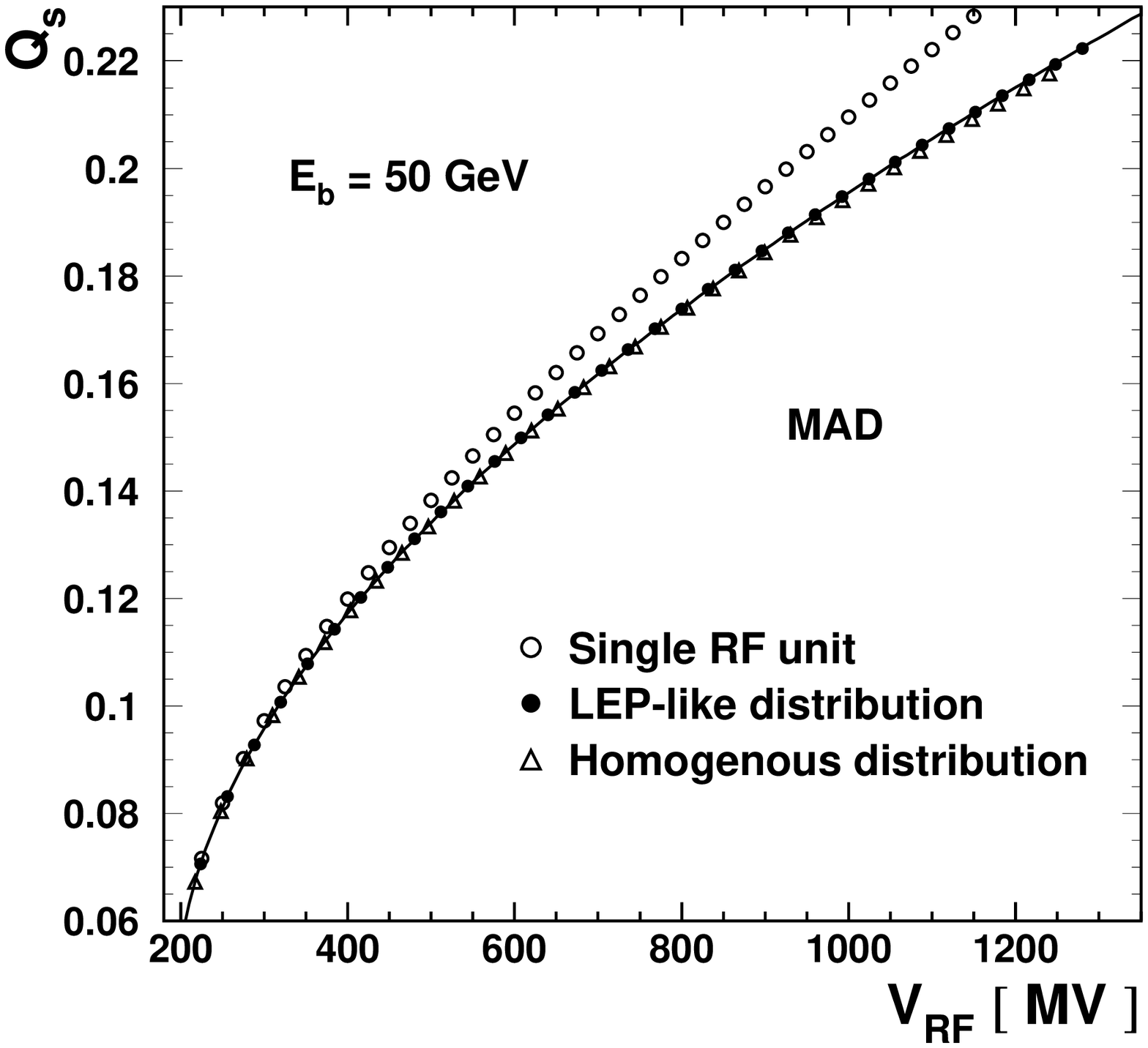,width=0.59\textwidth}
\caption[]{Synchrotron tune as a function of 
  total RF voltage as calculated with the MAD program for different RF
  configurations. The curve is a fit to the correct RF distribution
  using the model of equation~\ref{eq:finalfit}.}
\label{fig:madfit}
\end{center}
\end{figure}

The correction factor is determined separately for each experiment
by fitting the final $Q_s$ model (see expression~\ref{eq:finalfit}) 
to the low-energy
data for different values of $g$, to find the factor which results in
a beam energy in agreement with the energy model.
Care is taken to use the
same configuration of RF cavities and span of voltages at each energy
point, so that this correction factor is applicable to the high-energy point
of that experiment.
The uncertainty in $g$ is taken from the scatter in results over the low 
energy points, and the central value from the average.
The results for each experiment are shown in table~\ref{tab:v_calib}.
$g$ is typically found to be within a few percent of unity, 
with an uncertainty of $\sim$ 0.001.

\begin{table}
\begin{center}
\begin{tabular}{|l|c|}\hline
Fill  & $g$ \\ \hline
5128  & 0.9499 $\pm$ 0.0006 \\
6114  & 0.9805 $\pm$ 0.0009 \\
6338  & 1.0006 $\pm$ 0.0008 \\
8315  & 1.0054 $\pm$ 0.0009 \\
8445  & 1.0064 $\pm$ 0.0005 \\
8809  & 1.0030 $\pm$ 0.0002 \\ \hline
\end{tabular}
\caption[]{Fit results of the voltage calibration factor, $g$,
for the 6 fills used in the $Q_s$ energy measurement.}
\label{tab:v_calib}
\end{center}
\end{table}

Expression~\ref{eq:simmod} is derived assuming that the RF voltage
is distributed homogeneously around the accelerator.
In LEP, however, the cavities were 
concentrated in the four straight sections.  Investigations with the
MAD program~\cite{MAD} show that $Q_s$ has a dependence on this
distribution. This can be seen in figure~\ref{fig:madfit} which
shows $Q_s$ generated for a beam energy of
50~GeV with three different RF configurations: a typical case 
with the standard LEP RF distribution, 
a case where the same total voltage is concentrated at one point, 
and the limit of a homogeneous distribution where the
voltage is distributed over the whole ring.   The correct distribution 
can be adequately modelled by adding in expression~\ref{eq:simmod} a
term in $V_{\rm{RF}}^4$, controlled by a weighting coefficient $M$ of
order $10^{-7}$.  This
is illustrated by the superimposed curve in figure~\ref{fig:madfit}.

When analysing the data, the value for $M$ is taken from fits to the
appropriate MAD simulations.  Any residual imperfections in this
treatment are absorbed into the voltage calibration factor.

\vspace*{0.2cm}
\subsubsection{Total Energy Loss}
\label{sec:addeloss}

Expressions~\ref{eq:u0def2} and~\ref{eq:simmod} assume that the beam energy
is fully supplied by the dipole field, and that all the energy loss 
arises through synchrotron radiation in the dipoles.  As neither
assumption is wholly valid, the total energy loss $U_o$ as used in
equation~\ref{eq:simmod} has to be replaced by
\begin{equation}
  \tilde{U}_0 = \frac{r \, C_{\gamma}}{\rho}(\EbO)^{4} \,+\, 
\sum \Delta U,
\label{eq:eloss2}
\end{equation}
\nin
where \EbO\ refers to that part of the beam
energy defined by the dipoles alone, and
\begin{equation}
\sum \Delta U \, = \,(\Delta U_\rm{\Eb} + \Delta U_\rm{quad}) +
\Delta U_\rm{closed} + \Delta U_\sigma + \Delta U_\rm{cor} +
\Delta U_\rm{\,PML},
\nonumber
\end{equation}

\noindent
is the sum of all additional energy loss, which are explained in the 
following.
The factor $r$ represents a correction to
the inverse bending radius, and is discussed separately in 
section~\ref{sec:invbendrad}.

\vspace{0.3cm}
\nin
{\bf Quadrupole Effects}
\vspace{0.3cm}

\nin
As discussed in section~\ref{sec:eb_model}, the beam energy as set 
by the dipole field receives additional contributions, the most important 
of which is associated with off-centre trajectories in the quadrupoles.
According to relation~\ref{eq:u0def} the energy loss in the dipoles 
scales as $\Eb^2 \, B^2$, where
$B$ represents the dipole field. 
Therefore, in expression~\ref{eq:eloss2}
the familiar energy-to-the-fourth power term is specified as being associated
with the dipole field alone, and a correction $\Delta U_{\Eb}$,
is added, where $\Delta U_{\Eb} / {U_0} \, = 2 \, (\Delta \Eb / \Eb)$.

In addition to modifying the beam energy,  beam offsets in
the quadrupoles will result in synchrotron radiation in the quadrupoles
themselves. 
For a transverse offset of $(x_0, y_0)$ this contribution
to the turn-by-turn energy loss, $\Delta U_\rm{quad}$, goes as
$E_b^4 \, (x^2_0 \, + \, y^2_0)$.
A beam of energy 80~GeV at an offset of 0.5~mm will lose 
approximately 0.2~keV in each of the 850  quadrupoles.

\begin{figure}[htb]
\begin{center}
\epsfig{file=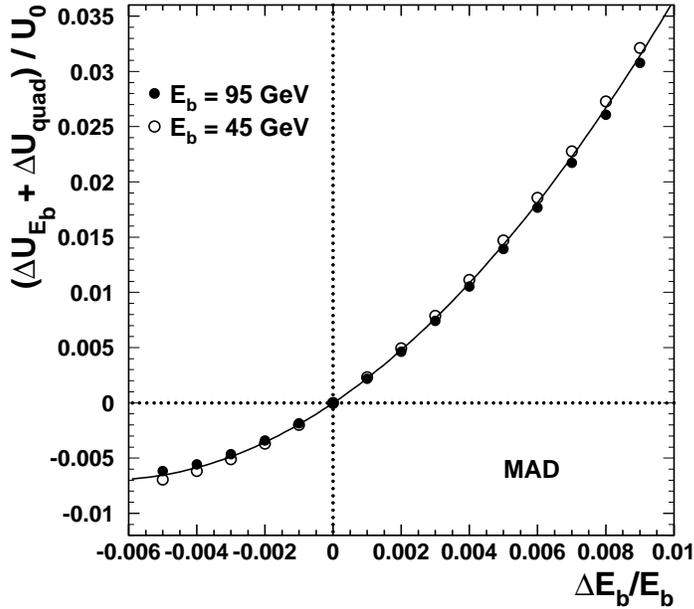,width=0.6\textwidth}
\caption[]{Variation of relative energy loss
with relative energy change induced by
beam offsets in the quadrupoles.  The results are calculated with MAD 
for two energy points.  A common parameterisation is superimposed.
Note that during operation the relative energy change from beam offsets is
typically $\le 10^{-4}$.}
\label{fig:qs_quadloss}
\end{center}
\end{figure}

The net effect of additional contributions to the beam energy,
and of synchrotron radiation in the quadrupoles, has been studied with
the MAD program.
Figure~\ref{fig:qs_quadloss} shows how the relative energy loss 
from both sources varies 
with relative \Eb\ changes induced by beam offsets.  As expected,
the dependence exhibits the superposition of a linear term,
associated with $\Delta U_{\Eb}$, and a quadratic term, coming from 
$\Delta U_\rm{quad}$.   The variation at high and low energy is sufficiently
similar to allow a common parameterisation.  This is then included in the
$Q_s$ model to account for the offsets caused by \fRF\
manipulations, earth tides,
and longer timescale
geological distortions, as tracked by the \fRFc\ evolution.

Even when the global effects coming from these sources 
are subtracted, there remain significant local 
offsets from quadrupole to quadrupole, with typical RMS of
0.5~mm.    These `closed orbit distortions' 
are logged in BPMs close to the defocusing quadrupoles, and can be
extrapolated to the focusing quadrupoles with knowledge of the local
betatron function.  These offsets are used to calculate the
additional energy loss, $\Delta U_\rm{closed}$, 
from all the quadrupoles around the machine.
Note that though fundamentally random in distribution, they contain a 
residual systematic component from the variation in horizontal
position arising from the RF sawtooth.

A final addition to the energy loss arising from the 
quadrupoles is $\Delta U_\rm{\sigma}$, a contribution 
caused by the finite beam size.   This is present even for
beams which have no offset, and is proportional to
$ E_b^4 \, (\sigma_x^2 \, + \, \sigma_y^2)$, where $\sigma_x$ and $\sigma_y$
are the horizontal and vertical beam sizes respectively.
MAD is used to calculate the beam size at each quadrupole, so 
that the energy loss from this source may also be included.

\vspace{0.4cm}
\nin
{\bf Other sources of Synchrotron Radiation}
\vspace{0.4cm}

\nin
Additional energy loss occurs through synchrotron radiation in the corrector
dipoles.  This contribution, $\Delta U_\rm{cor}$,
is calculated taking as input the RMS scatter in the 
logged values of the settings around the ring.

Synchrotron radiation in the sextupole magnets leads to negligible
energy loss.

\vspace{0.4cm}
\nin
{\bf Parasitic Mode Losses}
\vspace{0.4cm}

\nin
After the synchrotron radiation in the bending dipoles, the so-called
{\it parasitic mode losses}~\cite{ACCELBOOK2} 
are the largest contribution to the total energy
loss.  These arise from the impedance experienced by the beam from
resistance in the vacuum chamber walls and from resonator-like 
structures.

For each particle, the energy loss per turn from this source is 
$\Delta U_{\rm{\,PML}} \,=\, 2\pi\,e\,I_{b}\kappa_{||}\,/\,\omega_\rm{rev}$, 
where $I_b$ is the beam current,
and $\kappa_{||}$ the longitudinal loss factor, which in turn depends
on the longitudinal resistive impedance~\cite{HANDBOOK}. 
$\Delta U_\rm{\,PML}$ can be determined
from the data by including the parasitic mode loss in the $Q_s$ description
and fitting the model to the data at low energy over a range of different
beam currents. 

\begin{figure}[htb]
\begin{center}
\epsfig{file=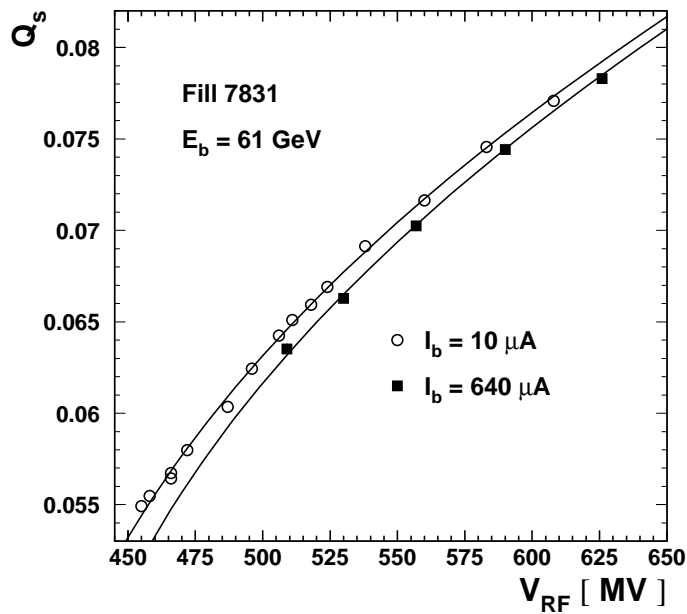,width=0.6\textwidth}
\caption[]{$Q_s$ as function of
  total RF voltage for two different bunch currents in fill 7831, showing
  the effect of the parasitic mode losses. 
  The curves are individual fits 
  to each dataset.}
\label{fig:qs_curr_zoom}
\end{center}
\end{figure}

Figure~\ref{fig:qs_curr_zoom} shows 
measurements of $Q_s$ as function of RF voltage at an energy of 61~GeV 
for two bunch currents of $10~{\rm \mu \rm A}$ and $640~{\rm \mu \rm A}$
for a fill in 1999.
The difference in behaviour due to parasitic mode losses is clearly 
visible.   A simultaneous fit to the data from five different bunch-current
values  yields a current
dependent energy loss
of $\Delta U_\rm{\,PML}/I_{\rm b} \, = \, (18.5  \pm 2.0)$~MeV/mA. 
This result is confirmed by the analysis of a second experiment, conducted in 
2000, which gives $\Delta U_\rm{PML}/I_{\rm b} \, = \, (20.7 \pm 3.1)$~MeV/mA.

The longitudinal loss factor is expected to have some weak dependence on
the bunch length, which itself varies with $Q_s$.  Fits to the
experiments with different bunch currents are not sensitive to
this variation, due to correlations with other parameters.   Therefore
in parameterising the parasitic mode loss in the energy fits for
a given dataset, a constant
value of $\Delta U_\rm{\,PML}/I_{\rm b}$ is assumed, 
and the approximation taken
account of in the error assignment.   
A dataset-dependent scaling factor is applied to account for the
differences in bunch length with optics setting.

\vspace{0.4cm}
\nin
{\bf Summary}
\vspace{0.4cm}

\begin{table}[htb]
\begin{center}
\begin{tabular}{|l |c|c|c|}\hline
Energy Loss Mechanism & \multicolumn{3}{|c|}{Energy Loss [MeV]} \\ \hline
& 50 GeV & 61 GeV & 80 GeV \\ \cline{2-4}
Offsets in quads ($\Delta U_\rm{\Eb} + \Delta U_\rm{quad}$)
                                      & -0.1 & -0.3 & -0.9 \\
Closed orbit distortions  ($\Delta U_\rm{closed}$)
                                      &  0.1 &  0.2 &  0.6 \\
Beam size ($\Delta U_\sigma$)         &  0.1 &  0.3 &  1.7 \\
Parasitic mode losses  ($\Delta U_\rm{\, PML}$)
                                      &  1.1 &  1.1 &  1.0 \\
Correctors ($\Delta U_\rm{cor}$)      &  0.1 &  0.1 &  0.3 \\ \hline
Total correction ($\sum \Delta U$)    &  1.3 &  1.4 &  2.7 \\ \hline
\end{tabular}
\caption[]{Estimates of the additional energy losses at three energy
points for a typical $Q_s$ experiment with bunch currents
of around 50 $\rm \mu A$.}
\label{tab:elossexample}
\end{center}
\end{table}

\nin
The additional contributions to the total energy loss and their sum, 
$\sum \Delta U$,
are listed in table~\ref{tab:elossexample} for three energy points in
a typical experiment.   The most important components are
the parasitic mode loss and the beam size.  The relative precision on these
corrections are estimated to be $\pm 20 \%$ and $\pm 10 \%$ respectively.
When fitting the complete $Q_s$ model to the data, a conservative
uncertainty of 0.5~MeV is assigned to $\sum \Delta U$ at all energy points.
$\sum \Delta U$ represents a $0.7 \times 10^{-3}$ relative correction
to the original energy loss estimates of expression~\ref{eq:u0def2} at
$\Eb = 50$~GeV, and $0.2 \times 10^{-3}$ at $\Eb = 80$~GeV.

\subsubsection{Correction to the Magnetic Inverse Bending Radius}
\label{sec:invbendrad}

\nin
The value of $\rho$, the average magnetic bending radius of LEP
which determines the energy loss in  expression~\ref{eq:u0def}, 
is taken from a calculation made with the MAD program.  This calculation, 
however, 
is based on an imperfect modelling of the dipole fringe fields,
and this has consequences for the energy loss.

\begin{figure}[htb]
\begin{center}
\epsfig{file=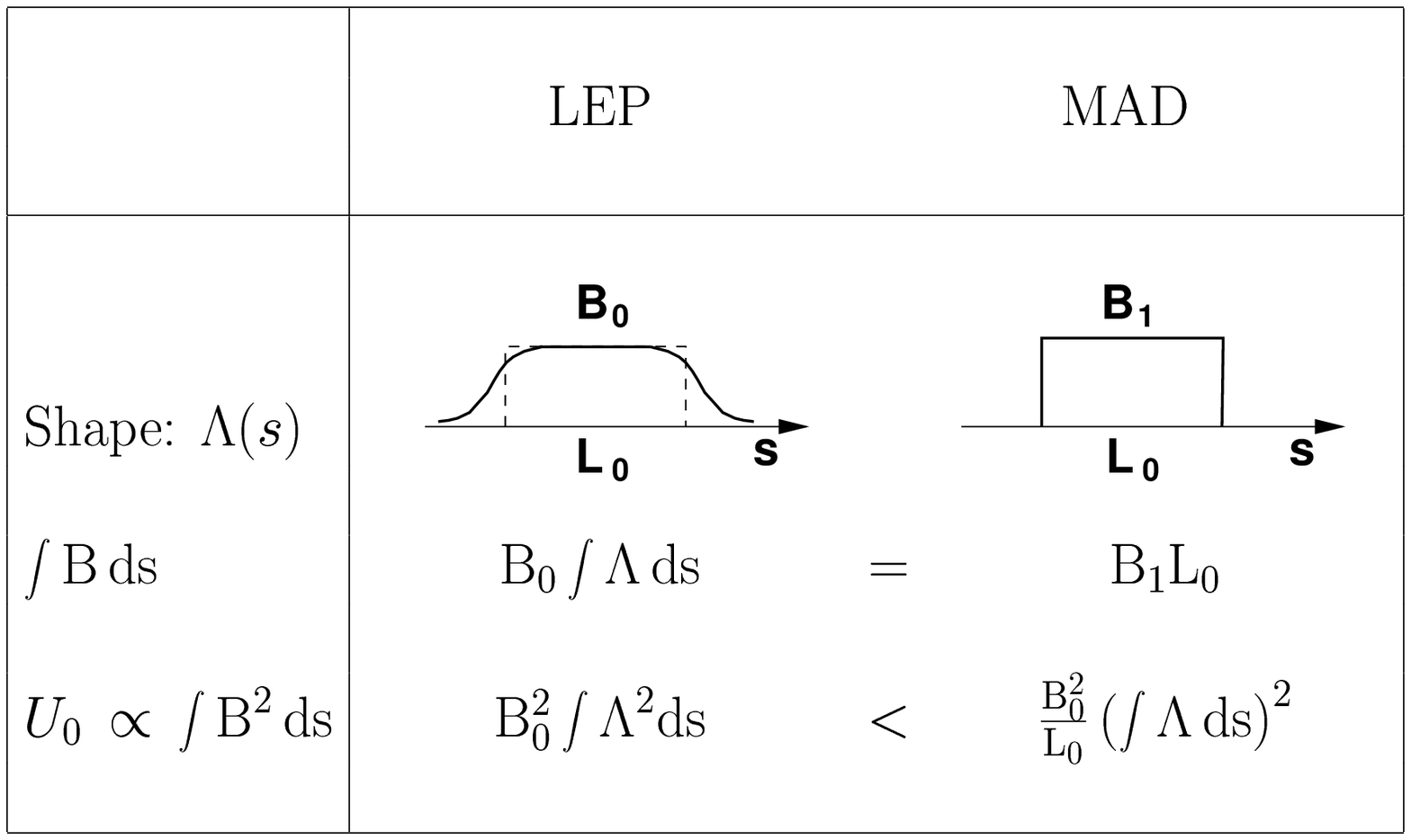,width=0.8\textwidth}
\caption[]{Calculation of the magnetic field integral and the integral over the
square of the magnetic field for a realistic magnet (LEP) and the 
treatment in the MAD program  (MAD).  
$\Lambda(s)$ represents the field distribution along the particle
trajectory; $L_0$ is the nominal magnet length;
$B_0$ is the peak field value of the realistic magnet;
$B_1$ in MAD is set to that
value which yields 
the field integral required
by the nominal beam energy.}
\label{fig:qs_fringe}
\end{center}
\end{figure}

The problem is illustrated schematically in figure~\ref{fig:qs_fringe}.
The magnetic field extends beyond the ends of each dipole, falling to
zero over a distance of the order of a meter.   The details of these
fringe fields cannot be modelled properly  by MAD.   Rather, a constant
field with zero fringe component 
is assumed for each magnet, with a magnitude tuned to agree
with the full field integral of the real dipoles.  
The energy loss, however, depends on the integral 
of the magnetic field amplitude squared,
and so is overestimated by the program.
A correction factor, $r$, is therefore present in equation~\ref{eq:eloss2}
to compensate for the MAD approximation.
 
The correction factor is determined by fitting the $Q_s$ model 
to all 1998 and 1999 datasets,
simultaneously minimising the global $\chi^2$ 
and the spread in the voltage calibration factors obtained for
one series of measurements as a function of $(1/\rho)$. 
The correction is found to be $0.9970 \pm 0.0005$.

\subsubsection{Non-Linear Synchrotron Oscillations in the 2000 Run}
\label{sec:nonlin}

\nin During the 2000 run it was only possible to achieve a measurable
$Q_s$ signal by significantly increasing the amplitude of the synchrotron 
oscillations through the application of timing jitter on the 
RF signal of selected cavities.
The expression for the oscillation frequency, equation~\ref{eq:omega2}, is 
written on the assumption that the amplitude is small.   The higher-order 
correction to this expression, necessary for the 2000 data,
is a term which shifts $Q_s$ by an amount 
$\Delta Q_s \, = \, -1/4 \,{\Delta \psi}^2 \,Q_s$, where $\Delta \psi$
is the oscillation amplitude.

%

When not accounted for, these non-linear effects lead
to an apparent energy dependence of the voltage calibration factor.
This behaviour was indeed observed in 2000.   To correct for
this bias, the parameter $\delta$ is included in the model,

\begin{equation}
Q_s^\rm{\,meas} \,=\,(1 \,+ \, \delta ) \, Q_s,
\label{eq:qsmeas}
\end{equation}           

\nin
in order to convert the measured frequency, $Q_s^\rm{\, meas}$, 
into a quantity appropriate for expression~\ref{eq:finalfit}.
$\delta$ is
extracted from a simultaneous fit to all the low-energy data points
in the 2000 run and found to be $\delta = - 0.0049 \pm 0.0016$.
This corresponds to an oscillation amplitude of $1.6$ bunch lengths,
which is compatible with what is expected from the excitation.

No evidence is  seen of non-linear behaviour in the data of earlier years,
and so for the 1998 and 1999 experiments $\delta$ is set to zero.   
An upper bound on the amplitude of natural longitudinal
oscillation comes from streak camera measurements made at LEP 1~\cite{STREAK},
which show an amplitude of $0.25$ bunch lengths.  Such an oscillation
would introduce a shift of $\delta = - 0.00012$.   In the analysis half of
this shift is applied, and half attributed as an uncertainty.

\subsubsection{The Final Parameterisation}

\begin{figure}
\begin{center}
\epsfig{file=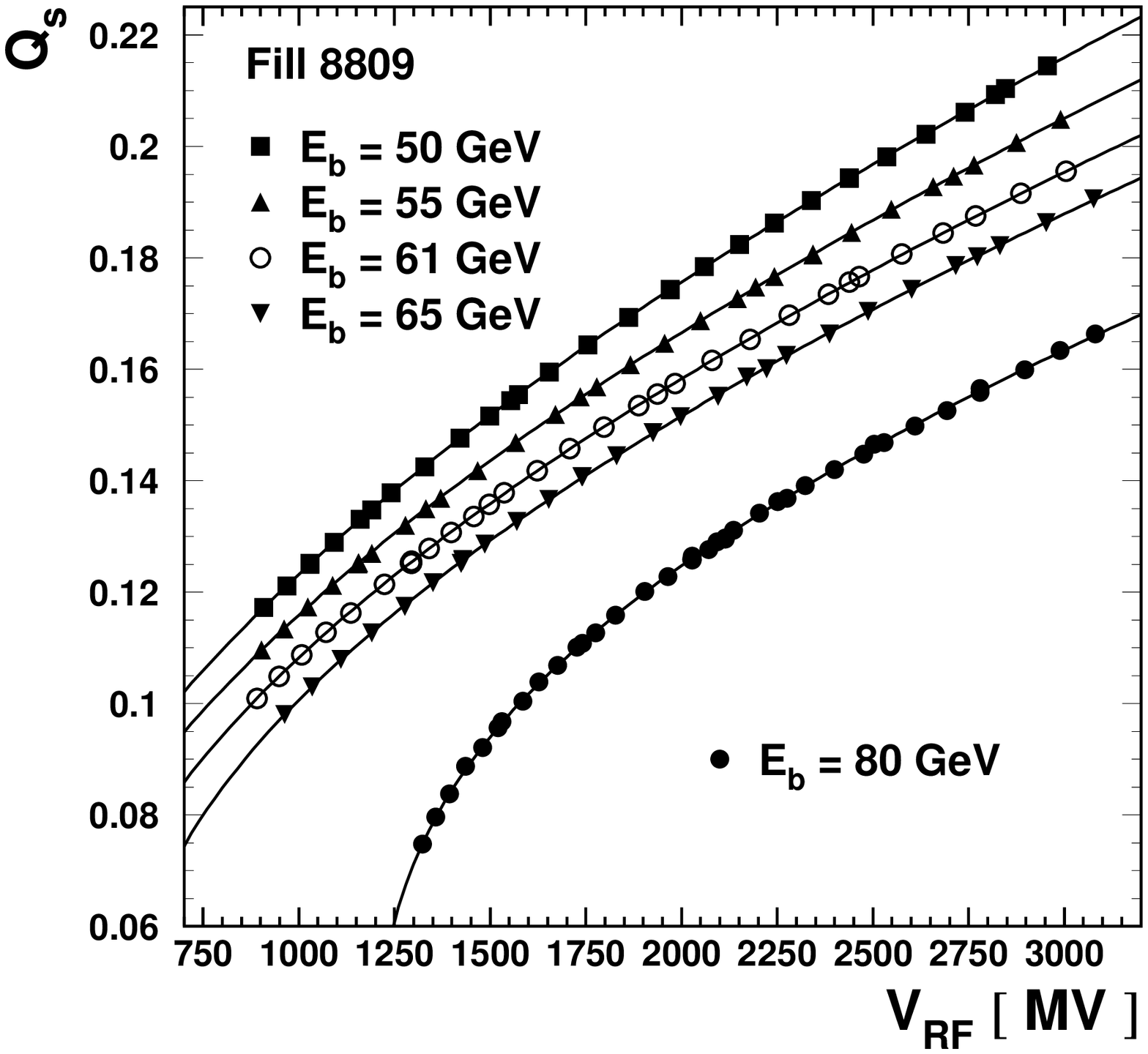,width=0.49\textwidth}
\epsfig{file=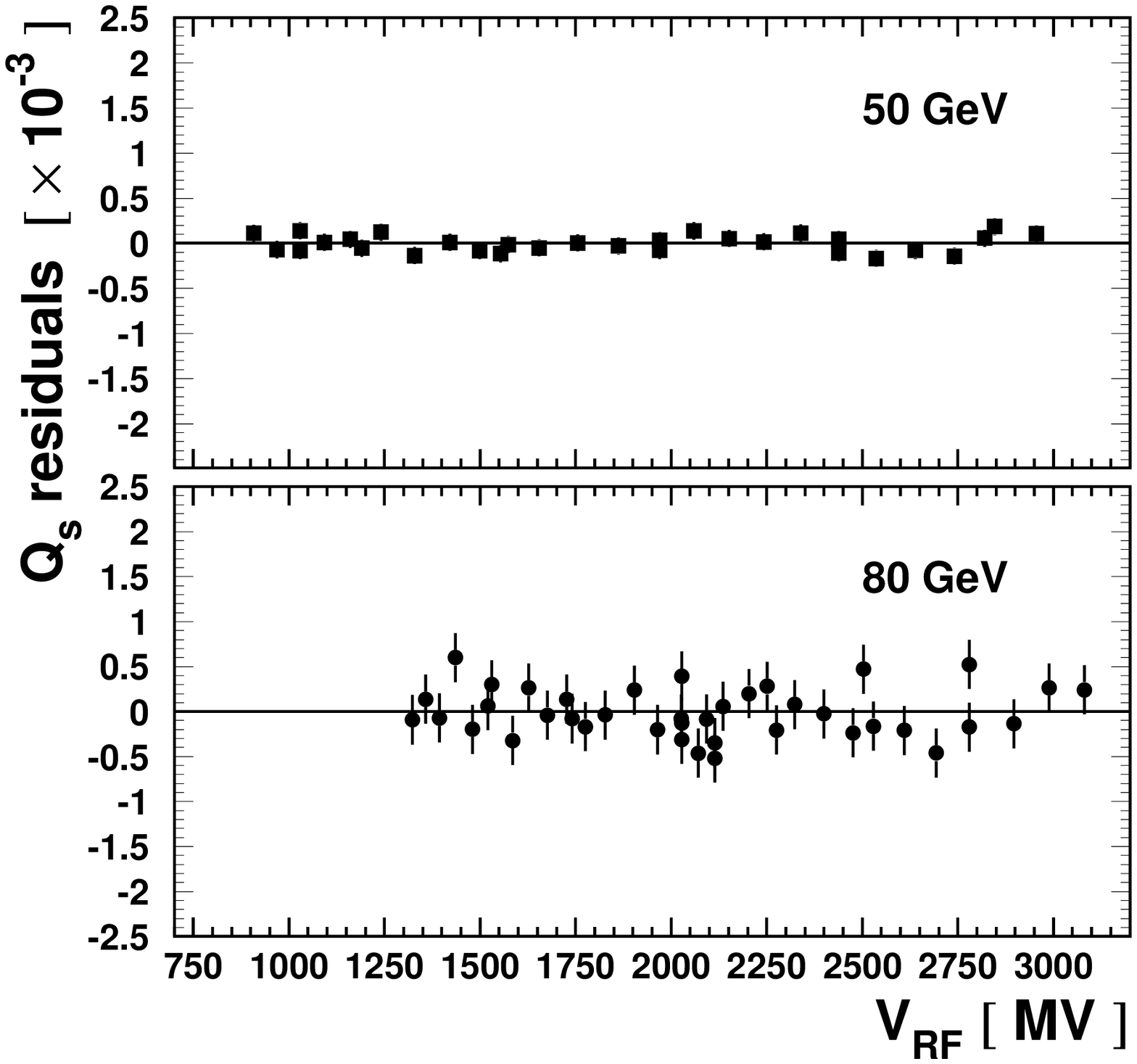,width=0.49\textwidth}
\caption[]{Data and fits from fill 8809.  Shown are the measurements
and fits as a function of $V_{\rm{RF}}$ at the five energy points,
together with the fit residuals at 
50~GeV and 80~GeV.}
\label{fig:qs_finalfit}
\end{center}
\end{figure}

Taking into account all the effects discussed,
the relationship between the measured synchrotron frequency,
the RF voltage and the beam energy 
can be expressed as  
\begin{equation}
   Q_s^4
            = \left( \frac{\alpha_c h}{2 \pi} \right)^{2}
              \left\{
                \frac{g^{2} e^{2} V^{2}_{RF}}{E^{2}_{b}} + M g^{4} V_{RF}^{4}
                - \frac{1}{E^{2}_{b}} 
                     ~\tilde{U}_{0}^{2}
              \right\},
\label{eq:finalfit}
\end{equation}
\nin
where 
$\tilde{U}_{0}$ is given by equation~\ref{eq:eloss2}.
This model describes the data well, as shown in
figure~\ref{fig:qs_finalfit}.  The increased scatter at 80~GeV
arises because the $Q_s$ signal is smaller at high energy, and therefore
is measured with less precision than at 50~GeV.
When applied to simulation data, good 
agreement is obtained between the extracted fit energy and the input
energy.


\subsection{Fit Results}

$\chi^2$ fits are made to each of the six high-energy datasets 
of 1998-2000.
The parameters fitted are
the difference between the preferred energy and the 
value from the model,
the additional contributions to the energy loss ($\sum \Delta U$), and 
the voltage calibration factor ($g$). $\sum \Delta U$ and $g$ are constrained
around their expected values with the uncertainties discussed in 
sections~\ref{sec:addeloss} and~\ref{sec:qsimprovedrf} respectively.
The input uncertainties on the individual $Q_s$ measurements are fixed from
the scatter in the fitted residuals.     

As in the case of the spectrometer measurements,  the procedure of 
normalising the analysis to low-energy reference points means that the
fit is only sensitive to non-linear systematics in $\EOnmr$,  
rather than uncertainties 
elsewhere in the model which scale with energy.  Therefore the
fitted differences are designated $\Eqs - \EOnmr$.  These are shown
in table~\ref{tab:qsresults}.  
In setting the error, 
the intrinsic precision on the $Q_s$ measurements, and the combined 
effect of the  uncertainty in $\sum \Delta U$ 
and $g$ have roughly equal weight.
Table~\ref{tab:qsresults} also lists explicit systematic error contributions
from other sources:

\begin{itemize}
\item{
The uncertainties associated with the correction to the inverse bending
radius ($r$) and with the effect of non-linear synchrotron oscillations
($\delta$) are determined by adjusting each parameter by its
assigned error, and re-evaluating the fits.  
The error induced 
by the uncertainty in $r$ is on average 7~MeV, but varies 
from experiment to experiment.  The non-linear synchrotron oscillation
correction introduces an error of $\sim$~40~MeV for the 2000
data, but is negligible for the earlier experiments.
}

\item{
In the row labelled `model imperfections' an error of 
4~MeV is assigned, to account for the fact that the values
for \EOnmr\ used in the fits come from an energy model with small
differences to that used to calculate the final physics energies.
A further contribution is 
added to this component for fill 5128, where the low-energy 
normalisation point,
at 66~GeV, is outside the range of comparison between the NMR model
and RDP.  This additional error is calculated through a polynomial fit 
to the $\Epol-\EOnmr$ residuals of figure~\ref{fig:banana},
which gives an negative offset of 7~MeV at this energy.
Fills 8445 and 8809 also include a 65~GeV energy point in the normalisation,
but the estimated uncertainty here is less than 2~MeV,
as in both cases three other low-energy points are used in the fit.
}
\item{
Finally, an estimated uncertainty of 1~\% in the momentum
compaction factor results in an error of 2~MeV for each
measurement, when propagated through the fit.
}
\end{itemize}

\begin{table}[htb]
\begin{center}
\begin{tabular}{|l|r|rr|rrr|} \hline
Year & \multicolumn{1}{|c|}{1998} & 
       \multicolumn{2}{|c|}{1999} &
       \multicolumn{3}{|c|}{2000} \\ \hline
Fill & 5128 & 6114 & 6338 & 8315 & 8445 & 8809 \\ \hline 
\Eb\  [GeV] &
       91   & 80   & 80   & 80   & 80   & 80   \\ \hline        
\Eqs~$-$~\EOnmr &
         3  &  -4  &  10  & -10  & -52  & -43  \\ \hline
Fit error &
        19  &  27  &  28  &  41  &  27  &  17  \\
Bending-radius error &
         3  &  12  &   9  &   7  &   4  &   8  \\
Non-linear oscillation error &
         1  &   3  &   3  &  45  &  26  &  48  \\
Model imperfections  &
         8  &   4  &   4  &   4  &   4  &   4  \\
Momentum compaction factor error &
         2  &   2  &   2  &   2  &   2  &   2  \\ \hline
Total error   &
        21  &  30  &  30  &  62  &  38  &  52  \\ \hline
\end{tabular}
\end{center}
\caption[]{Results of the $Q_s$ fit for the six experiments of 1998-2000.
Given is the difference between the fitted energy and the NMR model,
and the error assignment on this parameter, all in MeV.}
\label{tab:qsresults}
\end{table}

The results in table~\ref{tab:qsresults} can be combined to give
a single result for the $Q_s$ fits.
In making this combination
it is assumed that the bending-radius errors, the non-linear
oscillation errors, the momentum compaction factor errors 
and the two contributing sub-terms to
the model-imperfection errors are fully 
correlated between measurements.  
The fit errors are taken to be independent,  apart
from a 2~MeV component in common, which is the estimated
contribution coming from     
the fully correlated parasitic mode loss uncertainty.
Under these assumptions the $Q_s$ fits measure
the following offset with respect to the NMR model:
%
%
\begin{equation}
\Eqs - \EOnmr   =   -2.8 \, \pm \, 15.8 \: \rm{MeV},
\nonumber
\end{equation}

\noindent at a nominal \Eb\ of 85.2~GeV.  The $\chi^2$
of this combination is 2.7 for 5 degrees of freedom.

%% file: eb_comb.tex
\section{Combined Analysis of the \Eb\ Measurements}
\label{sec:ebcom}

The flux-loop, the spectrometer and the $Q_s$ fits provide three
independent tests of the NMR magnetic model.  Each method
measures an offset between the true energy and the NMR prediction
at one or more energies, as summarised in table~\ref{tab:extrap_input}.
All measurements are consistent with the NMR model.  
Under the assumption that all methods are measuring the
same quantity at different values of \Eb, the three sets of results 
may be combined to give an improved estimate of the offset
as a function of energy.  

\begin{table}[htb]
\begin{center}
\begin{tabular}{|l|c|c|c|c|} \hline
Method       &   \Eb\ [GeV]   &  $E_\rm{b}^\rm{MEAS} - \EOnmr$  [MeV]  &  
Correlation &  Period \\ \hline
& & & & \\
             &  
\multicolumn{1}{|l|}{ 72 } &  $-1.7 \pm 7.5$   &  &   \\
Flux-loop    & $\rightarrow$  & & 100 \% & 1997--2000 \\
             &
\multicolumn{1}{|r|}{ 106 } &  $-6.0 \pm 17.6$          &   & \\ 
& & & & \\ \hline
& & & & \\ 
  &  70         &  $-0.6 \pm 9.7$     & &   \\
Spectrometer & & & 75\% & 2000\\
             &  92.3       &  $-4.9 \pm 17.9$    & & \\ 
& & & & \\ \hline
& & & & \\ 
$Q_s$ vs $V_\rm{RF}$ &  85.2    & $-2.8 \pm 15.6$ & /  & 1998--2000 \\ 
& & & & \\  \hline
\end{tabular}
\end{center}
\caption[]{Summary of results from the tests of the NMR model.}
\label{tab:extrap_input}
\end{table}


An initial combination of the results in table~\ref{tab:extrap_input}
can be made
under the simple assumption that any non-linearity does not evolve 
significantly over the \Eb\ span of the LEP~2 datasets.
In performing this average,
the 70~GeV spectrometer point is discarded as being outside
the physics regime, the flux-loop results 
are represented by a single point at \Eb~=100~GeV and
any year-to-year variation in the NMR model is neglected.
The input values are very consistent within their assigned errors.
The result of this average is $-3.5 \pm 9.4~\rm{MeV}$.

As the NMR model is normalised to agree with the true energy around 50~GeV,
any offset observed at higher values of \Eb\ must have some energy
dependence.
Therefore all measurements have been included
in a two-parameter fit 
in order to determine the combined offset with respect to the  NMR model
as a function of energy.  
In order to use the data optimally,
the six $Q_s$ measurements, together with their
assigned covariances, are entered separately.   The range of flux-loop results
are represented by two fully-correlated measurements at 72~GeV 
and at 106~GeV.  

Because of small changes in probe positions, and magnet ageing, the
NMR calibration can vary from year to year. 
A test of the NMR model is therefore only valid for the year in which
it was performed. 
In practice however, with the exception of 1996, the year-to-year variation 
appears to be very small, with an upper bound given by the entries of 
table~\ref{tab:cal_stab}.
This possible variation is accommodated in the 
combination as follows.  The model, against which the offset and slope are
fitted, is considered to be that averaged over the years 1997--2000.
In order to account for a possible difference between this mean NMR-model 
and the year-specific models,
terms of $(2 \:\rm{MeV})^2$ are added to the error matrix for each of the 
$Q_s$ and  spectrometer entries,  with full correlations 
between measurements from the same year.
As the flux-loop analysis is based on data rather evenly distributed 
throughout 1997--2000, its errors are left unchanged.

\begin{figure}
\begin{center}
\epsfig{file=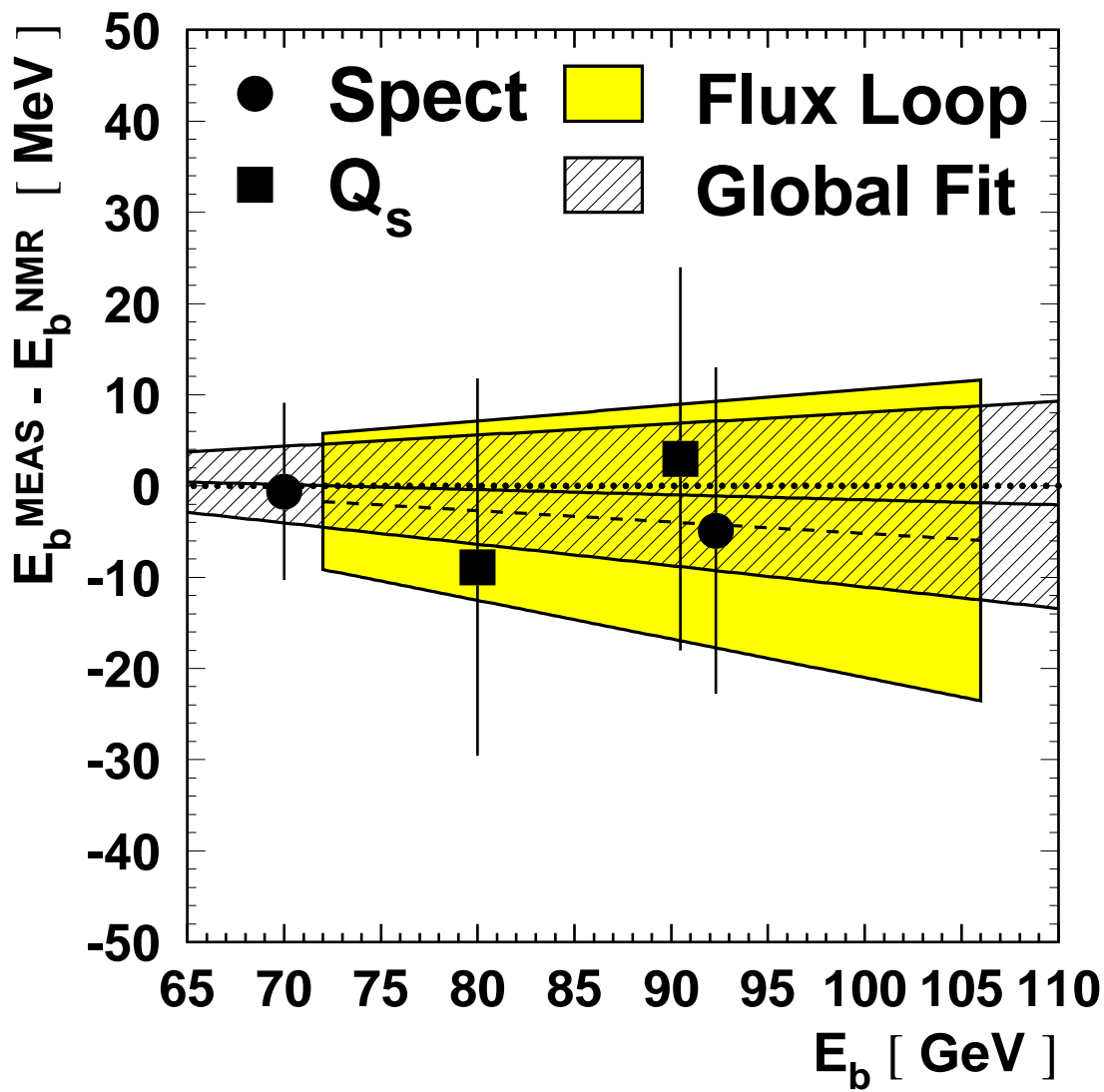,width=1.0\textwidth}
\end{center}
\caption[]{The difference between the measurements of \Eb\ and the NMR model prediction,
together with the results of the two-parameter fit.  In the fit, the $Q_s$ data
consist of five measurements at \Eb~=~80~GeV and one more precise measurement at
\Eb~=~90~GeV.  The 80~GeV measurements have been combined into a single point in
this figure.
Measurements made with the same method have correlations which are discussed in the text.}
\label{fig:ebcombres}
\end{figure}

The fit returns an offset at \Eb~=~100~GeV of $-1.5 \pm 9.6$~MeV 
and slope with \Eb\ of $-0.06 \pm 0.18$~MeV/GeV.  
The offset and accompanying error at the 
running points of LEP~2 are given in table~\ref{tab:ebcombres}.
The measurements and fit result are shown graphically in figure~\ref{fig:ebcombres}.
Because of correlations in the input data, the central value of the fit
in the regime of interest
is slightly higher than would be the case if all the measurements were
independent.

\begin{table}[htb]
\begin{center}
\begin{tabular}{|l|cccccccccc|} \hline
             & \multicolumn{10}{|c|}{ \Ecmn\ [GeV] } \\ \cline{2-11}
             & 161  & 172  & 183  & 189  & 192  & 196  & 200  & 202  & 205   & 207   \\ \cline{2-11}
\Eb\ offset [MeV] & -0.4 & -0.7 & -1.0 & -1.2 & -1.3 & -1.4 & -1.5 &  -1.6 &  -1.6 &  -1.7 \\
Error  [Mev]      &  6.1 &  7.1 &  8.0 &  8.6 &  8.8 &  9.2 &  9.5 &   9.7 &  10.0 &  10.2 \\ \hline
\end{tabular}
\end{center}
\caption[]{Results of the fit to the mean NMR-model, applied at the 
LEP~2 energy points.   (These results have been evaluated at the 
luminosity-weighted energies calculated
by the model, rather than the nominal \Ecm\ values displayed.)}
\label{tab:ebcombres}
\end{table}

The fit has been repeated with various subsets of the input data.
These include the flux-loop and spectrometer alone;
the flux-loop and $Q_s$ alone; the spectrometer and $Q_s$ alone;
the standard set without the contribution of the single 90~GeV 
$Q_s$ measurement; and the standard set without the 70~GeV spectrometer
input.
The results and errors, at three illustrative running points,
are shown in table~\ref{tab:ebcombvar}, together with the results
of the standard fit and the results of the flux-loop analysis.  
It can be seen that the three methods
have comparable weight and produce similar results.
The combined measurements of the  spectrometer and $Q_s$ 
provide a more precise result than the
more indirect method of the flux-loop.
The $Q_s$ data have significant weight in the fit even
when the single most precise measurement at 90~GeV is
excluded.
Dropping the 70~GeV spectrometer fit point from the combination
leads to a very small degradation in precision.
In all cases the results support the mean NMR-model over the full range
of energies.

\begin{table}[htb]
\begin{center}
\begin{tabular}{|l|cc|ccc|} \hline
                   &  Offset & Slope     & \multicolumn{3}{|c|}{Error [MeV] at \Ecmn\ of: }   \\
Variant             & [MeV]  & [MeV/GeV] &     161  GeV  &    200  GeV   &   207   GeV  \\ \hline
FL, Spect and $Q_s$ &  -1.5  &  -0.06    &      6.1      &     9.5       &    10.2      \\
FL and Spect        &  -5.1  &  -0.12    &      8.0      &    12.4       &    13.3      \\
FL and $Q_s$        &  -0.1  &  -0.03    &      6.9      &    10.8       &    11.5      \\
Spect and $Q_s$     &  -0.9  &  -0.02    &     11.2      &    14.5       &    15.1      \\
FL alone            &  -5.2  &  -0.13    &     10.1      &    15.7       &    16.8      \\
No 90~GeV $Q_s$     &  -1.8  &  -0.06    &      6.4      &    10.0       &    10.7      \\ 
No 70~GeV Spect     &  -1.7  &  -0.06    &      6.2      &     9.6       &    10.2      \\ \hline
\end{tabular}
\end{center}
\caption[]{Fitted offset at 100~GeV and slope with energy,
and errors on the offset at three illustrative LEP~2 energy points,
for various sub-sets of input data.  Also shown are the
input measurements for the flux-loop alone.
(Note that the errors have been evaluated at the luminosity 
weighted energies calculated
by the model, rather than the \Ecmn\ values displayed.)}
\label{tab:ebcombvar}
\end{table}

To go from the errors on the results of the mean NMR-model
to those appropriate for each energy point
an uncertainty of  2~MeV is
added, in common between the energy points
of 1999,  otherwise uncorrelated, to allow for
year-to-year variation in the model from
changing calibration coefficients.  The value of this component
is once more
motivated by the typical size of entries in the bottom row of 
table~\ref{tab:cal_stab}.

It is known that the change in calibration is 
much larger between 1996 and the later years, and 
so the exact results of the mean model fit are not 
applicable to the earlier dataset. 
Nevertheless, the fit demonstrates clearly that
the magnetic model procedure is not subject to any 
significant non-linearity.  Therefore, for the
two 1996 energy points
additional contributions of 
9.6~MeV and 10.3~MeV are
added to the mean model errors, these being the
{\it statistical} uncertainties on
the individual model for this year.
The values are derived from the observed RMS scatters
of the 16 individual values of \Enmri\ in
expression~\ref{eq:epolcor}, at collision energies of 161 and
172~GeV respectively.

The errors on \Ecmnmr\ associated with the NMR model for each energy
point of LEP~2 are presented in the first row of entries in 
table~\ref{tab:systsum}.   These are twice those values calculated 
for \Eb.

%% file: syst_sum.tex
\section{Summary of Results and Systematic Uncertainties}
\label{sec:systsum}
\subsection{High-Energy Analysis}

The results from the combination of the NMR tests
as presented in table~\ref{tab:ebcombres} are corrections
which must be applied to the output, \Ecmnmr, of the energy model.
The errors on these
corrected values come from considering the errors
from the NMR model, including year-to-year variation,
the  additional model uncertainties discussed in sections~\ref{sec:eb_model}
and~\ref{sec:ecm_ip}, and the small uncertainty arising
from the RDP measurement itself.   These errors are given
in table~\ref{tab:systsum} for each nominal energy point.
The dominant component for all years is the NMR model
error, apart from in 2000 where the BFS uncertainty is
more important.  The next largest
contribution comes from the error on the RF sawtooth,
which is 8--10~MeV.
Table~\ref{tab:systcor} shows the
accompanying correlation matrix.  For almost all energy points
taken after 1996 the correlation is close to 100\%.
The correlation between the 2000 errors and those of the earlier
years is less because of the BFS.

\begin{table}
\begin{center}
\begin{tabular}{|l|r|r|r|r|r|r|r|r|r|r|} \hline\
\Ecmn\  [GeV] &
\multicolumn{1}{|c|}{161} & 
\multicolumn{1}{|c|}{172} & 
\multicolumn{1}{|c|}{183} & 
\multicolumn{1}{|c|}{189} & 
\multicolumn{1}{|c|}{192} & 
\multicolumn{1}{|c|}{196} & 
\multicolumn{1}{|c|}{200} & 
\multicolumn{1}{|c|}{202} & 
\multicolumn{1}{|c|}{205} & 
\multicolumn{1}{|c|}{207} \\  \hline 
 NMR model            & 22.8 & 25.0 & 16.5 & 17.6 & 18.1 & 18.8 & 19.5 & 19.8 & 20.4 & 20.7 \\
 RDP                  &  1.0 &  1.0 &  1.0 &  1.0 &  1.0 &  1.0 &  1.0 &  1.0 &  1.0 &  1.0 \\
 $f^\rm{RF}_c$        &  0.0 &  0.0 &  5.4 &  5.6 &  5.8 &  5.8 &  6.0 &  6.0 &  0.0 &  0.0 \\
 $\alpha_c$           &  0.3 &  0.4 &  3.5 &  4.4 &  4.4 &  5.2 &  4.7 &  3.0 &  2.3 &  1.4 \\
 $\Delta \Eb$ in fill &  1.0 &  1.0 &  1.0 &  1.0 &  1.0 &  1.0 &  1.0 &  1.0 &  1.0 &  1.0 \\
 Hcor/BFS             &  1.6 &  1.8 &  3.4 &  4.6 &  0.6 &  1.0 &  0.2 &  0.6 & 28.6 & 34.4 \\
 QFQD                 &  1.4 &  1.4 &  0.6 &  0.6 &  0.6 &  0.8 &  0.8 &  0.8 &  0.8 &  0.8 \\
 RF sawtooth          & 10.0 & 10.0 &  8.0 &  8.0 &  8.0 & 10.0 & 10.0 & 10.0 & 10.0 & 10.0 \\
 $e^+e^-$ difference  &  4.0 &  4.0 &  4.0 &  4.0 &  4.0 &  4.0 &  4.0 &  4.0 &  4.0 &  4.0 \\
 Dispersion           &  2.0 &  2.0 &  2.0 &  2.0 &  2.0 &  2.0 &  2.0 &  2.0 &  2.0 &  2.0 \\ \hline
 Total                & 25.4 & 27.4 & 20.3 & 21.6 & 21.6 & 23.2 & 23.7 & 23.7 & 36.9 & 41.7 \\ \hline
\end{tabular}
\end{center}
\caption[]{Summary of systematic errors on \Ecmnmr, in MeV, at all nominal energy points.}
\label{tab:systsum}
\end{table}

\begin{table}
\begin{center}
\begin{tabular}{c|cccccccccc}       
\Ecmn\        [GeV] & 161 & 172 & 183 & 189 & 192 & 196 & 200 & 202 & 205 & 207 \\ \hline
   161 &   1.00 &   1.00 &   0.57 &   0.56 &   0.57 &   0.57 &   0.58 &   0.58 &   0.38 &   0.34 \\
   172 &   1.00 &   1.00 &   0.58 &   0.57 &   0.58 &   0.59 &   0.59 &   0.59 &   0.39 &   0.35 \\
   183 &   0.57 &   0.58 &   1.00 &   0.94 &   0.95 &   0.95 &   0.95 &   0.94 &   0.57 &   0.51 \\
   189 &   0.56 &   0.57 &   0.94 &   1.00 &   0.94 &   0.94 &   0.94 &   0.93 &   0.57 &   0.50 \\
   192 &   0.57 &   0.58 &   0.95 &   0.94 &   1.00 &   1.00 &   1.00 &   0.99 &   0.58 &   0.52 \\
   196 &   0.57 &   0.59 &   0.95 &   0.94 &   1.00 &   1.00 &   1.00 &   0.99 &   0.58 &   0.52 \\
   200 &   0.58 &   0.59 &   0.95 &   0.94 &   1.00 &   1.00 &   1.00 &   1.00 &   0.59 &   0.52 \\
   202 &   0.58 &   0.59 &   0.94 &   0.93 &   0.99 &   0.99 &   1.00 &   1.00 &   0.59 &   0.53 \\
   205 &   0.38 &   0.39 &   0.57 &   0.57 &   0.58 &   0.58 &   0.59 &   0.59 &   1.00 &   0.99 \\
   207 &   0.34 &   0.35 &   0.51 &   0.50 &   0.52 &   0.52 &   0.52 &   0.53 &   0.99 &   1.00 \\
\end{tabular}
\end{center}
\caption[]{Correlation matrix for errors on \Ecmnmr\ at all nominal energy points.}
\label{tab:systcor}
\end{table}

\subsection{Uncertainty for \Zz\ Runs and Lower-Energy Data}

The energy  model has been used to calculate collision
energies for the fills at the \Zz\ resonance,
scheduled
for the purposes of providing calibration data for the experiments.
As the energy scale of 
\Zz\ running is directly set by RDP, the NMR model
is no longer a source of uncertainty.
Most other error sources are also 
smaller at these energies.
An upper bound of 10~MeV can be assigned as the 
total error on \Ecmnmr\ for \Zz\ operation during 
the LEP~2 programme.

The uncertainty on \Ecmnmr\ for the 130-136~GeV running in 1997 is 
conservatively assumed to be the same as for the higher-energy operation in
that year,  and so it is set to 20~MeV.

%% file: ecm_spread.tex
\section{Centre-of-mass Energy Spread}
\label{sec:ecmspread}

The spread in centre-of-mass energy is relevant for evaluating
the width of the W boson, which is about 2~GeV and is 
measured with the full LEP~2 
dataset with a statistical precision of 
around 70~MeV~\cite{EWWGREP}.   The spread of the beam
energy, $\sigma_{\Eb}$, 
varies as $\Eb^2$,  with an optics-dependent correction
associated with any RF frequency shift.   The value of the spread
has been calculated to accompany each energy record distributed to
the experiments.  In order to obtain the centre-of-mass energy spread,
$\sigma_{\Ecm}$,
it is necessary to multiply  $\sigma_{\Eb}$ by $\sqrt{2}$.
 The luminosity-weighted values of $\sigma_{\Ecm}$ are shown
in table~\ref{tab:ecmspread} for each nominal energy point.
The decrease in $\sigma_{\Ecm}$ seen for 202~GeV and above
is because of the smaller frequency shifts applied at these running
points. 

\begin{table}[htb]
\begin{center}
\begin{tabular}{|c|c|} \hline
\Ecmn\ [GeV] & $\sigma_{\Ecm}$ [MeV] \\ \hline
161 & $144 \pm 7$ \\
172 & $165 \pm 8$ \\
183 & $218 \pm11$ \\
189 & $236 \pm12$ \\
192 & $255 \pm13$ \\
196 & $265 \pm13$ \\
200 & $264 \pm13$ \\
202 & $250 \pm12$ \\
205 & $236 \pm 24$ \\
207 & $235 \pm 24$ \\ \hline 
\end{tabular}
\caption[]{Luminosity-weighted centre-of-mass energy spreads, 
$\sigma_{\Ecm}$.
}
\label{tab:ecmspread}
\end{center}
\end{table}

Measurements of the longitudinal
bunch length at an interaction point, performed with the 1996 and 1997 data,
have been used in conjunction with the measured values of  $Q_s$ to
make indirect determinations of $\sigma_{\Eb}$, as reported
in~\cite{LEP2ECAL}.  These results agree well with the calculated values
given to the experiments.   

The error on the calculated energy spread in 1996--1999 is estimated to be
about $5\%$, fully correlated between years.  This value is assigned 
from the differences observed with respect to the result of the 
analytic calculation,
when a simulation of the photon emission
process is implemented.  
The error is 10\% in 2000,  
because of additional uncertainties associated with the BFS.
The corresponding uncertainty on the W width is negligible.

%% file: conclusions.tex
\section{Conclusions}
\label{sec:conclude}

The method of energy determination,  based on the NMR magnetic model
calibrated through resonant depolarisation,  has enabled the collision
energies to be calculated for all LEP~2 running.    Three independent
methods have been used to verify the linearity of this calibration at high
energy.   Uncertainties on other ingredients in the energy model
have been assigned, benefitting from the detailed understanding acquired
during the LEP~1 Z resonanace scans, and from subsequent measurements.
The total uncertainty for each energy point is presented in
table~\ref{tab:systsum}, and the corresponding correlation matrix is
given in table~\ref{tab:systcor}.  For the majority of the data, collected 
in the 1997-1999 runs, the relative uncertainty is 
$1.1 - 1.2 \times 10^{-4}$. For the operation in 2000 this error rises
to $2.0 \times 10^{-4}$ at the highest energies. This increase is driven 
by the uncertainty associated with the spreading of the bending field 
applied in order to raise the maximum beam energy.

The error induced on \Mw\ from the collision energy uncertainty depends on the
point-to-point correlations, the relative
statistical uncertainties at these points,  and the correlations in
the other systematic errors contributing to \Mw.    With the presently
available preliminary results~\cite{EWWGREP} the error on \Mw\ from
the collision energy is determined to be 
around 10~MeV/$c^{\rm{2}}$~\cite{CHRISP}.
This contribution is small compared with the statistical uncertainty on
the \Mw\ measurement.

\section*{Acknowledgements}

We are grateful for the careful work and help of many people in
SL Division, which was essential in making measurements for the
energy calibration. The design, construction and installation of
the spectrometer involved the efforts and expertise of numerous 
engineers and technicians now in TS Division.

We also acknowledge the support of the  
Particle Physics and Astronomy Research Council,  UK, and
the National Science Foundation, USA.